\documentclass[10pt,oneside,a4paper,onecolumn]{elsarticle}
\pdfoutput=1
\usepackage{times}
\usepackage{fullpage}

\usepackage{gensymb}
\usepackage{tikz}
\usepackage{alltt}
\usepackage{upquote}
\usepackage{float}
\usepackage{array}
\usepackage{pgfplots}
\usetikzlibrary{positioning}
\usepackage{graphicx}
\usepackage[toc,page]{appendix}
\usepackage{amsmath,mathtools}
\usetikzlibrary{calc,decorations.markings,fit,shapes,chains,arrows}
\usepackage{amsthm}
\usepackage{amsfonts}
\usepackage{enumitem}
\usepackage{subcaption}
\usepackage{caption}
\usepackage{xcolor}
\usepackage{scalerel}
\usepackage[hidelinks=true]{hyperref}
\usepackage{algorithm}
\usepackage{algorithmic}
\usepackage{multicol}
\usepackage{mathtools}
\usepackage{url}

\tikzset{
  block/.style={rectangle, draw, rounded corners, text centered,text width = 16em, minimum height = 2em},
  line/.style={draw, -latex'}
  }
\tikzset{
  block2/.style={text centered,text width = 22em, minimum height = 2em},
  line/.style={draw, -latex'}
  }
\tikzset{
  block3/.style={rectangle, draw, rounded corners, text centered,text width = 19em, minimum height = 1em},
  line/.style={draw, -latex'}
  }
\tikzset{
  block4/.style={rectangle, draw, rounded corners, text centered,text width = 15em, minimum height = 1em},
  line/.style={draw, -latex'}
  }
\tikzset{
  blockNS/.style={rectangle, draw, fill=black!20, rounded corners, text centered,text width = 20em, minimum height = 2em, label={center:Navier-Stokes solver}},
  line/.style={draw, -latex'}
  }
\tikzset{
  blockAC/.style={rectangle, draw, fill=black!20, rounded corners, text centered,text width = 20em, minimum height = 2em, label={center:Allen-Cahn solver}},
  line/.style={draw, -latex'}
  }
\tikzset{  
  decision/.style = {diamond, draw, minimum width=4cm, minimum height=0.2cm},
  line/.style={draw, -latex'}
  }

\newcolumntype{M}[1]{>{\centering\arraybackslash}m{#1}}
\newcolumntype{N}{@{}m{0pt}@{}}

\tikzset{>=latex}
\graphicspath{ }
\def\@author#1{\g@addto@macro\elsauthors{\normalsize%
    \def\baselinestretch{1}%
    \upshape\authorsep#1\unskip\textsuperscript{%
      \ifx\@fnmark\@empty\else\unskip\sep\@fnmark\let\sep=,\fi
      \ifx\@corref\@empty\else\unskip\sep\@corref\let\sep=,\fi
      }%
    \def\authorsep{\unskip,\space}%
    \global\let\@fnmark\@empty
    \global\let\@corref\@empty
    \global\let\sep\@empty}%
    \@eadauthor={#1}
}

\graphicspath{{Pictures_v2/}}

\begin{document}
\begin{frontmatter}
\title{An adaptive variational procedure for the conservative and positivity preserving Allen-Cahn phase-field model}
\author[nus]{Vaibhav Joshi}
\ead{vaibhav.joshi16@u.nus.edu}

\author[nus]{Rajeev K. Jaiman\corref{cor1}}
\ead{mperkj@nus.edu.sg}
\cortext[cor1]{Corresponding author}
\address[nus]{Department of Mechanical Engineering, National University of Singapore, Singapore 119077}

\begin{abstract}
\noindent In this paper, we present an adaptive variational procedure for unstructured meshes to capture fluid-fluid interfaces in two-phase flows. 
The two phases are modeled by the phase-field finite element formulation, which involves the conservative Allen-Cahn equation coupled with the incompressible Navier-Stokes equations. 
The positivity preserving variational formulation is designed to maintain the bounded and stable solution of the Allen-Cahn equation. 
For the adaptivity procedure, we consider the residual-based error estimates for the underlying differential equations of the two-phase system.
In particular, the adaptive refinement/coarsening is carried out by the newest vertex bisection algorithm by evaluating the residual error indicators based on the error estimates of the Allen-Cahn equation. 
The coarsening algorithm avoids the storage of the tree data structures for the hierarchical mesh, thus providing the ease of  numerical implementation.
Furthermore, the proposed nonlinear adaptive partitioned procedure aims at reducing the amount of coarsening while maintaining the convergence properties of the underlying nonlinear coupled differential equations.
We investigate the adaptive phase-field finite element scheme through the spinodal decomposition in a complex curved geometry and the volume-conserved interface motion driven 
by the mean curvature flow for two circles in a square domain.
We then assess the accuracy and efficiency of the proposed procedure by modeling the free-surface motion in a sloshing tank. 
 In contrast to the non-adaptive Eulerian grid counterpart, we demonstrate that the mesh adaptivity remarkably reduces the degrees of freedom and the computational cost by nearly half for similar accuracy. The mass loss in the Allen-Cahn equation via adaptivity process is also  reduced by nearly three times compared to the non-adaptive mesh. 
Finally, we apply the adaptive numerical framework to solve the application of a dam-breaking problem with topological changes. 

\end{abstract}
\end{frontmatter}

\section{Introduction}
Multiphase flow is a highly complex multi-scale nonlinear phenomenon, which is encountered in various applications ranging from small-scale droplet interactions \cite{Khatavkar}, microstructural phase evolution in materials \cite{Fried}, flows in oil and natural gas pipelines \cite{jfm_phasefield_mit} to large scale free-surface ocean waves and wave breaking \cite{jcp_breakingwave}. Such kinds of fluid flows can be quite complex to understand and analyze via physical experiments and theoretical ways. The kinematic and dynamic effects of these flows can directly impact the design conditions of industrial systems, which makes their study essential from both experimental and computational standpoint. Of particular interest to the present study is the two-phase modeling of air-water interface with its application to free-surface motion.

While the physical experiments to understand these phenomena can be expensive to set-up and investigate, numerical simulations offer an economical choice. Some of the challenges for the two-phase modeling of air-water interface are the complexity in the representation of the interface, the mass conservation, and the treatment of the discontinuities in the physical quantities such as density, viscosity, pressure and velocity across the interface. Numerical treatment of the modeling of two-phase flows can be broadly categorized into interface tracking \cite{Unverdi,donea} and interface capturing techniques \cite{Osher, Hirt, Kim}. 
The interface tracking methods such as the arbitrary Eulerian Lagrangian (ALE) are  accurate for flexible moving boundaries with small deformations, but 
they pose some difficulties to use when the topology of the interface changes (such as breaking and merging) dramatically.  Furthermore, there is a need for significant remeshing when 
there is a large deformation of fluid-fluid interface.
While interface capturing methods can handle breaking and merging of the interface in a simple manner compared to interface tracking, the absence of prior knowledge of the location of the convecting fluid interface forces the user to refine the computational grid throughout the domain. This in turn increases the number of unknowns, increasing the computational cost for a desired spatial accuracy of the underlying fluid-fluid interface problem. Refining the computational mesh adaptively with the evolving interface during the run-time is a promising way to reduce the degrees of freedom and to increase accuracy and efficiency of interface capturing methods. In this work, we consider the interface-driven adaptivity to minimize errors during the stabilized finite element computations of two-phase flows using the Navier-Stokes and Allen-Cahn phase field equations.

Some of the refinement strategies are mesh movement or repositioning ($r$-methods), mesh enrichment ($h/p$-methods) and adaptive remeshing ($m$-methods) \cite{Lohner_adaptivity}. Among these methods, $h$-refinement is particularly attractive and deals with the addition of extra degrees of freedom near a region of interest in the computational domain and $p$-refinement elevates the order of approximation by using higher order polynomials. The $h$-refinement is preferred for the current study since it maintains conservation of the physical quantity and can be easily vectorized and parallelized. It can be carried out by either subdivision of the elements into equal parts or by recursive subdivision of the largest edge side of the element. To optimize the degrees of freedom and underlying mesh, the error indicators based on \textit{a posteriori} residual estimates are constructed. The indicators can be based on gradients of the concerned quantities or the residuals of the equation being solved. Some of the commonly used error estimators are residual-based, hierarchical-basis error estimates \cite{Bank} and Zienkiewicz-Zhu \cite{ZZ} error estimators. A review of the mathematical theory behind the error estimates derived for a variety of equations can be found in \cite{Verfurth_1, Verfurth_2, Verfurth}. Basic strategies and algorithms for adaptive methods from a computational perspective are discussed in \cite{Schmidt}.

Of particular interest to the present study is the $h$-refinement adaptivity scheme for the phase-field Allen-Cahn equation to capture the fluid-fluid interface in complex two-phase flows. Some of the works dealing with the error estimates for the non-conservative Allen-Cahn equation are \cite{Kessler, Bartels_1, Georgoulis, Bartels_2, Feng2005, Zhang_2}. The mesh refinement of the conservative Allen-Cahn equation coupled with Navier-Stokes equations has been the topic of some of the recent works \cite{Vasconcelos, Zhang}. One of the recent works in \cite{Vasconcelos} utilized the open-source library \cite{libMesh} for the adaptive mesh refinement, which utilizes a tree data structure to store the refinement hierarchy while a mesh distribution strategy was used in \cite{Zhang}. A tree-type data structure for the hierarchical mesh can cover a good amount of the computational storage space. The algorithm of \cite{Chen_3} avoids such data structures thereby easing the implementation of the algorithm. Moreover, most of the past literature has dealt with adaptivity algorithms using structured meshes, restricting the use of unstructured grids for practical geometries. While maintaining the ease of implementation, the current work is an attempt to address these issues and propose a stable and efficient adaptive finite element algorithm for unstructured grids with minimal data structures.

The present study builds upon our previous work of  \cite{AC_JCP}, where a conservative and energy stable variational scheme for the Navier-Stokes 
and Allen-Cahn system is presented. The recently proposed positivity preserving variational (PPV) technique \cite{PPV} is employed to maintain 
the stable and bounded solution of the coupled nonlinear differential system without any operator splitting of Allen-Cahn phase field equation.
In the present work, we successfully integrate the incompressible Navier-Stokes and the Allen-Cahn equations and implement
the mesh adaptivity process based on the residual error estimates of the Allen-Cahn equation to capture the interface between the two phases. 
To conserve the mass for evolving interface via the Allen-Cahn equation, we consider the variational formulation by means of time and space dependent Lagrange multiplier. An adaptive algorithm based on the newest vertex bisection method \cite{Chen_2} is employed for unstructured triangular meshes. This algorithm avoids the data structure to store the information about the refinement and coarsening nodes. The present procedure referred to as nonlinear adaptive variational partitioned (NAVP)  solves the underlying coupled differential equations in a partitioned manner while maintaining positivity in the solution. The convergence criterion is aimed to reduce the amount of coarsening while maintaining the convergence properties of the coupled discretized equations for the incompressible two-phase flow dynamics. 

This article makes a significant contribution to the development of a stable, robust and general nonlinear adaptive variational procedure for solving 
the coupled Navier-Stokes and Allen-Cahn equations to model two-phase flows via phase-field modeling.
 For arbitrary unstructured meshes, the novel features of the proposed adaptive procedure are: positivity preservation, energy stability, and mass conservation.
We perform the implicit temporal discretizations of the system which helps to decouple the Navier-Stokes and the Allen-Cahn equations. In addition to using larger time steps, the implicit formulation allows to deal with the fluid-fluid interface problem in a straightforward manner via partitioned iterative solution strategy. The proposed NAVP procedure aims at reducing the error introduced by the mesh adaptivity followed by the reduction of the nonlinear errors while solving the fully-coupled Navier-Stokes and the Allen-Cahn equations. 
The present adaptivity procedure has two salient features: (i) it delays the coarsening step to the end of the nonlinear iterations, thereby reducing the nonlinear errors, and
(ii) the absence of the tree data structure for the refinement/coarsening procedure reduces the memory requirements and provides the ease of implementation.
We first investigate the generality and the energy stability of the proposed scheme on an unstructured grid over a domain of complex geometry via spinodal decomposition while the mass conservation property is studied by a simple test of volume conserved motion driven by the mean curvature flow.
To verify the stability and robustness of our approach, a well-known benchmark of free-surface motion in a  sloshing tank is considered to perform a series of detailed numerical tests. 
The effectiveness of the adaptive algorithm is assessed and several recommendations are made based on the numerical experiments performed.  
Finally, we demonstrate the applicability of the adaptive variational scheme for topological changes through a well-known dam break problem. 

The organization of the paper is as follows. Section \ref{formulation} briefly describes the governing equations and their fully-discrete variational formulations. The details of the proposed adaptive variational procedure are discussed in Section \ref{algorithm}.  While the numerical tests are performed to assess the effectiveness and robustness of the 
proposed adaptive scheme in Section \ref{tests},  
the scheme is applied to a widely used dam break problem in Section \ref{damBreak}. Finally, the key findings from the study are summarized in Section \ref{conclusion}. \ref{error_estimates} briefly describes an analytical proof of the error estimates for the Allen-Cahn phase-field equation.

\section{Variational formulation}
\label{formulation}
We first review the governing equations for the phase-field formulation of two-phase flows at the continuum level and their respective variational formulations associated with the coupled Navier-Stokes and Allen-Cahn differential equations.

\subsection{The incompressible Navier-Stokes equations for two-phase flows}
The unsteady Navier-Stokes equations for a viscous two-phase incompressible flow on a physical domain $\Omega(t)$ are
\begin{align} \label{NS}
	\rho(\phi)\frac{\partial {\boldsymbol{u}}}{\partial t} + \rho(\phi){\boldsymbol{u}}\cdot\nabla{\boldsymbol{u}} &= \nabla\cdot {\boldsymbol{\sigma}} + \mathbf{SF}(\phi) + \boldsymbol{b}(\phi),&&\mathrm{on}\  \Omega(t),\\
	\nabla\cdot{\boldsymbol{u}} &= 0,&&\mathrm{on}\ \Omega(t),
\end{align}
where ${\boldsymbol{u}} = {\boldsymbol{u}}(\boldsymbol{x},t)$ denotes the fluid velocity defined for each spatial point $\boldsymbol{x} \in \Omega(t)$, $\rho(\phi)$ is the fluid density, $\boldsymbol{b}(\phi)$ is the body force applied on the fluid, $\mathbf{SF}(\phi)$ is the singular force acting at the interface which ensures the pressure jump across the interface \cite{Brackbill} and ${\boldsymbol{\sigma}}$ is the Cauchy stress tensor for a Newtonian fluid, given as ${\boldsymbol{\sigma}} = -{p}\boldsymbol{I} + \mu(\phi)( \nabla{\boldsymbol{u}}+ (\nabla{\boldsymbol{u}})^T)$,
where ${p}$ denotes the fluid pressure and $\mu(\phi)$ is the dynamic viscosity of the fluid. The physical parameters of the fluid such as $\rho$ and $\mu$ depend on the order parameter $\phi$ as
\begin{align}
	\rho(\phi) = \frac{1+\phi}{2}\rho_1 + \frac{1-\phi}{2}\rho_2, \qquad
	\mu(\phi) = \frac{1+\phi}{2}\mu_1 + \frac{1-\phi}{2}\mu_2, \label{dens}
\end{align}
where $(\cdot)_i$ denotes the value of its argument on the phase $i$ of the fluid. The order parameter $\phi$ is solved by the Allen-Cahn equation. The body force $\boldsymbol{b}$ is taken as the gravity force as $\boldsymbol{b}(\phi) = \rho(\phi)\boldsymbol{g}$, where  $\boldsymbol{g}$ is the acceleration due to gravity.

Consistent with the formulation of \cite{AC_JCP}, we employ the generalized-$\alpha$ time integration scheme \cite{Gen_alpha} for the time stepping. Consider $\mathcal{S}^\mathrm{h}$ as the space of trial solution which satisfy the Dirichlet boundary condition and $\mathcal{V}^\mathrm{h}$ as the space of test functions which vanish on the Dirichlet boundary. The stabilized variational form of the flow equations can be written as: find $[ {\boldsymbol{u}}_\mathrm{h}^\mathrm{n+\alpha}, {p}_\mathrm{h}^\mathrm{n+1}] \in\mathcal{S}^\mathrm{h}$ such that $\forall [\boldsymbol{\psi}_\mathrm{h}, q_\mathrm{h}] \in \mathcal{V}^\mathrm{h}$,
\begin{align}
	&\int_{\Omega} \rho(\phi) ( \partial_t{\boldsymbol{u}}_\mathrm{h}^\mathrm{n+\alpha_m} + {\boldsymbol{u}}_\mathrm{h}^\mathrm{n+\alpha} \cdot \nabla{\boldsymbol{u}}_\mathrm{h}^\mathrm{n+\alpha})\cdot \boldsymbol{\psi}_\mathrm{h} \mathrm{d\Omega} + \int_{\Omega} {\boldsymbol{\sigma}}_\mathrm{h}^\mathrm{n+\alpha} : \nabla \boldsymbol{\psi}_\mathrm{h} \mathrm{d\Omega} - \int_{\Omega} \mathbf{SF}^\mathrm{n+\alpha}_\mathrm{h}(\phi)\cdot\boldsymbol{\psi}_\mathrm{h} \mathrm{d\Omega}+ \int_{\Omega} q_\mathrm{h}(\nabla\cdot {\boldsymbol{u}}_\mathrm{h}^\mathrm{n+\alpha}) \mathrm{d\Omega}\nonumber \\
	&+ \displaystyle\sum_\mathrm{e=1}^\mathrm{n_{el}}\int_{\Omega_\mathrm{e}} \frac{\tau_\mathrm{m}}{\rho(\phi)} (\rho(\phi){\boldsymbol{u}}_\mathrm{h}^\mathrm{n+\alpha}\cdot \nabla\boldsymbol{\psi}_\mathrm{h}+ \nabla q_\mathrm{h} )\cdot \boldsymbol{\mathcal{R}}_\mathrm{m}({\boldsymbol{u}},{p}) \mathrm{d\Omega_e} + \displaystyle\sum_\mathrm{e=1}^\mathrm{n_{el}}\int_{\Omega_\mathrm{e}} \nabla\cdot \boldsymbol{\psi}_\mathrm{h}\tau_\mathrm{c}\rho(\phi) \boldsymbol{\mathcal{R}}_\mathrm{c}(\boldsymbol{u}) \mathrm{d\Omega_e}\nonumber \\
	&= \int_{\Omega} \boldsymbol{b}(t^\mathrm{n+\alpha})\cdot \boldsymbol{\psi}_\mathrm{h} \mathrm{d\Omega} + \int_{\Gamma_\mathrm{h}} \boldsymbol{h}\cdot \boldsymbol{\psi}_\mathrm{h} \mathrm{d\Gamma},
\end{align}
where the terms in the first line correspond to the Galerkin terms of the momentum and the continuity equations. The second line represents the Petrov-Galerkin stabilization terms for the momentum and continuity equations. Equal-order polynomial bases are used for the velocity and pressure fields given by the momentum and continuity equations, respectively. Here, $\boldsymbol{h}$ is the corresponding Neumann boundary condition for the Navier-Stokes equations. The element-wise residual of the momentum and the continuity equations denoted by $\boldsymbol{\mathcal{R}}_\mathrm{m}$ and $\boldsymbol{\mathcal{R}}_\mathrm{c}$ respectively are given by
\begin{align}
	\boldsymbol{\mathcal{R}}_\mathrm{m}({\boldsymbol{u}},{p}) &= \rho(\phi)\partial_t{\boldsymbol{u}}_\mathrm{h}^\mathrm{n+\alpha_m} + \rho(\phi){\boldsymbol{u}}_\mathrm{h}^\mathrm{n+\alpha} \cdot \nabla{\boldsymbol{u}}_\mathrm{h}^\mathrm{n+\alpha} - \nabla \cdot {\boldsymbol{\sigma}}_\mathrm{h}^\mathrm{n+\alpha} - \mathbf{SF}^\mathrm{n+\alpha}_\mathrm{h}(\phi) - \boldsymbol{b}(t^\mathrm{n+\alpha}), \\
	\boldsymbol{\mathcal{R}}_\mathrm{c}(\boldsymbol{u}) &= \nabla\cdot\boldsymbol{u}^\mathrm{n+\alpha}_\mathrm{h}.
\end{align}
The stabilization parameters $\tau_\mathrm{m}$ and $\tau_\mathrm{c}$ are the least-squares metrics added to the element-level integrals in the stabilized formulation \cite{Hughes_X,France_II} and are defined as
\begin{align}
	\tau_\mathrm{m} = \bigg[ \bigg( \frac{2}{\Delta t}\bigg)^2 + {\boldsymbol{u}}_\mathrm{h}\cdot \boldsymbol{G}{\boldsymbol{u}}_\mathrm{h} + C_I \bigg(\frac{\mu(\phi)}{\rho(\phi)}\bigg)^2 \boldsymbol{G}:\boldsymbol{G}\bigg] ^{-1/2},\quad
	\tau_\mathrm{c} = \frac{1}{\mathrm{tr}(\boldsymbol{G})\tau_\mathrm{m}}, \label{tau_c}
\end{align}
where $C_I$ is a constant derived from the element-wise inverse estimate \cite{Hughes_inv_est}, $\boldsymbol{G}$ is the element contravariant metric tensor and $\mathrm{tr(\boldsymbol{G})}$ in Eq.~(\ref{tau_c}) denotes the trace of the contravariant metric tensor. The stabilization in the variational form provides stability to the velocity field in convection dominated regimes of the fluid domain and circumvents the Babu$\mathrm{\check{s}}$ka-Brezzi condition which needs to be satisfied by any standard mixed Galerkin method \cite{Johnson}. The element metric tensor $\boldsymbol{G}$ deals with different element topology for different mesh discretization and has been greatly studied in the literature \cite{Johnson,Hsu,Tezduyar_1,Hughes_V,Tezduyar_stab}.
\subsection{The positivity preserving Allen-Cahn formulation}
In this section, we present the variational formulation of Allen-Cahn 
phase-field equation, which is a time-dependent 
phenomenological reaction-diffusion equation 
based on the stochastic Ginzburg-Landau free-energy model \cite{allen_cahn}.
Owing to a positive excess of surface free energy associated
with the phase boundaries, net diffusion is adjusted in such
a way that the total area of the phase boundaries is reduced and 
the interfacial velocity is proportional to the thermodynamic driving force, 
whereby the proportionality constant is mobility.
In this phenomenological theory, the thermodynamic driving force 
is the product of the local mean curvature 
of the phase boundary and the excess surface free energy per
unit area of the phase boundary. 
 In addition 
to good variational properties, an implicit link with the mean curvature flow 
makes the Allen-Cahn phase-field model very attractive 
for moving interface and
free boundary problems arising from various fluid dynamic applications.
%

\subsubsection{Strong form}
In the present study, we consider the convective form of the conservative Allen-Cahn equation with a Lagrange multiplier to conserve mass:
\begin{equation}
\left.
\begin{aligned} \label{CAC}
	\partial_t\phi + \boldsymbol{u}\cdot\nabla\phi - \gamma\big(\varepsilon^2\nabla^2\phi + F'(\phi) - \beta(t)\sqrt{F(\phi)}\big) = 0,\ \ &\mathrm{on}\ \Omega(t),\qquad \\
	\frac{\partial \phi}{\partial n}\bigg|_{\Gamma} = \mathbf{n}\cdot\nabla\phi = 0,\ \ &\mathrm{on}\ \Gamma,\\
	\phi |_{t=0} = \phi_0,	\ \ &\mathrm{on}\ \Omega(t), 
\end{aligned}
\right\}
\end{equation}
where $\boldsymbol{u}$ is the convection velocity, $\gamma$ is a relaxation factor with units of $[T^{-1}]$ selected as $1$ in the present study, $\varepsilon$ is a parameter related to the interfacial thickness and $\mathbf{n}$ is the outward normal to the boundary $\Gamma$. $F'(\phi)$ is the derivative of the double well potential function which is taken as $F(\phi)=\frac{1}{4}(\phi^2-1)^2$. $\beta(t)$ is the time dependent part of the Lagrange multiplier which can be derived using the incompressible flow condition of divergence-free velocity and the given boundary conditions as
\begin{align} \label{eqn:beta}
	\beta(t) = \frac{\int_\Omega F'(\phi)\mathrm{d}\Omega}{\int_\Omega \sqrt{F(\phi)}\mathrm{d}\Omega}.
\end{align}
Consider $K(\phi) = 0.5(\phi^3/3 - \phi)$ and the Lagrange multiplier is written in such a way that
\begin{align} \label{K_cons}
	\int_\Omega K(\phi) \mathrm{d}\Omega =\ \mathrm{constant}.
\end{align}
We approximate $F'(\phi)$ and $\sqrt{F(\phi)}$ with the help of mid-point approximation to have an energy-stable scheme \cite{Du, AC_JCP} as:
\begin{align} \label{midpt_approx}
	F'(\phi) = \frac{F(\phi^\mathrm{n+1}) - F(\phi^\mathrm{n})}{\phi^\mathrm{n+1} - \phi^\mathrm{n}}, \quad K'(\phi) = \sqrt{F(\phi)} = \frac{K(\phi^\mathrm{n+1}) - K(\phi^\mathrm{n})}{\phi^\mathrm{n+1} - \phi^\mathrm{n}}.
\end{align}

\subsubsection{Semi-discrete variational form}
As shown in \cite{AC_JCP}, the semi-discrete Allen-Cahn equation can be recast in the form of a convection-diffusion-reaction (CDR) equation as follows:
\begin{align}
	G(\partial_t{\phi}^{\mathrm{n}+\alpha_\mathrm{m}}, \phi^{\mathrm{n}+\alpha}) = \partial_t{\phi}^{\mathrm{n}+\alpha_\mathrm{m}} + \boldsymbol{u}\cdot\nabla\phi^{\mathrm{n}+\alpha} - k\nabla^2\phi^{\mathrm{n}+\alpha} + s\phi^{\mathrm{n}+\alpha} - f = 0,
\end{align}
where
\begin{align}
	\mathrm{Convection\ velocity} &= \boldsymbol{u},\\
	\mathrm{Diffusion\ coefficient} &= k = \varepsilon^2,\\
	\mathrm{Reaction\ coefficient} &= s = \frac{1}{4}\bigg[ \frac{(\phi^\mathrm{n+\alpha})^2}{\alpha^3} - \bigg(\frac{3}{\alpha^3} - \frac{4}{\alpha^2}\bigg)\phi^\mathrm{n+\alpha}\phi^\mathrm{n} + \bigg( \frac{3}{\alpha^3} - \frac{8}{\alpha^2} + \frac{6}{\alpha} \bigg) (\phi^\mathrm{n})^2 - \frac{2}{\alpha} \bigg]\nonumber \\ &- \frac{\beta(t)}{2}\bigg[ \frac{\phi^\mathrm{n+\alpha}}{3\alpha^2} + \frac{1}{3}\bigg( -\frac{2}{\alpha^2} + \frac{3}{\alpha} \bigg)\phi^\mathrm{n} \bigg],\\
	\mathrm{Source\ term} &= f = -\frac{1}{4}\bigg[ \bigg(-\frac{1}{\alpha^3} +\frac{4}{\alpha^2} - \frac{6}{\alpha} + 4 \bigg)(\phi^\mathrm{n})^3 + \bigg( \frac{2}{\alpha} - 4\bigg)\phi^\mathrm{n} \bigg] \nonumber \\
	&+ \frac{\beta(t)}{2}\bigg[ \frac{1}{3}\bigg( \frac{1}{\alpha^2} - \frac{3}{\alpha} + 3 \bigg)(\phi^\mathrm{n})^2 - 1\bigg].
\end{align}

We next present the finite element formulation of the Allen-Cahn phase field equation. In particular, the positivity preserving stabilized variational form can be stated as: find $\phi_\mathrm{h}(\boldsymbol{x},t^{\mathrm{n}+\alpha}) \in \mathcal{S}^\mathrm{h}$ such that $\forall w_\mathrm{h} \in \mathcal{V}^\mathrm{h}$,
\begin{align} \label{AC_variational}
	&\int_\Omega \bigg( w_\mathrm{h}\partial_t{\phi}_\mathrm{h} + w_\mathrm{h}(\boldsymbol{u}\cdot\nabla\phi_\mathrm{h}) + \nabla w_\mathrm{h}\cdot(k\nabla\phi_\mathrm{h} ) + w_\mathrm{h}s\phi_\mathrm{h} - w_\mathrm{h}f \bigg) \mathrm{d}\Omega \nonumber \\
	&+ \displaystyle\sum_\mathrm{e=1}^\mathrm{n_{el}}\int_{\Omega_\mathrm{e}}\bigg( \big(\boldsymbol{u}\cdot\nabla w_\mathrm{h} \big)\tau \big( \partial_t{\phi}_\mathrm{h} + \boldsymbol{u}\cdot\nabla\phi_\mathrm{h} - \nabla\cdot(k\nabla\phi_\mathrm{h}) + s\phi_\mathrm{h} -f \big) \bigg) \mathrm{d}\Omega_\mathrm{e}  \nonumber \\
	&+ \displaystyle\sum_\mathrm{e=1}^\mathrm{n_{el}}\int_{\Omega_\mathrm{e}} \chi \frac{|\mathcal{R}(\phi_\mathrm{h})|}{|\nabla\phi_\mathrm{h}|}k_s^\mathrm{add} \nabla w_\mathrm{h}\cdot \bigg( \frac{\boldsymbol{u}\otimes \boldsymbol{u}}{|\boldsymbol{u}|^2} \bigg) \cdot \nabla\phi_\mathrm{h} \mathrm{d}\Omega_\mathrm{e} + \sum_\mathrm{e=1}^\mathrm{n_{el}} \int_{\Omega_\mathrm{e}}\chi \frac{|\mathcal{R}(\phi_\mathrm{h})|}{|\nabla \phi_\mathrm{h}|} k^\mathrm{add}_{c} \nabla w_\mathrm{h} \cdot \bigg( \mathbf{I} - \frac{\boldsymbol{u}\otimes \boldsymbol{u}}{|\boldsymbol{u}|^2} \bigg) \cdot \nabla\phi_\mathrm{h} \mathrm{d}\Omega_\mathrm{e} \nonumber \\
	&= 0, 	
\end{align}

While in Eq.~(\ref{AC_variational}), the first line represents the Galerkin terms,
the second line consists of linear stabilization terms in the form of SUPG stabilization with the stabilization parameter $\tau$ which is given by \cite{Hughes_X}
\begin{align} \label{tau_eqn}
	\tau &= \bigg[ \bigg( \frac{2}{\Delta t}\bigg)^2 + \boldsymbol{u}\cdot \boldsymbol{G}\boldsymbol{u} + 9k^2 \boldsymbol{G}:\boldsymbol{G}+ s^2\bigg] ^{-1/2},
\end{align}
where $\boldsymbol{G}$ is the element contravariant metric tensor. $\mathcal{R}(\phi_\mathrm{h})$ is the residual of the Allen-Cahn equation given as
\begin{align}
	\mathcal{R}(\phi_\mathrm{h}) = \partial_t\phi_\mathrm{h} + \boldsymbol{u}\cdot\nabla\phi_\mathrm{h} - \nabla\cdot(k\nabla\phi_\mathrm{h}) + s\phi_\mathrm{h} -f. 	
\end{align}

\subsubsection{Positivity condition}
We next describe the positivity condition in the context of Galerkin finite element formulation for the convection-dominated problems.
Here, the terms in the third line of Eq.~(\ref{AC_variational}) reflect the nonlinear positivity preserving stabilization. These stabilization terms are derived for the multi-dimensional convection-diffusion-reaction equation in \cite{PPV} by establishing the positivity condition of the element level matrix of the variationally discretized equation. The positivity preserving condition for the Eq.~(\ref{CAC}) can be defined as follows. Consider a simplified form of Eq.~(\ref{CAC}) with only convection effects:
 \begin{align}
	\partial_t \phi + \boldsymbol{u}\cdot\nabla\phi = 0.
\end{align}
The finite element Galerkin approximation for an explicit scheme of a one-dimensional element between $i-1$ and $i$ can be written as
\begin{align}
	\frac{\phi^\mathrm{n+1}_{i}-\phi^\mathrm{n}_{i}}{\Delta t} =  -\int_{\Omega} w_\mathrm{h} (u\frac{\partial\phi_\mathrm{h}}{\partial x}) \mathrm{d}\Omega = -\int_\Omega N^T u \frac{\partial N}{\partial x} \mathrm{d}\Omega \begin{bmatrix} \phi^\mathrm{n}_{i-1} \\ \phi^\mathrm{n}_{i}\end{bmatrix} = -\frac{u}{2h}\begin{bmatrix} -1 & 1\\ -1 & 1\end{bmatrix}\begin{bmatrix} \phi^\mathrm{n}_{i-1} \\ \phi^\mathrm{n}_{i}\end{bmatrix},
\end{align}
where $N$ is the row-vector of Lagrangian linear shape functions for one-dimensional elements satisfying the partition of unity property.
Therefore, after assembly of the elements for a uniform grid, the finite element based stencil can be expressed as follows
\begin{align} \label{Galerkin_approx}
	\frac{\phi^\mathrm{n+1}_{i}-\phi^\mathrm{n}_{i}}{\Delta t} = -\frac{u}{2h}\big( \phi^\mathrm{n}_{i+1} - \phi^\mathrm{n}_{i-1} \big),
\end{align}
which is the same as the central difference scheme. 
Any scheme which can be written in the following form
\begin{align}
	\frac{\phi^\mathrm{n+1}_{i}-\phi^\mathrm{n}_{i}}{\Delta t} = C^+ (\phi^\mathrm{n}_{i+1} - \phi^\mathrm{n}_{i}) - C^- (\phi^\mathrm{n}_{i}-\phi^\mathrm{n}_{i-1}),
\end{align}
satisfies the positivity preserving property if the coefficients $C^+$ and $C^-$ satisfy \cite{Harten}
\begin{align}
	C^+\geq 0,\qquad C^-\geq 0,\qquad C^++C^- \leq 1.
\end{align}
We observe that the Galerkin approximation (Eq.~(\ref{Galerkin_approx})) does not satisfy this condition. The scheme has to be modified by adding the stabilization terms to satisfy the positivity property, as suggested by \cite{Harten} for finite-difference approximations . These bounds for the PPV formulation for the above problem have been shown for some particular cases in \ref{positivity_derivation}. The definition for the positivity preservation can be generalized to the implicit matrix form of the scheme by transforming the left-hand-side matrix $\boldsymbol{A} = \{a_{ij}\}$ to an M-matrix which ensures the positivity and convergence with the following properties \cite{Kuzmin}
\begin{align}
	a_{ii} &> 0, \forall i,\\
	a_{ij} &\leq 0, \forall j\neq i,\\
	\sum_{j} a_{ij} &= 0, \forall i.
\end{align}
This transformation to an M-matrix is carried out by the addition of the discrete upwind matrix. Satisfaction of the above properties renders the variational scheme positivity preserving and monotone \cite{FCT}. The nonlinear property of the variational scheme is a result of the factor $\chi |\mathcal{R}(\phi_\mathrm{h})|/|\nabla\phi_\mathrm{h}|$ which acts as a limiter to the upwinding near the regions of high solution gradients.  Several test cases have been performed to assess the effectiveness of the PPV technique in \cite{PPV}. The details of the derivation of the added diffusions $k_s^\mathrm{add}$, $k_c^\mathrm{add}$ and $\chi$ can be found in \cite{PPV}, which are given by
\begin{align}
	\chi &= \frac{2}{|s|h + 2|\boldsymbol{u}|},\\
	k_s^\mathrm{add} &= \mathrm{max} \bigg\{ \frac{||\boldsymbol{u}| - \tau|\boldsymbol{u}|s|h}{2} - (k + \tau|\boldsymbol{u}|^2) + \frac{sh^2}{6}, 0 \bigg\},\\
	k_c^\mathrm{add} &= \mathrm{max} \bigg\{ \frac{|\boldsymbol{u}|h}{2} - k + \frac{sh^2}{6}, 0 \bigg\},
\end{align} 
where $|\boldsymbol{u}|$ is the magnitude of the convection velocity and $h$ is the characteristic element length defined in \cite{PPV}. 
Our PPV-based finite element method for the phase-field Allen-Cahn equation solves the nonlinear CDR equation without 
operator splitting, while satisfying the positivity condition and removing the operator splitting error.
In contrast, the method in \cite{shin_kim2014} utilizes an operator splitting approach and divides the Allen-Cahn equation into the heat diffusion equation and a simpler nonlinear differential equation.
It was shown in \cite{AC_JCP} that the above variational scheme for the Allen-Cahn equation with $\boldsymbol{u}=\mathbf{0}$ is unconditionally energy stable with a modified discrete free energy functional given by
\begin{align} \label{energy}
	\mathrm{E}(\phi) = \int_\Omega \bigg( \frac{1}{2}(\varepsilon^2 + \chi \frac{|\mathcal{R}(\phi_\mathrm{h})|}{|\nabla\phi_\mathrm{h}|}k_c^\mathrm{add})|\nabla\phi|^2 + F(\phi) \bigg) \mathrm{d}\Omega,
\end{align}
i.e., the free energy is decreasing with time in the discrete sense for the above functional in the PPV-based scheme for the Allen-Cahn equation.
We next present the nonlinear adaptive partitioned procedure based on the aforementioned PPV technique for the Allen-Cahn equation.

\section{Adaptive variational Allen-Cahn formulation}
\label{algorithm}
In this section, we present our nonlinear adaptive partitioned procedure for implicitly discretized Navier-Stokes and Allen-Cahn equations. 
The goal of the proposed procedure is the reduction of the nonlinear errors while solving the Navier-Stokes and the Allen-Cahn equations along with the reduction of the error indicator due to adaptivity. Since the underlying equations involved in the phase-field formulation are inherently nonlinear,  the nonlinear convergence plays a crucial role in the mass conservation and accuracy of the solution. The errors introduced due to the mesh refinement/coarsening also profoundly impact the solution accuracy. Therefore, the primary goal of the proposed algorithm is to reduce the errors introduced due to mesh adaptivity within a tolerance limit followed by the reduction in the nonlinear errors of the underlying equations. This will be elaborated in the subsequent subsections.
We first define some of the errors used to quantify the solution obtained by solving the Navier-Stokes, the Allen-Cahn equations and the errors characterizing the adaptive algorithm.

The increments in the velocity, pressure and the order parameter at each time step are evaluated by Newton-Raphson type nonlinear iterations. The linearized system for the incompressible Navier-Stokes equations is given by
\begin{align} \label{LS_NS}
	\begin{bmatrix}
		\boldsymbol{K}_\Omega &  & \boldsymbol{G}_\Omega \\
 & \\
	       -\boldsymbol{G}^T_\Omega &  &\boldsymbol{C}_\Omega
	\end{bmatrix} 
	\begin{Bmatrix}
		\Delta \boldsymbol{u}^\mathrm{n+\alpha_f} \\
\\
		\Delta p^\mathrm{n+1}
	\end{Bmatrix}
	= \begin{Bmatrix} 
		\widetilde{\boldsymbol{\mathcal{R}}}_\mathrm{m}(\boldsymbol{u},p) \\
\\
		\widetilde{\boldsymbol{\mathcal{R}}}_\mathrm{c}(\boldsymbol{u})
	  \end{Bmatrix}
\end{align}
where $\boldsymbol{K}_\Omega$ is the stiffness matrix of the momentum equation consisting of inertia, convection, diffusion and stabilization terms, $\boldsymbol{G}_\Omega$ is the discrete gradient operator, $\boldsymbol{G}^T_\Omega$ is the divergence operator and $\boldsymbol{C}_\Omega$ is the pressure-pressure stabilization term. Here, $\Delta \boldsymbol{u}$ and $\Delta p$ are the increments in velocity and pressure, respectively and $\widetilde{\boldsymbol{\mathcal{R}}}_\mathrm{m}(\boldsymbol{u},p)$ and $\widetilde{\boldsymbol{\mathcal{R}}}_\mathrm{c}(\boldsymbol{u})$ represent the weighted residuals of the stabilized momentum and continuity equations respectively. Let the vector containing the increments of velocity and pressure be denoted by $\Delta \boldsymbol{X}$ and the vector of the respective updated quantities at $t^\mathrm{n+1}_{\mathrm{(k)}}$ be represented as $\boldsymbol{X}^\mathrm{n+1}_{\mathrm{(k)}}$, $\mathrm{k}$ being the nonlinear iteration index. We define the error in solving the Navier-Stokes equations as
\begin{align}
	e_{NS} = \frac{||\Delta \boldsymbol{X}||}{||\boldsymbol{X}^\mathrm{n+1}_{\mathrm{(k)}}||}.
\end{align}  
Similarly, the Allen-Cahn equation can be linearized as follows:
\begin{align} \label{LS_AC}
	\begin{bmatrix}
		\boldsymbol{K}_{AC}
	\end{bmatrix} 
	\begin{Bmatrix}
		\Delta \phi^\mathrm{n+\alpha}
	\end{Bmatrix}
	= \begin{Bmatrix} 
		\widetilde{\mathcal{R}}(\phi)
	  \end{Bmatrix}
\end{align}
where $\boldsymbol{K}_{AC}$ consists of the inertia, convection, diffusion, reaction and stabilization terms and $\widetilde{\mathcal{R}}(\phi)$ represents the weighted residual for the stabilized conservative Allen-Cahn equation. Similar to the error evaluation for the Navier-Stokes equations, the numerical error in solving the Allen-Cahn equation can be written as
\begin{align}
	e_{AC} = \frac{||\Delta \phi^\mathrm{n+\alpha}||}{||\phi^\mathrm{n+1}_{\mathrm{(k)}}||}.
\end{align} 
These error expressions are utilized for the convergence criteria of the nonlinear iterations.

\subsection{Residual error estimates for the conservative Allen-Cahn equation}
The Galerkin variational formulation of the Navier-Stokes and the Allen-Cahn equations is  simply the differential operator multiplied by the weighting function
and integrated by parts as appropriate, which tends to minimize the residual of the equations in a chosen set of weighting functions. 
When the solution of the underlying differential equation is smooth, as measured
relative to the differential equation on the given mesh, the variational error tends to be small at convergence.  
If the solution exhibits oscillations near sharp gradients due to dominant convection and reaction effects on an under-resolved mesh, the residual of the equation has a significantly large value. This residual needs to be minimized to obtain an accurate and close-to-converged solution, defined by the user-defined tolerance limit. Therefore, the mesh has to be refined in those areas to capture the sharp gradients in the solution. This suggests using the residual of the differential equation as an indicator of the error for the adaptive mesh algorithm.

For the quantification of the error on which our adaptive algorithm will be based, the residual error estimates are derived for the Galerkin discretization of the Allen-Cahn equation. Consider the domain $\Omega$ which consists of elements $\Omega_\mathrm{e}$, chosen such that $\Omega = \cup_\mathrm{e=1}^\mathrm{n_{el}} \Omega_\mathrm{e}$ and $\emptyset = \cap_\mathrm{e=1}^\mathrm{n_{el}} \Omega_\mathrm{e}$ where $\mathrm{n}_\mathrm{el}$ is the number of elements. Let $\Gamma$ be the Lipschitz continuous boundary of the domain $\Omega$, $\Gamma_D$ and $\Gamma_N$ be the Dirichlet and Neumann boundaries of $\Omega$ respectively such that $\Gamma = \Gamma_D \cup \Gamma_N$. Furthermore, let $\mathcal{E}$ denote the set of edges for all the elements in the domain, $\mathcal{E}_\Gamma$, $\mathcal{E}_{\Gamma_D}$ and $\mathcal{E}_{\Gamma_N}$ be the set of edges on the boundary $\Gamma$, $\Gamma_D$ and $\Gamma_N$ respectively. Consider $\mathcal{E}_{\Omega_\mathrm{e}}$ to be the set of edges of an element $\Omega_\mathrm{e}$ and $\mathcal{E}_\Omega$ be the set of all the interior edges of $\Omega$. The Galerkin terms of Eq.~(\ref{AC_variational}) can be written after integration by parts as
\begin{align}
&\int_\Omega w_\mathrm{h}\partial_t\phi_\mathrm{h} \mathrm{d}\Omega + \int_\Omega w_\mathrm{h}(\boldsymbol{u}\cdot\nabla\phi_\mathrm{h}) \mathrm{d}\Omega +\int_{\Omega} \nabla w_\mathrm{h}\cdot(k \nabla\phi_\mathrm{h}) \mathrm{d}\Omega + \int_\Omega w_\mathrm{h}s\phi_\mathrm{h} \mathrm{d}\Omega - \int_\Omega w_\mathrm{h}f\mathrm{d}\Omega \nonumber \\
	 =& \displaystyle\sum_\mathrm{e=1}^\mathrm{n_{el}} \int_{\Omega_\mathrm{e}} w_\mathrm{h} \mathcal{R}_{\Omega_\mathrm{e}}(\phi_\mathrm{h}) \mathrm{d}\Omega_\mathrm{e} + \displaystyle\sum_{E\in \mathcal{E}} \int_{E} w_\mathrm{h} \mathcal{R}_{E}(\phi_\mathrm{h}) \mathrm{d}E
\end{align}
where $\mathcal{R}_{\Omega_\mathrm{e}}$ and $\mathcal{R}_{E}$ are the element and edge based residuals given as, 
\begin{align}
	\mathcal{R}_{\Omega_\mathrm{e}} &= \partial_t\phi_\mathrm{h} + \boldsymbol{u}\cdot\nabla\phi_\mathrm{h} - k\nabla^2\phi_\mathrm{h} + s\phi_\mathrm{h} - f,\\
	\mathcal{R}_{E} &= \begin{cases}
		-\mathbb{J}_{E}(\mathbf{n}_{E}\cdot k\nabla\phi_\mathrm{h}),\ \ \ \ \ &\mathrm{if}\ E\in \mathcal{E}_\Omega \\
		-\mathbf{n}_{E}\cdot k\nabla\phi_\mathrm{h},\ \ \ \ \ &\mathrm{if}\ E\in \mathcal{E}_{\Gamma_N} \\
		0,\ \ &\mathrm{if}\ E\in \mathcal{E}_{\Gamma_D}
		\end{cases}
\end{align}
where $\mathbb{J}_E(\cdot)$ is the jump of the argument across the element edge $E$ and $\mathbf{n}_E$ is the normal to the edge $E$.

After some algebraic manipulations and using inequalities (see \ref{error_estimates} for detailed steps), we establish the expression of the error estimate as
\begin{align}
	||\phi - \phi_\mathrm{h}|| \leq c^* \bigg\{ \displaystyle\sum_\mathrm{e=1}^\mathrm{n_{el}} h^2_{\Omega_\mathrm{e}}||\mathcal{R}_{\Omega_\mathrm{e}}||^2_{\Omega_\mathrm{e}} + \displaystyle\sum_{E \in \mathcal{E}} h_{E} ||\mathcal{R}_{E}||^2_{E}   \bigg\}^{1/2}||w||_{H^1(\Omega)},
\end{align} 
where $h_{\Omega_\mathrm{e}}$ is the diameter of the element $\Omega_\mathrm{e}$, $h_E$ is the length of the edge $E$ and $c^*$ is some arbitrary constant.
\begin{align}
	||\phi - \phi_\mathrm{h}|| \leq c^* \bigg\{ \displaystyle\sum_\mathrm{e=1}^\mathrm{n_{el}} h^2_{\Omega_\mathrm{e}}||\mathcal{R}_{\Omega_\mathrm{e}}||^2_{\Omega_\mathrm{e}} + \displaystyle\sum_\mathrm{e=1}^\mathrm{n_{el}} \displaystyle\sum_{E \in \mathcal{E}_{\Omega_\mathrm{e}}\cap \mathcal{E}} h_{E} ||\mathcal{R}_{E}||^2_{E}  \bigg\}^{1/2}||w||_{H^1(\Omega)}.
\end{align}
Let $\eta_{\Omega_\mathrm{e}}$ denote the error indicator of the triangular element $\Omega_\mathrm{e}$ on the mesh. The error estimate can be recast as
\begin{align}
	||\phi - \phi_\mathrm{h}|| &\leq \eta = \bigg( \displaystyle\sum_\mathrm{e=1}^\mathrm{n_{el}} \eta_{\Omega_\mathrm{e}}^2 \bigg)^{1/2},\label{etaR1}\\
	\eta^2_{\Omega_\mathrm{e}} &=  h^2_{\Omega_\mathrm{e}}||\mathcal{R}_{\Omega_\mathrm{e}}||^2_{\Omega_\mathrm{e}} + \displaystyle\sum_{E \in \mathcal{E}_{\Omega_\mathrm{e}}\cap \mathcal{E}} h_E ||\mathcal{R}_E||^2_E, \label{etaR2} 
\end{align}
where $\eta$ is the error estimate and $\eta_{\Omega_\mathrm{e}}$ is the error indicator for the element $\Omega_\mathrm{e}$.

\subsection{The adaptive algorithm}
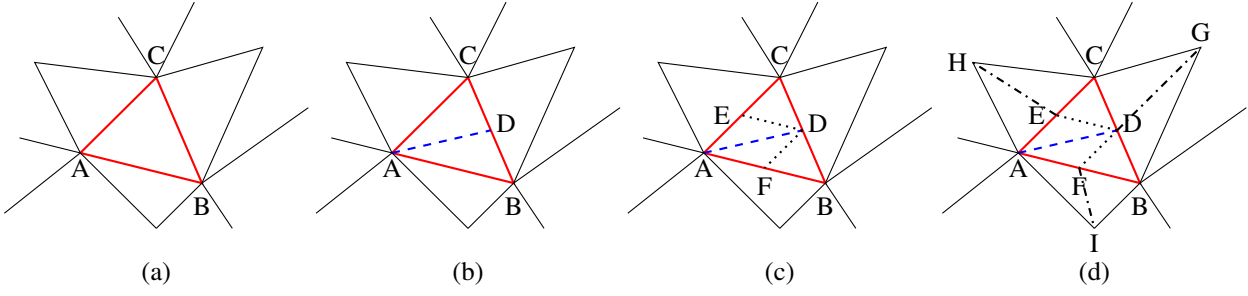
\begin{figure}
\centering
\hspace{-1cm}
\begin{tikzpicture}[decoration={markings,mark=at position 1.0 with {\arrow{>}}},scale=2]
	\draw [thick,red](0,0)--(0.8,-0.2)--(0.5,0.5)--(0,0);
	\draw (0,-0.1) node {A};
	\draw (0.8,-0.35) node {B};
	\draw (0.5,0.65) node {C};
	\draw (0.8,-0.2)--(1.2,0.7)--(0.5,0.5);
	\draw (0,0)--(-0.3,0.6)--(0.5,0.5);
	\draw (0.5,0.5)--(0.25,0.9);
	\draw (0.5,0.5)--(0.75,1.0);
	\draw (0,0)--(-0.5,-0.4);
	\draw (0,0)--(0.5,-0.5);
	\draw (0.5,-0.5)--(0.8,-0.2);
	\draw (0,0)--(-0.4,0.1);
	\draw (0.8,-0.2)--(1.0,-0.5);
	\draw (0.8,-0.2)--(1.5,0.3);
	\draw (0.5,-0.8) node {(a)};
\end{tikzpicture}
\begin{tikzpicture}[decoration={markings,mark=at position 1.0 with {\arrow{>}}},scale=2]	
	\draw [thick,red](0,0)--(0.8,-0.2)--(0.5,0.5)--(0,0);
	\draw (0,-0.1) node {A};
	\draw (0.8,-0.35) node {B};
	\draw (0.5,0.65) node {C};
	\draw [thick,blue,dashed](0,0)--(0.65,0.15);
	\draw (0.75,0.2) node{D};
	\draw (0.8,-0.2)--(1.2,0.7)--(0.5,0.5);
	\draw (0,0)--(-0.3,0.6)--(0.5,0.5);
	\draw (0.5,0.5)--(0.25,0.9);
	\draw (0.5,0.5)--(0.75,1.0);
	\draw (0,0)--(-0.5,-0.4);
	\draw (0,0)--(0.5,-0.5);
	\draw (0.5,-0.5)--(0.8,-0.2);
	\draw (0,0)--(-0.4,0.1);
	\draw (0.8,-0.2)--(1.0,-0.5);
	\draw (0.8,-0.2)--(1.5,0.3);
	\draw (0.5,-0.8) node {(b)};
\end{tikzpicture}
\begin{tikzpicture}[decoration={markings,mark=at position 1.0 with {\arrow{>}}},scale=2]	
	\draw [thick,red](0,0)--(0.8,-0.2)--(0.5,0.5)--(0,0);
	\draw (0,-0.1) node {A};
	\draw (0.8,-0.35) node {B};
	\draw (0.5,0.65) node {C};
	\draw [thick,blue,dashed](0,0)--(0.65,0.15);
	\draw (0.75,0.2) node{D};
	\draw [thick,dotted](0.65,0.15)--(0.25,0.25);
	\draw (0.12,0.25) node {E};
	\draw [thick,dotted](0.65,0.15)--(0.4,-0.1);
	\draw (0.4,-0.22) node {F};
	\draw (0.8,-0.2)--(1.2,0.7)--(0.5,0.5);
	\draw (0,0)--(-0.3,0.6)--(0.5,0.5);
	\draw (0.5,0.5)--(0.25,0.9);
	\draw (0.5,0.5)--(0.75,1.0);
	\draw (0,0)--(-0.5,-0.4);
	\draw (0,0)--(0.5,-0.5);
	\draw (0.5,-0.5)--(0.8,-0.2);
	\draw (0,0)--(-0.4,0.1);
	\draw (0.8,-0.2)--(1.0,-0.5);
	\draw (0.8,-0.2)--(1.5,0.3);
	\draw (0.5,-0.8) node {(c)};
\end{tikzpicture}
\begin{tikzpicture}[decoration={markings,mark=at position 1.0 with {\arrow{>}}},scale=2]	
	\draw [thick,red](0,0)--(0.8,-0.2)--(0.5,0.5)--(0,0);
	\draw (0,-0.1) node {A};
	\draw (0.8,-0.35) node {B};
	\draw (0.5,0.65) node {C};
	\draw [thick,dashed,blue](0,0)--(0.65,0.15);
	\draw (0.75,0.2) node{D};
	\draw [thick,dotted](0.65,0.15)--(0.25,0.25);
	\draw (0.12,0.25) node {E};
	\draw [thick,dotted](0.65,0.15)--(0.4,-0.1);
	\draw (0.4,-0.22) node {F};
	\draw [thick,dashdotted](0.25,0.25)--(-0.3,0.6);
	\draw (-0.4,0.6) node {H};
	\draw [thick,dashdotted](0.4,-0.1)--(0.5,-0.5);
	\draw (0.5,-0.6) node {I};
	\draw [thick,dashdotted](0.65,0.15)--(1.2,0.7);
	\draw (1.2,0.8) node {G};
	\draw (0.8,-0.2)--(1.2,0.7)--(0.5,0.5);
	\draw (0,0)--(-0.3,0.6)--(0.5,0.5);
	\draw (0.5,0.5)--(0.25,0.9);
	\draw (0.5,0.5)--(0.75,1.0);
	\draw (0,0)--(-0.5,-0.4);
	\draw (0,0)--(0.5,-0.5);
	\draw (0.5,-0.5)--(0.8,-0.2);
	\draw (0,0)--(-0.4,0.1);
	\draw (0.8,-0.2)--(1.0,-0.5);
	\draw (0.8,-0.2)--(1.5,0.3);
	\draw (0.5,-0.8) node {(d)};
\end{tikzpicture}
\caption{A schematic of the newest vertex bisection algorithm for refinement: (a) marked element for refinement, (b) bisection of the largest edge, (c) bisection of the other marked edges, (d) inclusion of more elements to remove hanging nodes.} 
\label{NVB}
\end{figure}

Given the error indicators for each element $\eta_{\Omega_\mathrm{e}}$ calculated from the error estimates of the Allen-Cahn equation, we employ the newest vertex bisection algorithm \cite{Chen_2} for the mesh adaptivity. The algorithm first marks the elements and the edges based on the error indicators and then proceeds to refine or coarsen the elements. The marking of the elements for refinement  is carried out by the D\"{o}rfler criterion \cite{Dorfler},  which finds the minimum set of elements $\Omega^M$ such that
\begin{align}
	\theta \displaystyle\sum_{\Omega_\mathrm{e}\in \Omega} \eta_{\Omega_\mathrm{e}}^2 \leq \displaystyle\sum_{\Omega_\mathrm{e}\in \Omega^M} \eta_{\Omega_\mathrm{e}}^2,
\end{align} 
for a user-defined $\theta \in (0,1)$. This criterion therefore selects the elements for refinement which have a large contribution to the error estimate. A schematic for the refinement algorithm is shown in Fig. \ref{NVB}. Suppose the element ABC is selected for refinement. The edges of the element (AB, BC and CA) are then marked for the refinement process. The bisection algorithm starts with the largest edge among the marked edges (BC) by addition of a new node (D), followed by the bisection of the other edges (AB and AC). At this point, a new array is created which indicates the nodes which are added by the bisection (can be coarsened) or which belong to the initial grid (cannot be coarsened). This simplifies the implementation of the algorithm since we do not need a tree structure containing the information about the parent and children elements apart from the mentioned array. Once the marked edges are refined, there would be some hanging nodes (D, E and F in Fig. \ref{NVB}(c)). Some extra elements are then created to avoid these hanging nodes (Fig. \ref{NVB}(d)). This completes the refinement algorithm for a particular element. A similar bisection process is carried out for all the marked elements. Once all the marked elements are refined, the algorithm proceeds to solve the underlying equation on the updated refined mesh. The interpolation of the variables while refining is carried out by setting the value of the newest added node equal to the half of the sum of the values at the nodes of the two extremes of the corresponding edge, for example, in Fig. \ref{NVB}(b), the value of a variable at D will be half the sum of its values at B and C.

Based on the residual error indicators, the elements with small errors can be coarsened using the node-based coarsening algorithm in a similar way with the help of the created array containing the information about the added nodes by bisection. Certain nodes are marked which are considered to be ``good" for coarsening to ensure that the conformity of the mesh is not compromised. Nodes are then removed based on the D\"{o}rfler criterion with $\theta_c$ as the user-defined coarsening parameter. Further details about the coarsening algorithm can be found in \cite{Chen_3}.

\subsection{The nonlinear adaptive variational partitioned (NAVP) procedure}
We now present the algorithm for the partitioned iterative coupling of the incompressible Navier-Stokes and the Allen-Cahn equations. Within a predictor-corrector process,  
implicit time discretizations are adopted for efficiency and robustness of the partitioned coupling for general multiphase problems with complex geometries. Nonlinear iterations are employed to minimize the partitioning errors between the two full-discrete forms of the coupled differential equations.
\begin{algorithm}
  \caption{Nonlinear adaptive variational partitioned (NAVP)  procedure for implicit Navier-Stokes and Allen-Cahn solvers}
  \label{algorithm_1}
  \begin{algorithmic}[1]
  	\STATE Given $\boldsymbol{u}^0$, $p^0$, $\phi^0$ on $\mathcal{T}^0$ \\
	\STATE Loop over time steps, $\mathrm{n}=0,1,\cdots$ \\
	\STATE \quad Start from known variables $\boldsymbol{u}^\mathrm{n}$, $p^\mathrm{n}$, $\phi^\mathrm{n}$ on $\mathcal{T}^\mathrm{n}$\\
	\STATE \quad Predict the solution on $\mathcal{T}^\mathrm{n+1}_\mathrm{(0)} = \mathcal{T}^\mathrm{n}$: \\
	\STATE \qquad\quad $\boldsymbol{u}^\mathrm{n+1}_{(0)}=\boldsymbol{u}^\mathrm{n}$\\ 
	\STATE \qquad\quad $p^\mathrm{n+1}_{(0)} = p^\mathrm{n}$ \\ 
	\STATE \qquad\quad $\phi^\mathrm{n+1}_{(0)} = \phi^\mathrm{n}$\\
	\STATE \quad Loop over the nonlinear iterations, $\mathrm{k}=0,1,\cdots ,nIterMax$ until convergence \\
	\STATE \qquad\quad Solve Navier-Stokes equations for updated values of $\boldsymbol{u}^\mathrm{n+1}_\mathrm{(k)}$ and $p^\mathrm{n+1}_\mathrm{(k)}$ on $\mathcal{T}^\mathrm{n+1}_\mathrm{(k)}$ and evaluate $e_{NS}$\\
	\STATE \qquad\quad Solve Allen-Cahn equation for updated values of $\phi^\mathrm{n+1}_\mathrm{(k)}$ on $\mathcal{T}^\mathrm{n+1}_\mathrm{(k)}$ and evaluate $e_{AC}$\\
	\STATE \qquad\quad Evaluate the error estimator $\eta$\\
	\STATE \qquad\quad \textbf{if}\ ($e_{NS} \leq tol_{NS}$) \textbf{and} ($e_{AC} \leq tol_{AC}$) \textbf{and} ($\eta \leq tol_{R}$) \textbf{then}\\
	\STATE \qquad\qquad refine = 0; coarsen = 1; break = 1\\
	\STATE \qquad\quad \textbf{elseif}\ ($e_{NS} > tol_{NS}$) \textbf{and} ($e_{AC} > tol_{AC}$) \textbf{and} ($\eta \leq tol_{R}$) \textbf{then}\\
	\STATE \qquad\qquad refine = 0; coarsen = 0\\
	\STATE \qquad\quad \textbf{elseif}\ ($\eta > tol_{R}$) \textbf{then}\\
	\STATE \qquad\qquad refine = 1; coarsen = 0\\
	\STATE \qquad\quad \textbf{elseif}\ ($\mathrm{k} = nIterMax$) \textbf{then}\\
	\STATE \qquad\qquad refine = 0; coarsen = 1; break = 1\\
	\STATE \qquad\quad \textbf{endif}\\
	\STATE \qquad\quad \textbf{if}\ ($\mathrm{n}_\mathrm{el} > nElemMax$) \textbf{and} ( $\mathrm{k}=1$ \textbf{or} $\mathrm{k}=nIterMax$ ) \textbf{then}\\
	\STATE \qquad\qquad refine = 0; coarsen = 1\\
	\STATE \qquad\quad \textbf{elseif}\ ($\mathrm{n}_\mathrm{el} > nElemMax$) \textbf{and} ($\mathrm{k}\neq 1$) \textbf{and} (break $\neq$ 1) \textbf{then}\\
	\STATE \qquad\qquad refine = 0; coarsen = 0\\
	\STATE \qquad\quad \textbf{endif}\\
	\STATE \qquad\quad \textbf{if}\ (coarsen = 1) \textbf{then}\\
	\STATE \qquad\qquad coarsen the mesh\\
	\STATE \qquad\quad \textbf{endif}\\
	\STATE \qquad\quad \textbf{if}\ (refine = 1) \textbf{then}\\
	\STATE \qquad\qquad refine the mesh\\
	\STATE \qquad\quad \textbf{endif}\\
	\STATE \qquad\quad Satisfy boundary conditions on the new mesh $\mathcal{T}^\mathrm{n+1}_\mathrm{(k+1)}$\\
	\STATE \qquad\quad \textbf{if}\ (break = 1) \textbf{then}\\
	\STATE \qquad\qquad break the nonlinear interation loop\\
	\STATE \qquad\quad \textbf{endif}\\
	\STATE \quad Copy the solution of current time step to previous time step on $\mathcal{T}^\mathrm{n+1}_\mathrm{(k+1)}$ 	
  \end{algorithmic}
\end{algorithm}

The proposed algorithm for the coupled Navier-Stokes and Allen-Cahn solver is summarized in Algorithm \ref{algorithm_1}. Consider the velocity $\boldsymbol{u}^\mathrm{n}$, pressure $p^\mathrm{n}$ and the order parameter $\phi^\mathrm{n}$ given at the discretized points at time $t^\mathrm{n}$ on a grid $\mathcal{T}^\mathrm{n}$ (line 3). The variables are predicted for the next time step before the start of the nonlinear iterations to $\boldsymbol{u}^\mathrm{n+1}_{(0)}$, $p^\mathrm{n+1}_{(0)}$ and $\phi^\mathrm{n+1}_{(0)}$ (lines 4-7). In the first step of a nonlinear iteration $\mathrm{k}$, the Navier-Stokes equations are solved using Eq.~(\ref{LS_NS}) and the velocity and pressure values at $\mathrm{n+1}$ are updated. The error corresponding to the Navier-Stokes equations $e_{NS}$ is also evaluated at this step. The updated velocity is then transferred to the Allen-Cahn solver which solves Eq.~(\ref{LS_AC}) and updates the order parameter $\phi^\mathrm{n+1}_{(\mathrm{k})}$. The error $e_{AC}$ is also evaluated for the error in solving the Allen-Cahn equation. Then, the error estimator $\eta$ is computed using Eqs.~(\ref{etaR1}) and (\ref{etaR2}). At this step, the convergence criterion is checked based on which it is decided if the mesh is to be coarsened/refined (lines 12-20). The tolerances for the convergence criteria are set by the user, whereby $tol_{NS}$, $tol_{AC}$ and $tol_R$ denote the tolerances for the Navier-Stokes, the Allen-Cahn equations and the error estimator,  respectively. 

The convergence criteria are constructed such that the mesh will be coarsened only at the last nonlinear iteration and it will be refined until the criteria for $tol_R$ is satisfied. In the algorithm, while $nIterMax$ denotes the maximum number of nonlinear iterations specified by the user,  $nElemMax$ represents  the number of maximum elements that can exist in the domain. We coarsen the grid if the number of elements $\mathrm{n}_\mathrm{el}$ exceeds this limit (lines 21-25). This ensures that there is no refining/coarsening in the intermediate nonlinear iterations to capture the nonlinearities of the underlying equations. Based on the outcome of convergence criteria, the mesh is refined/coarsened (lines 26-31) after which the boundary conditions on the new mesh $\mathcal{T}^\mathrm{n+1}_\mathrm{(k+1)}$ are satisfied. After the end of the nonlinear iterations, the solver updates the variables and proceeds to the next time step.
While satisfying the convergence criteria, the primary objective is to reduce the residual errors due to the adaptive algorithm or the error indicator $\eta$, for which the mesh will be refined. Once the refinement criterion is satisfied, the solver will tend to reduce the individual errors corresponding to the Navier-Stokes and the Allen-Cahn equations by iterating through the nonlinear iterations while maintaining the same mesh resolution. This captures the nonlinearities and maintains the convergence properties of the underlying equations. Based on the proposed algorithmic procedure, we  next present the numerical tests of increasing complexity to assess the effectiveness of the partitioned adaptive scheme.

\section{Convergence and performance study}
\label{tests}

\subsection{Spinodal decomposition in a complex curved geometry}
In this subsection, we demonstrate our adaptive phase-field finite element formulation for the spinodal decomposition in a complicated curved domain. The motivation of this test is to demonstrate the generality and the energy stability of our method based on unstructured grid and the body-fitted formulation for complex geometric boundaries 
for two-phase transport found in numerous micro-fluidics and oil/gas applications.

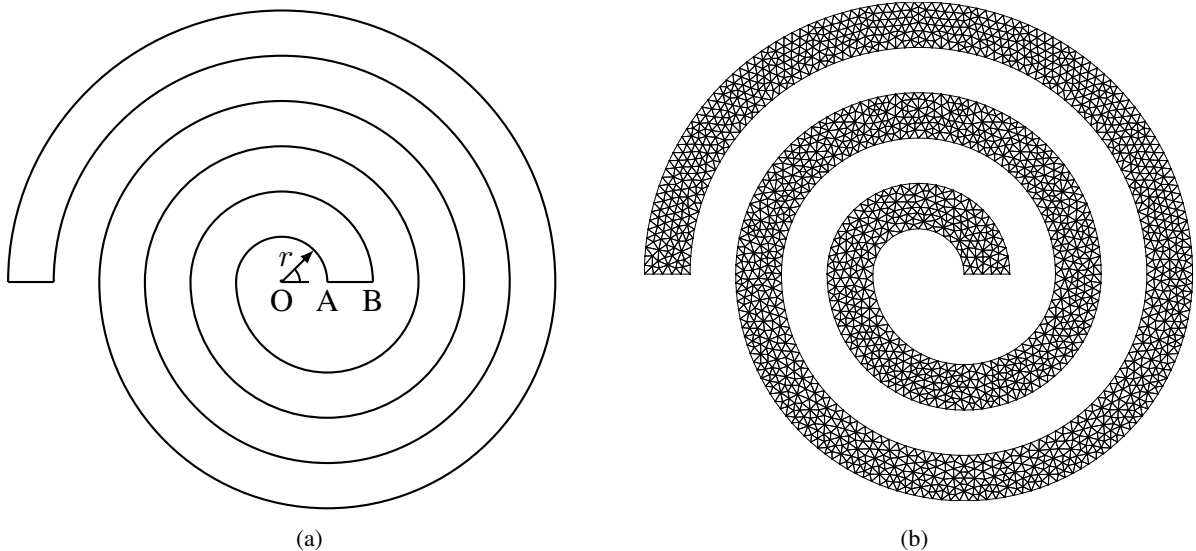
\begin{figure}
\centering
\begin{subfigure}[b]{0.5\textwidth}
\begin{tikzpicture}[decoration={markings,mark=at position 1.0 with {\arrow{>}}},scale=1.2, every node/.style={scale=1.2}]
	\draw [thick](0,0) arc (0:180:0.5);
	\draw [thick](0.5,0) arc (0:180:1);
	\draw [thick](-1,0) arc (180:360:1);
	\draw [thick](-1.5,0) arc (180:360:1.5);
	\draw [thick](1,0) arc (0:180:1.5);
	\draw [thick](1.5,0) arc (0:180:2);
	\draw [thick](-2,0) arc (180:360:2);
	\draw [thick](-2.5,0) arc (180:360:2.5);
	\draw [thick](2,0) arc (0:180:2.5);
	\draw [thick](2.5,0) arc (0:180:3);
	\draw [thick](0,0)--(0.5,0);
	\draw [thick](-3,0)--(-3.5,0);
	\draw (0,-0.2) node {A};
	\draw (0.5,-0.2) node {B};
	\draw (-0.5,0) node{$\mathbf{\cdot}$};
	\draw (-0.5,-0.2) node {O};
	\draw [->,thick] (-0.5,0) to (-0.1464,0.3535);
	\draw [thick] (-0.5,0)--(-0.2,0);
	\draw [thick](-0.3,0) arc (0:45:0.2);
	\draw (-0.45,0.25) node {$r$};
\end{tikzpicture}
	\caption{}
\end{subfigure}%
\begin{subfigure}[b]{0.5\textwidth}
		\includegraphics[trim={2cm 2cm 2cm 0.2cm},clip,width=8cm]{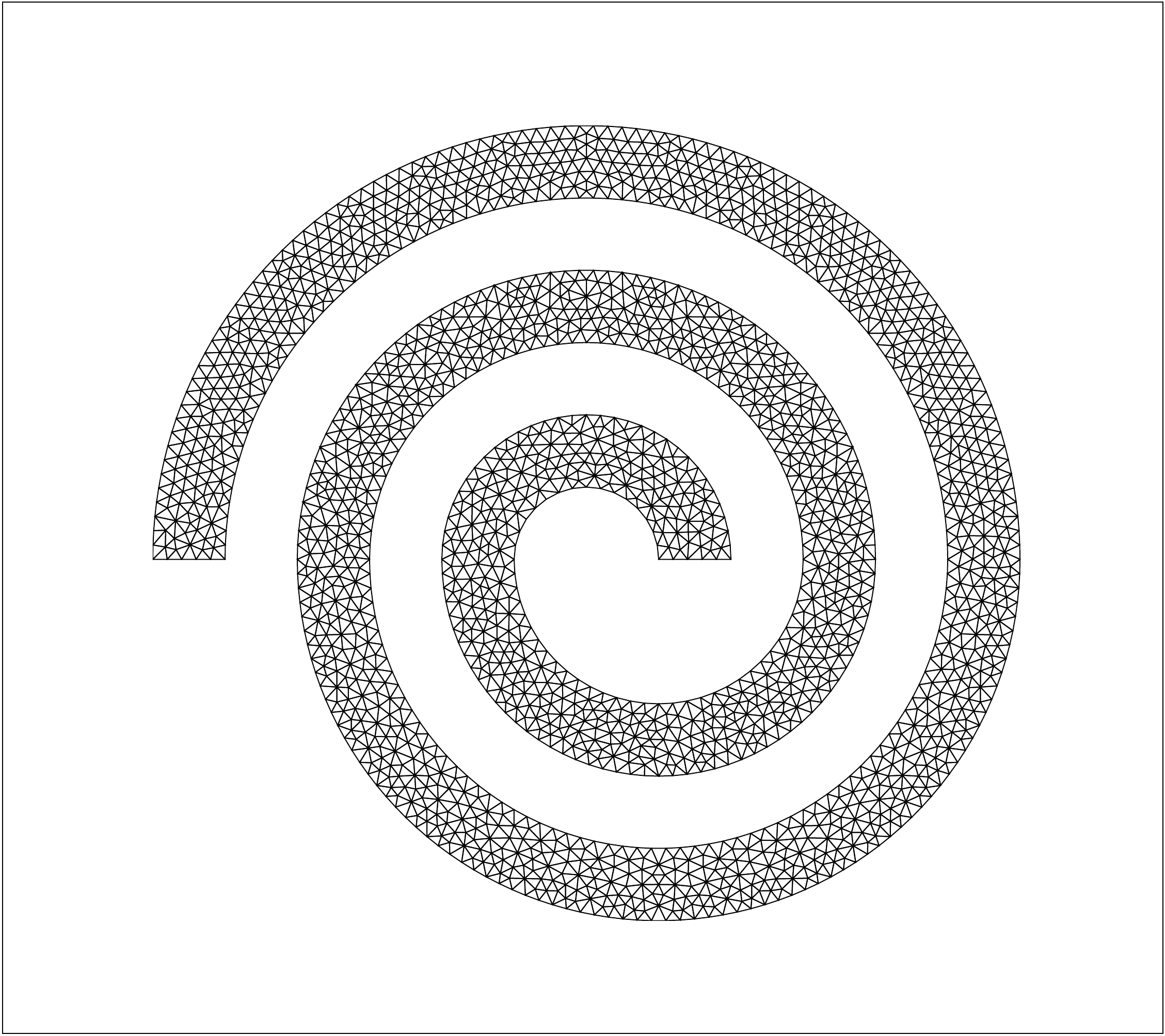}		
	\caption{}
	\end{subfigure}
\caption{Spinodal decomposition in a spiral domain: (a) schematic diagram showing the computational domain and (b) initial triangular mesh at $t=0$. In (a), the coordinates of O, A and B are $(0,0)$, $(0.5,0)$ and $(1,0)$ respectively and $r$ denotes the radius of the spiral curve varying with angle $\theta$ as given in Eq.~(\ref{spiral_curve}).} 
\label{sp_decomp_1}
\end{figure}

\begin{figure}
\centering
		\includegraphics[trim={3cm 0 0cm 0cm},clip,width=17cm]{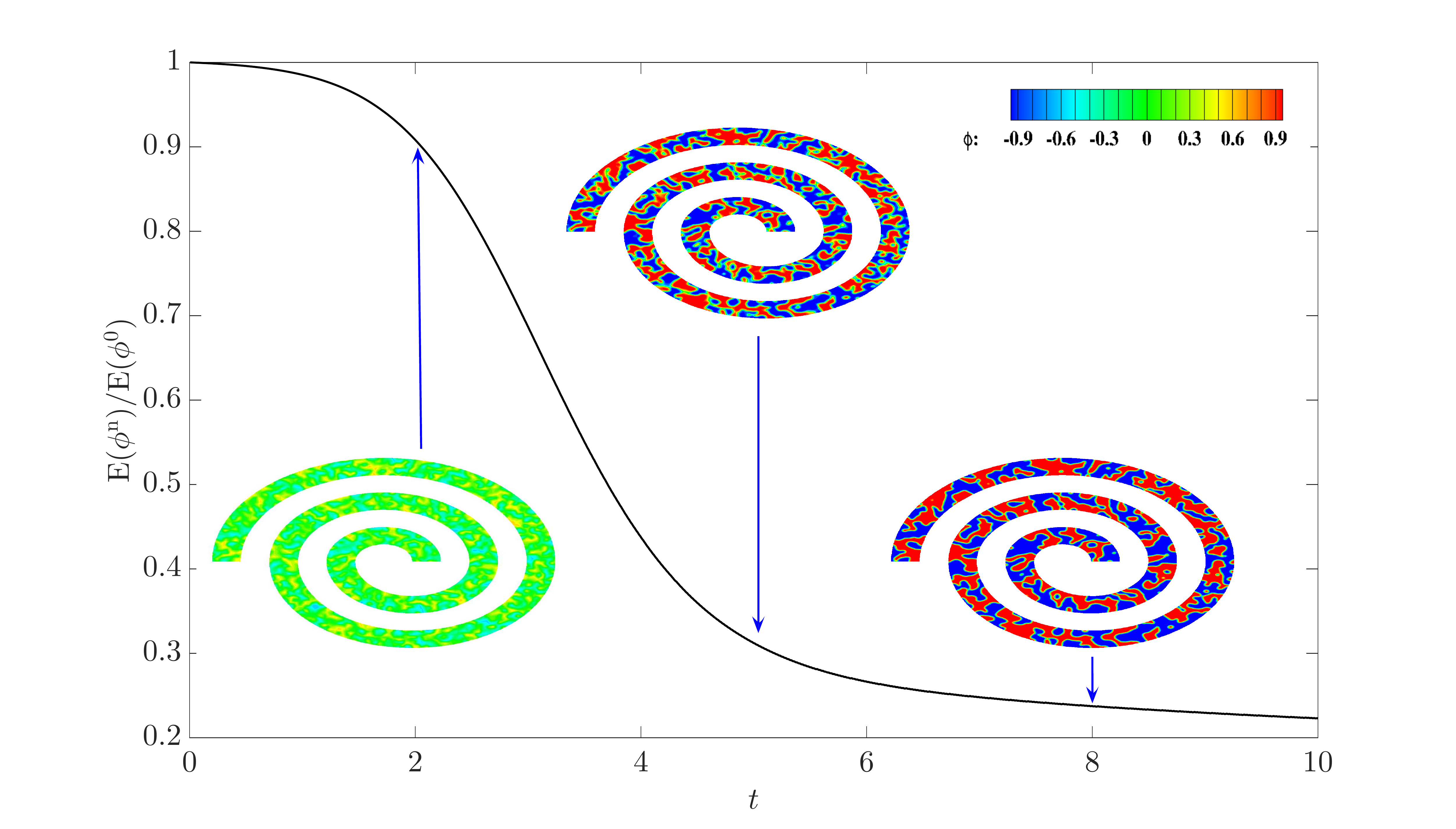}		
\caption{Decay of the total discrete energy with time for the spinodal decomposition in a spiral domain. The contours of the order parameter are shown at different time instances $t\in [2, 5, 8]$. The total energy decreases with time making the scheme energy stable.} 
\label{sp_decomp_2}
\end{figure}

\begin{figure}
\centering
	\begin{subfigure}[b]{0.5\textwidth}
		\includegraphics[trim={2cm 2cm 2cm 0.2cm},clip,width=8cm]{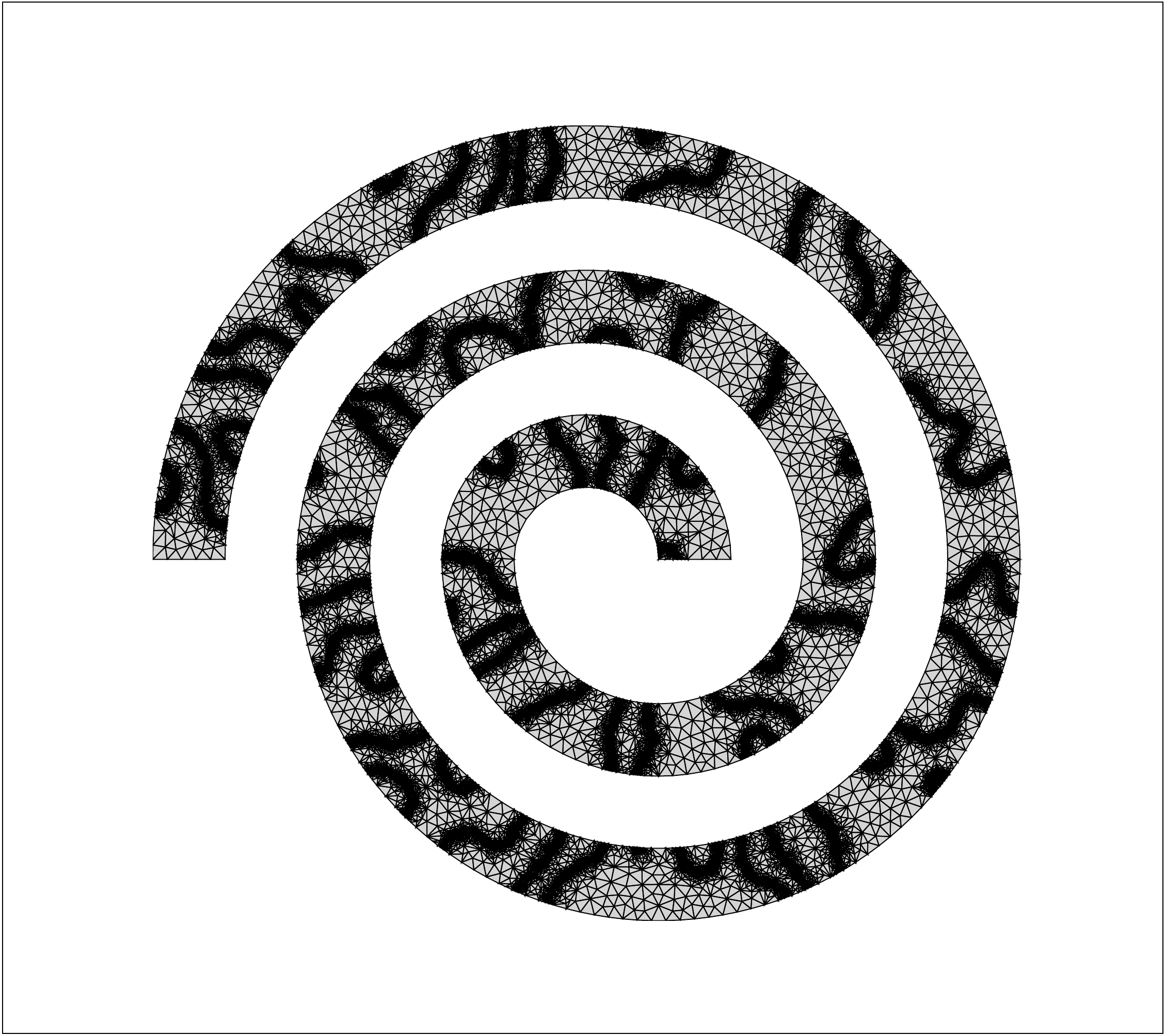}		
	\caption{}
	\end{subfigure}%
	\begin{subfigure}[b]{0.5\textwidth}
		\includegraphics[trim={2cm 2cm 2cm 0.2cm},clip,width=8cm]{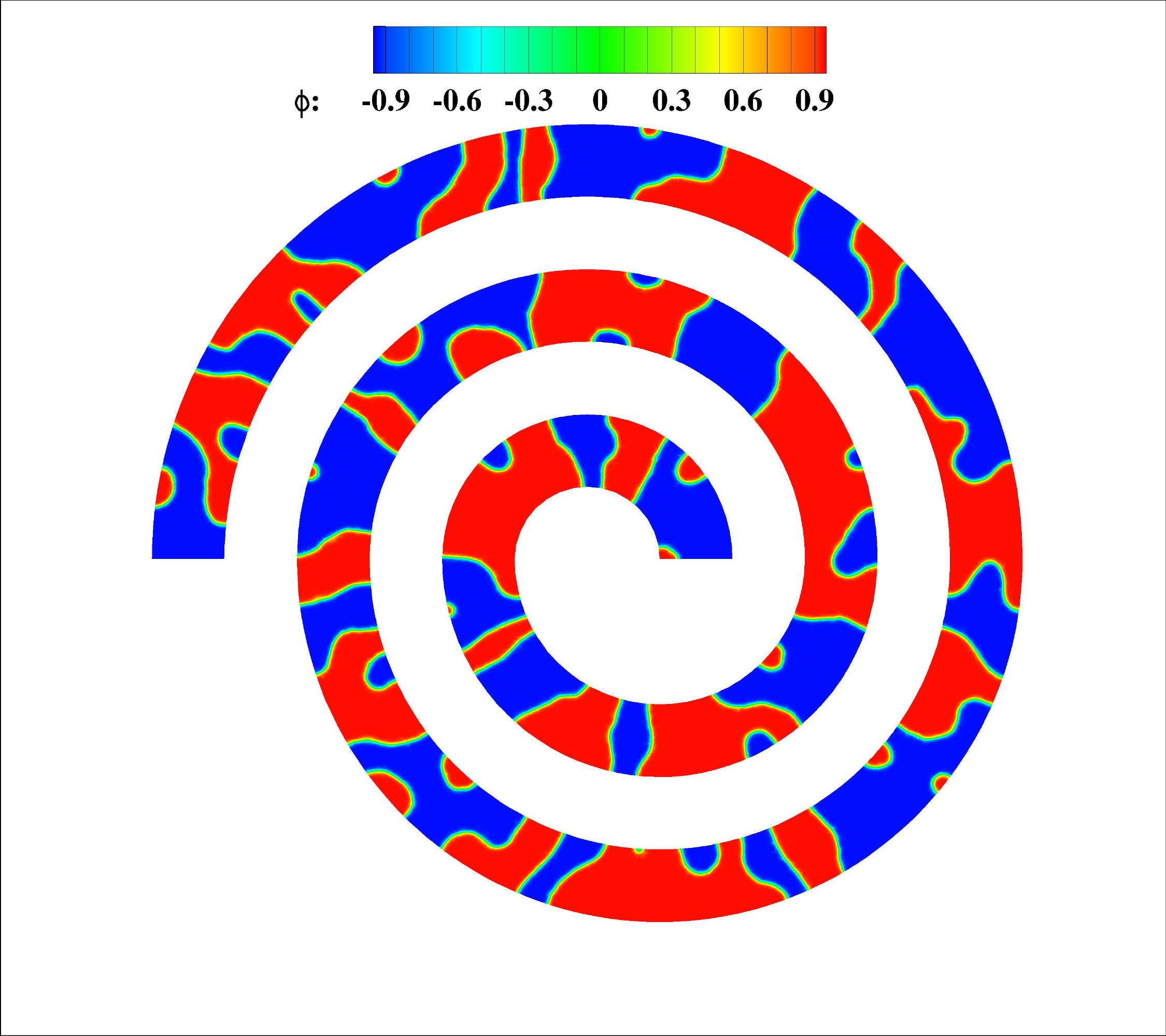}		
	\caption{}
	\end{subfigure}
\caption{Solution of the spinodal decomposition in a spiral domain at $t=100$: (a) the adapted mesh and (b) the contours of the order parameter $\phi$.} 
\label{sp_decomp_3}
\end{figure}

We take the complex geometry as a spiral curve made from semicircles of increasing radius, which is given as:
\begin{align} \label{spiral_curve}
	c(r,\theta) = (r\mathrm{cos}(\theta),r\mathrm{sin}(\theta)), 
\end{align}
where $\theta \in [0,5\pi]$ and $r$ is the radius of the semicircle which varies as $0.5+(n-1)0.5$ with $n=1,2,3,4,5$ corresponding to the intervals $[0,\pi]$,$[\pi,2\pi]$,$[2\pi,3\pi]$,$[3\pi,4\pi]$ and $[4\pi,5\pi]$ respectively with center of the semicircle alternatively varying between $(0,0)$ and $(0.5,0)$ with the intervals. The width of the curve is $0.5$. The computational domain with an unstructured triangular mesh discretization is shown in Fig. \ref{sp_decomp_1}. The initial condition of the order parameter is considered as
\begin{align}
	\phi(x,y,0) = 0.1\mathrm{rand},
\end{align}
where $\mathrm{rand}$ are the random values from the uniform distribution in the interval $[-1,1]$. The interface thickness parameter is $\varepsilon=0.01$. With $\Delta t=0.01$, the simulation is run for $10000$ time steps. The user-defined convergence tolerances are selected as $tol_{NS}=tol_{AC}=5\times 10^{-4}$ and $tol_R=10^{-2}$. The refining and coarsening parameters are $\theta=0.5$ and $\theta_c=0.05$, respectively. The random initial condition falls under the chemical spinodal region of the double well potential curve which has a higher free-energy compared to the equilibrium phases. This leads to a phase separation due to the spinodal decomposition which tends to minimize the free energy functional. We track the variation in the discrete free energy given by Eq.~(\ref{energy}) through the spinodal decomposition process in Fig. \ref{sp_decomp_2}. The energy functional at time $t^\mathrm{n}$ is given by
\begin{align}
	\mathrm{E}(\phi^\mathrm{n}) = \int_\Omega \bigg( \frac{1}{2}(\varepsilon^2 + \chi \frac{|\mathcal{R}(\phi_\mathrm{h})|}{|\nabla\phi_\mathrm{h}|}k_c^\mathrm{add})|\nabla\phi^\mathrm{n}|^2 + F(\phi^\mathrm{n}) \bigg) \mathrm{d}\Omega.
\end{align} 
The trend clearly depicts the decreasing energy with a large gradient flow at the initial phase separation region while the rate of decrease becomes slower as the coarsening of the phases begin. The adaptive mesh at $t=100$ of the computational domain with the contours of the order parameter are depicted in Fig. \ref{sp_decomp_3}.


\subsection{Volume conserved motion by curvature}
\begin{figure}
\centering
	\begin{subfigure}[b]{0.5\textwidth}
		\includegraphics[trim={10cm 0 11.8cm 0cm},clip,width=7cm]{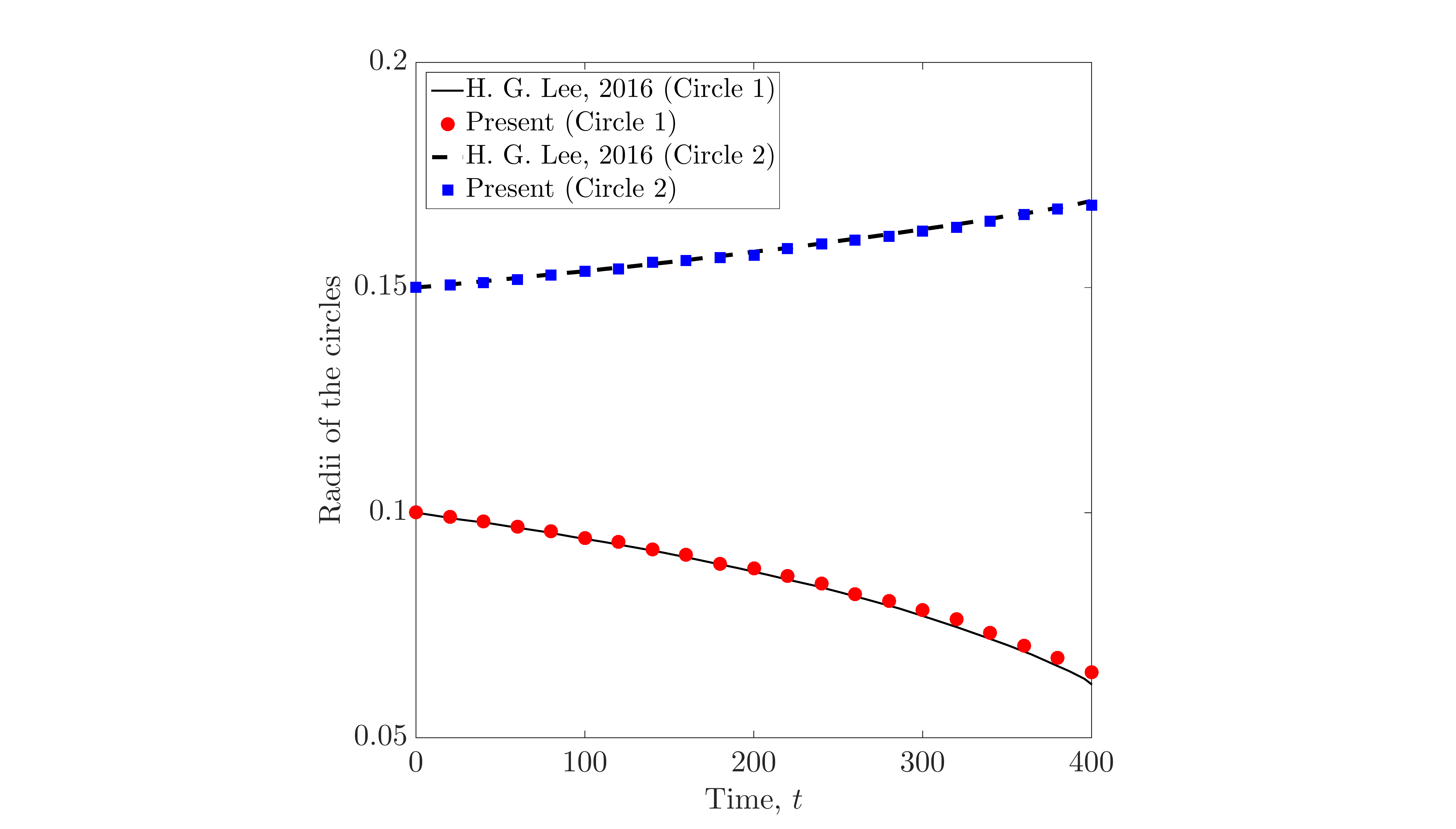}		
	\caption{}
	\end{subfigure}%
	\begin{subfigure}[b]{0.5\textwidth}
		\includegraphics[trim={10cm 0 11.8cm 0cm},clip,width=7cm]{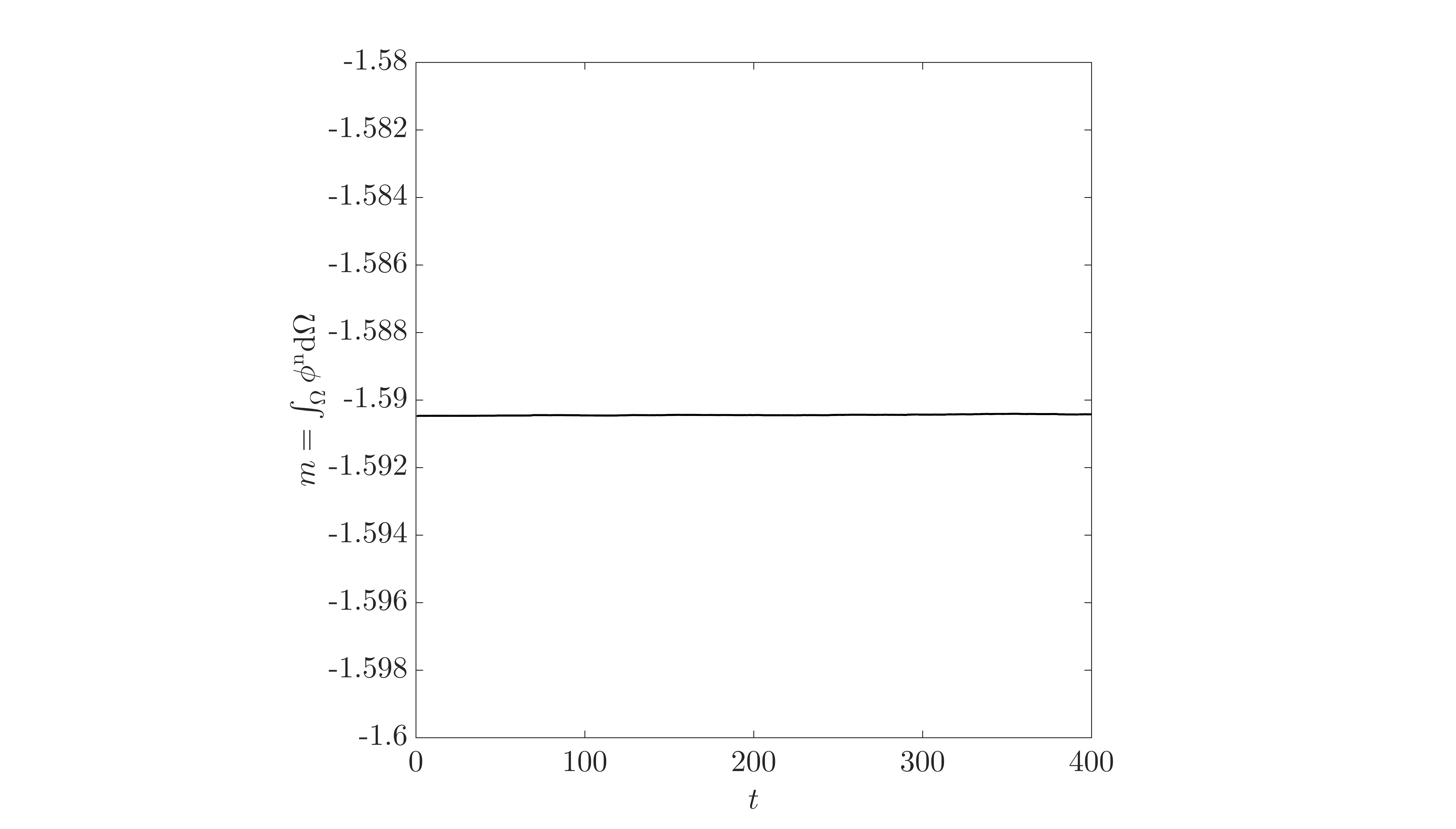}		
	\caption{}
	\end{subfigure}
\caption{Assessment of volume conservation during mean curvature flow: (a) validation of the evolution of the radii of the two circles with the literature \cite{H_G_Lee}, and (b) the variation of the total mass of the order parameter $\phi$ as a function of time.} 
\label{ac_val_2}
\end{figure}

We next illustrate the effectiveness of the standalone Allen-Cahn solver for the conservation of volume (or mass) of the order parameter $\phi$ .  
The computational domain consists of a square domain $[0,1]\times [0,1]$ and the periodic boundary conditions are imposed on all the boundaries. Based on the work of \cite{H_G_Lee}, the initial condition for this problem is considered as:
\begin{align}
	\phi(x,y,0) = 1 + \mathrm{tanh}\bigg( \frac{R_1 - \sqrt{(x-0.25)^2 + (y-0.25)^2}}{\sqrt{2}\varepsilon} \bigg) + \mathrm{tanh}\bigg( \frac{R_2 - \sqrt{(x-0.57)^2 + (y-0.57)^2}}{\sqrt{2}\varepsilon} \bigg)
\end{align}
where $R_1=0.1$ and $R_2=0.15$ are the radii of the two circles centered at $(0.25,0.25)$ and $(0.57,0.57)$, respectively. The interface thickness parameter is chosen as $\varepsilon=0.005$ with the error tolerances as $tol_{NS}=tol_{AC}=5\times 10^{-4}$ and $tol_R=3\times 10^{-4}$. The refining and coarsening parameters are $\theta=0.5$ and $\theta_c=0.05$,  respectively.
\begin{figure}
\centering
	\begin{subfigure}[b]{0.5\textwidth}
\qquad
\begin{tikzpicture}[decoration={markings,mark=at position 1.0 with {\arrow{>}}},scale=5.9]
	\draw (0,0) -- (1,0)-- (1,1) -- (0,1) -- cycle;
	\draw[fill={rgb:black,1;white,2}, fill opacity=0.4] (0.25,0.25) circle (0.1cm);
	\draw[fill={rgb:black,1;white,2}, fill opacity=0.4] (0.57,0.57) circle (0.15cm);
	\draw (0.25,0.25) node(A){$\Omega_1$};
	\draw (0.57,0.57) node(B){$\Omega_1$};
	\node [below=0.3cm of A]{Circle 1};
	\node [below=0.6cm of B]{Circle 2};
	\draw (0.75,0.25) node{$\Omega_2$};
	\draw[thick,postaction={decorate}] (0,0) to (0.2,0);
	\draw[thick,postaction={decorate}] (0,0) to (0,0.2);
	\draw (0.2,0) node[anchor=north]{X};
	\draw (0,0.2) node[anchor=east]{Y};
	\draw[postaction={decorate}] (1.05,0.5) to (1.05,1);
	\draw[postaction={decorate}] (1.05,0.5) to (1.05,0);
	\draw (1.01,0) -- (1.1,0);
	\draw (1.01,1) -- (1.1,1);
	\draw (1.05,0.5) node[anchor=west]{$1$};
	\draw[postaction={decorate}] (0.5,1.05) to (1,1.05);
	\draw[postaction={decorate}] (0.5,1.05) to (0,1.05);
	\draw (0,1.01) -- (0,1.1);
	\draw (1,1.01) -- (1,1.1);
	\draw (0.5,1.05) node[anchor=south]{$1$};
\end{tikzpicture}
	\caption{}
	\end{subfigure}%
	\begin{subfigure}[b]{0.5\textwidth}
		\includegraphics[trim={0.2cm 0.2cm 0.2cm 0.2cm},clip,width=8.25cm]{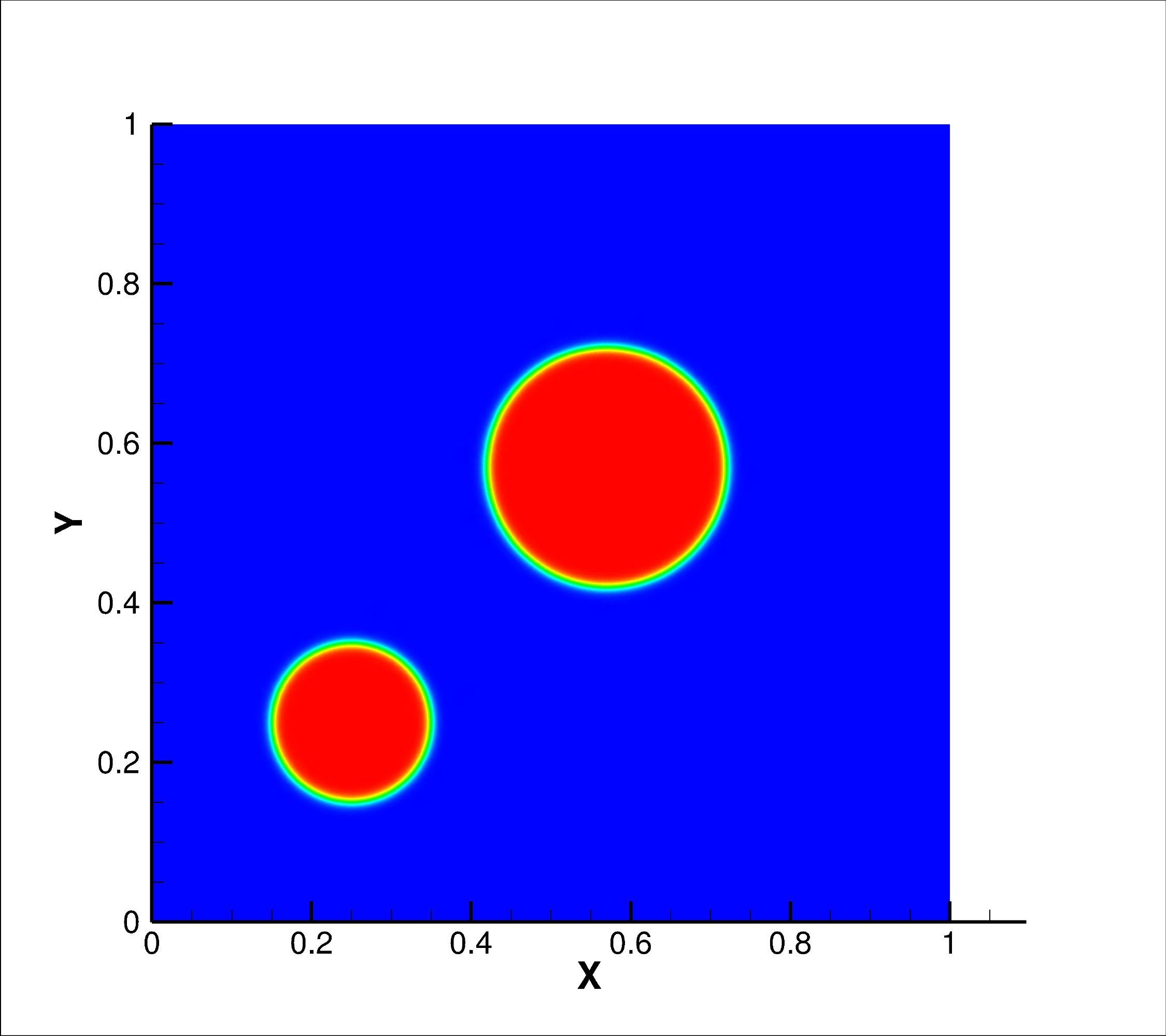}		
	\caption{}
	\end{subfigure}

	\begin{subfigure}[b]{0.5\textwidth}
	\ \ \
		\includegraphics[trim={0.2cm 0.2cm 0.2cm 0.2cm},clip,width=8.5cm]{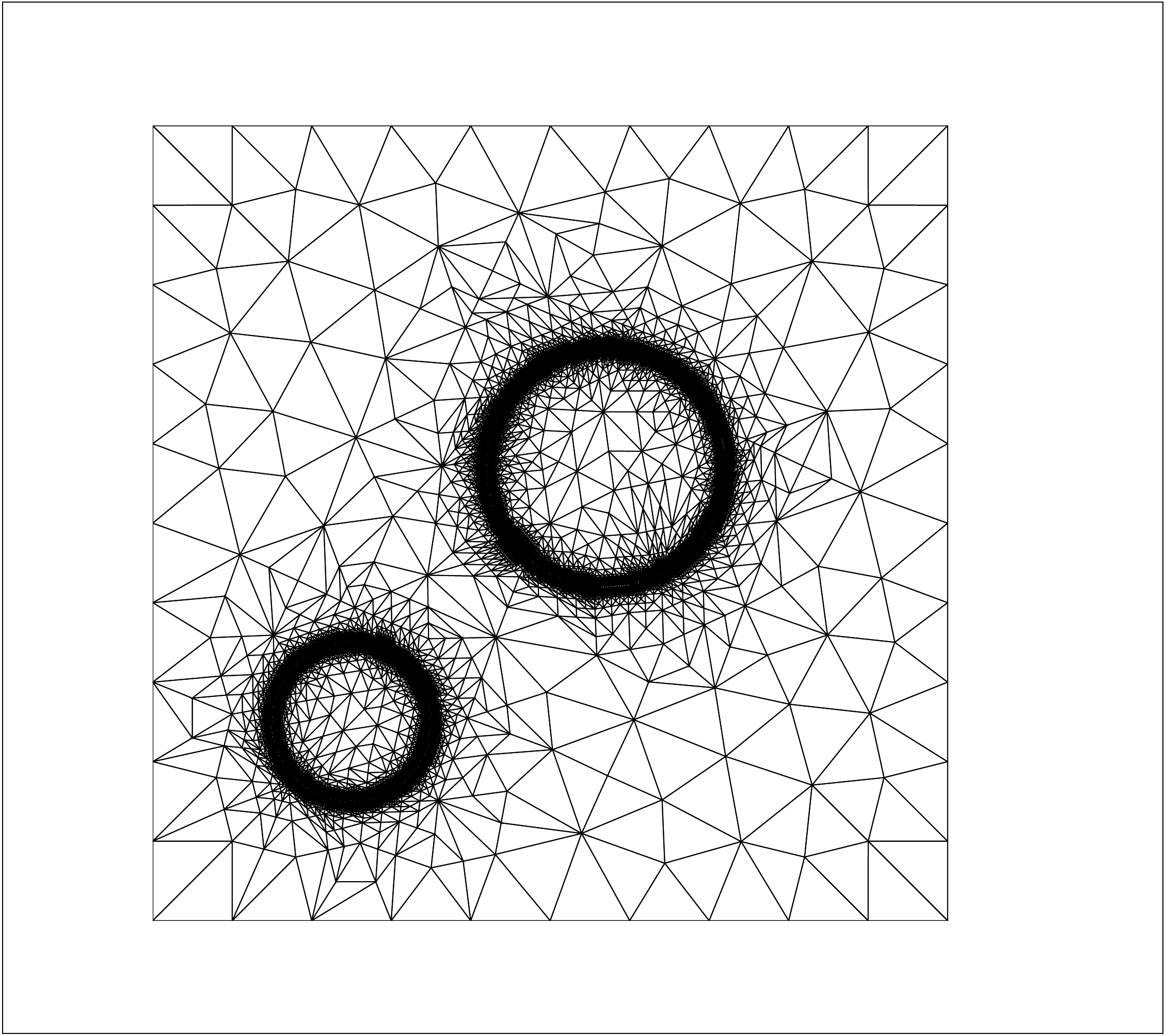}
	\caption{}
	\end{subfigure}%
	\begin{subfigure}[b]{0.5\textwidth}
		\includegraphics[trim={0.2cm 0.2cm 0.2cm 0.2cm},clip,width=8.5cm]{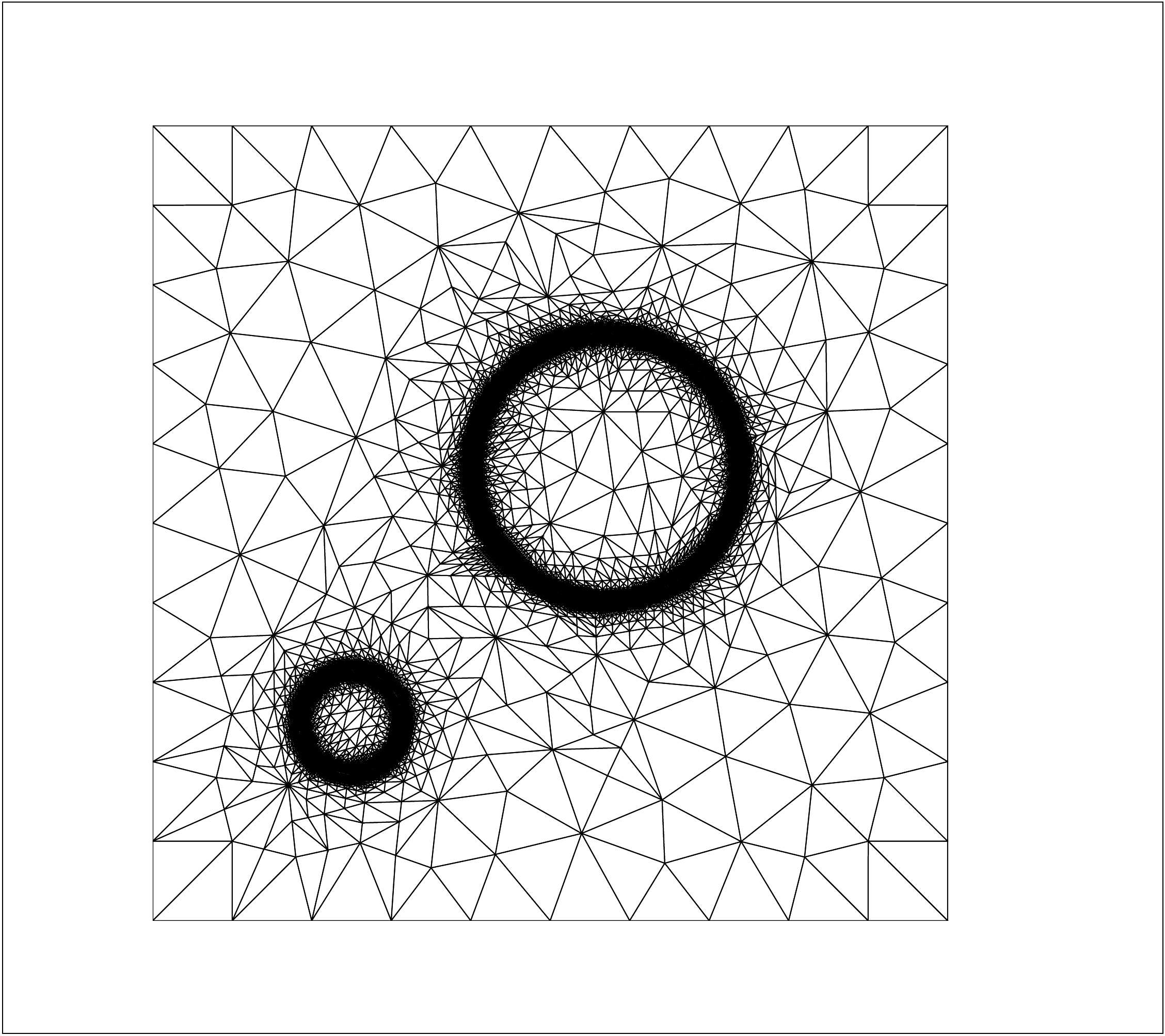}
	\caption{}
	\end{subfigure}
\caption{Volume conservation during the interface evolution according to mean curvature flow: (a) schematic diagram showing the computational domain, (b) the contour plot of $\phi$ at $t=0$, the adaptive mesh at (c) $t=0$, and (d) $t=400$.  In (a), $\Omega_1$ and $\Omega_2$ are the two phases with periodic boundary conditions imposed on all the sides. } 
\label{ac_val}
\end{figure}
The mesh is refined iteratively along the initial profile of the order parameter based on the gradients of the initial condition before the start of the time loop. The time step size in the present simulation is $0.4$ with the final time $t=400$. The problem set-up with the evolution of the adaptive mesh of the two circles is shown in Fig. \ref{ac_val}. The variation of the radii is compared with the literature in Fig. \ref{ac_val_2}(a). The results are in very close agreement with the reference. The variation of the total mass of the order parameter as a function of time is also shown in Fig. \ref{ac_val_2}(b). The mass of the order parameter at a particular time step $t^\mathrm{n}$ is defined as:
\begin{align} \label{mass_def}
	m = \int_\Omega \phi^\mathrm{n} \mathrm{d}\Omega.
\end{align}
 Quantitatively, the fraction of the change in mass over the total time and the initial mass is evaluated as $3.2066\times 10^{-5}$ ($\approx 0.003\%$).

\subsection{Liquid sloshing in a tank}
\begin{figure}
\centering
\hspace{-1cm}
	\begin{subfigure}[b]{0.31\textwidth}
\begin{tikzpicture}[decoration={markings,mark=at position 1.0 with {\arrow{>}}},scale=4]
	\draw[fill=black!10] plot[smooth] coordinates {(0,0.91) (0.1,0.9149) (0.2,0.9291) (0.3,0.9512) (0.4,0.9791) (0.5,1.01) (0.6,1.0409) (0.7,1.0688) (0.8,1.0909) (0.9,1.1051) (1.0,1.11)} -- (1,0) -- (0,0) -- (0,0.91);
	\draw (0,0.91) -- (0,1.5)-- (1,1.5) -- (1,1.11);
	\draw (0.5,0.75) node[anchor=south](A){$\Omega_1$};
	\node [below = 0.0cm of A]{$(\rho_1, \mu_1)$};
	\draw (0.5,1.25) node[anchor=south](B){$\Omega_2$};
	\node [below = 0.0cm of B]{$(\rho_2, \mu_2)$};
	\draw (0.2,0.92) node[anchor=south]{$\Gamma$};
	\draw[thick,postaction={decorate}] (0,0) to (0.2,0);
	\draw[thick,postaction={decorate}] (0,0) to (0,0.2);
	\draw (0.2,0) node[anchor=north]{X};
	\draw (0,0.2) node[anchor=east]{Y};
	\draw[postaction={decorate}] (1.05,0.5) to (1.05,1.5);
	\draw[postaction={decorate}] (1.05,0.5) to (1.05,0);
	\draw (1.01,0) -- (1.1,0);
	\draw (1.01,1.5) -- (1.1,1.5);
	\draw (1.05,0.75) node[anchor=west]{$1.5$};
	\draw[postaction={decorate}] (0.5,1.55) to (1,1.55);
	\draw[postaction={decorate}] (0.5,1.55) to (0,1.55);
	\draw (0,1.51) -- (0,1.6);
	\draw (1,1.51) -- (1,1.6);
	\draw (0.5,1.55) node[anchor=south]{$1$};
	\draw (0.5,-0.15) node{};
\end{tikzpicture}
	\caption{}
	\end{subfigure}%
	\hspace{0.4cm}
	\begin{subfigure}[b]{0.37\textwidth}
		\includegraphics[trim={0.5cm 0.2cm 9cm 0.2cm},clip,width=5.1cm]{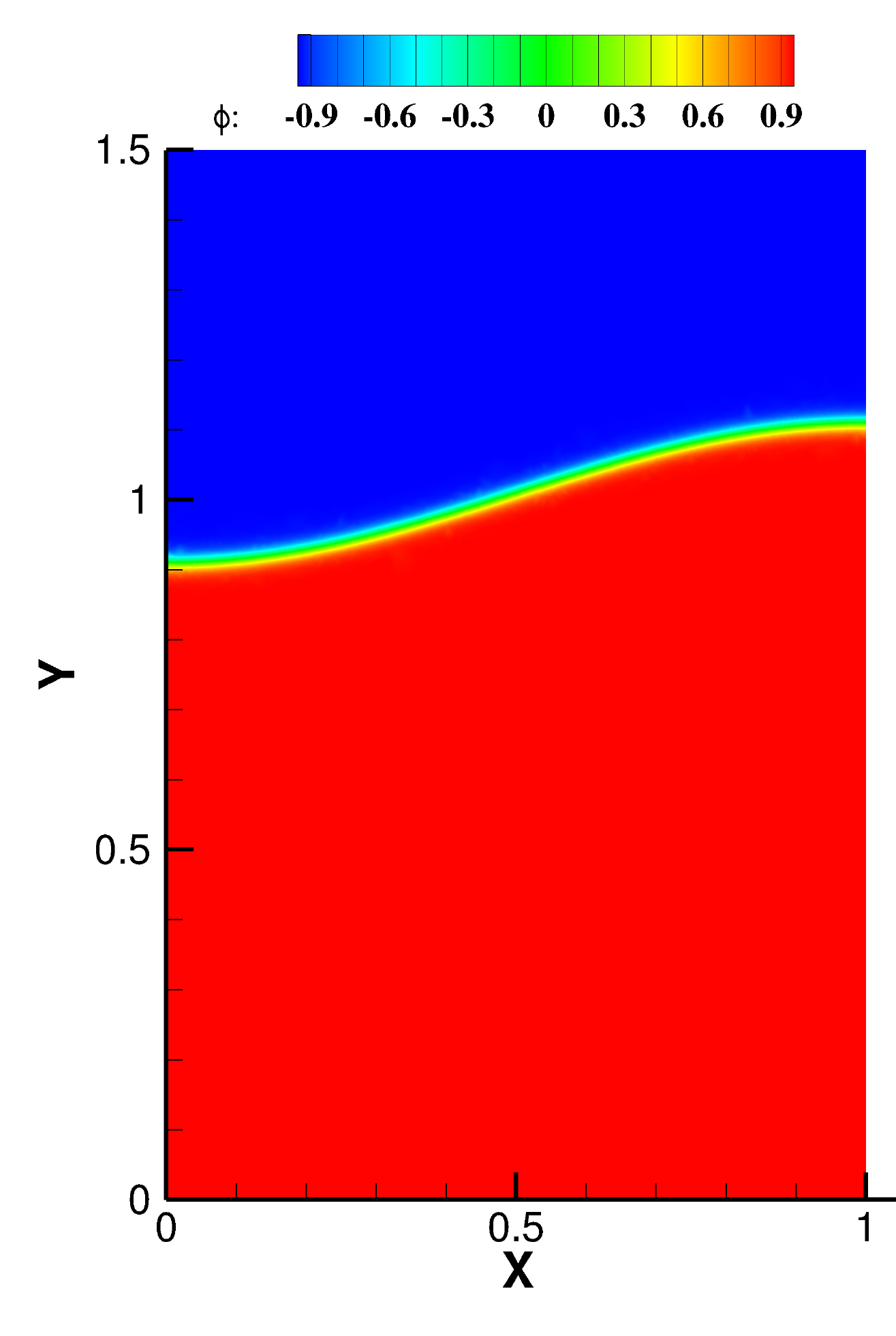}
	\caption{}
	\end{subfigure}
	\hspace{-1cm}
	\begin{subfigure}[b]{0.37\textwidth}
		\includegraphics[trim={0.5cm 0.2cm 9cm 0.2cm},clip,width=5.1cm]{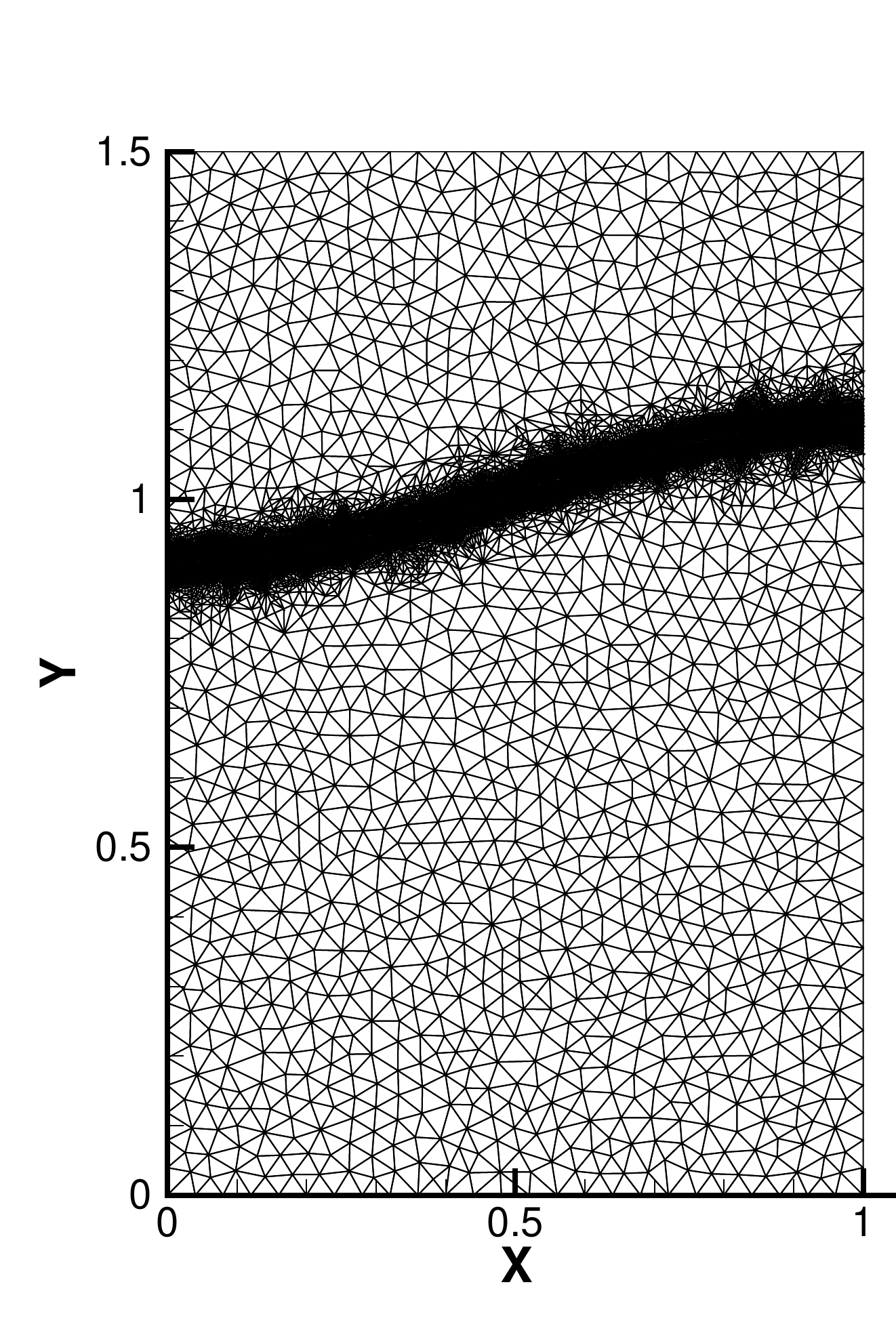}
	\caption{}
	\end{subfigure}
	\hspace{-2cm}
\caption{Sloshing in a rectangular tank: (a) schematic diagram showing the computational domain,  (b) contour plot of the order parameter $\phi$ at $t=0$, and (c) initial refined mesh employed for the simulation. In (a), $\Omega_1$ and $\Omega_2$ are the two fluid phases with densities $\rho_1=1000$ and $\rho_2=1$, viscosities $\mu_1=1$, $\mu_2=0.01$ and acceleration due to gravity $\boldsymbol{g}=(0,-1,0)$ and slip boundary condition is satisfied along all the boundaries of the domain.} 
\label{ST_1}
\end{figure}
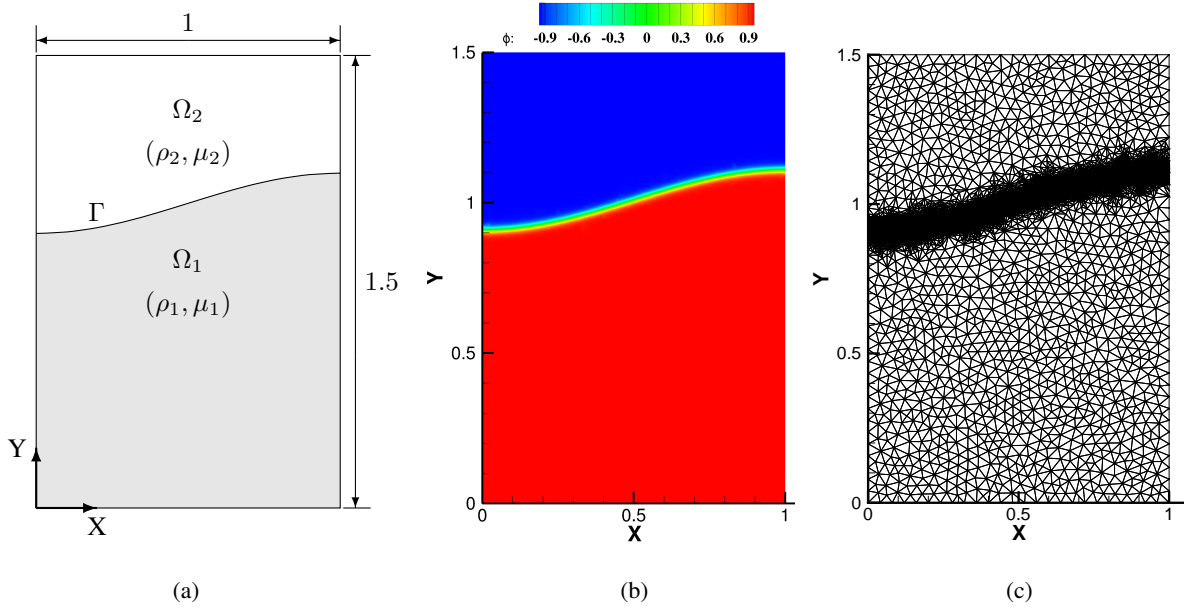

In this subsection, we study the results of the numerical tests conducted for a liquid sloshing tank problem. A rectangular domain $\Omega \in [0,1] \times [0,1.5]$ is discretized with three-node triangles of varying sizes for the simulation. It consists of two phases of a fluid with different densities and viscosities, viz. $\rho_1 = 1000$, $\rho_2 = 1$, $\mu_1 = 1$, $\mu_2 = 0.01$ with the acceleration due to gravity as $\boldsymbol{g} = (0, -1, 0)$. Surface tension effects are neglected. 
The initial condition for the order parameter $\phi$ is defined as:
\begin{align}
	\phi(x,y,0) = - \mathrm{tanh}\bigg( \frac{ y - (1.01 + 0.1\mathrm{sin}((x - 0.5)\pi)) }{\sqrt{2}\varepsilon}\bigg)
\end{align}
Slip boundary condition is satisfied along all the boundaries. The schematic of the problem is given in Fig. \ref{ST_1}(a) with the contour plot of $\phi$ of the initial condition in Fig. \ref{ST_1}(b).
Each test case is set up on an initial grid of element size $\Delta x$. The mesh is refined iteratively along the initial profile of the order parameter based on the gradients of the initial condition before the start of the time loop. A typical initial refined mesh with $\Delta x=0.04$ is shown in Fig. \ref{ST_1}(c). For the detailed analysis of the problem, we perform a series of experiments to assess some of the measured quantities such as mass conservation, the number of degrees of freedom, the computational cost (elapsed time), the error in solving the Allen-Cahn equation ($e_{AC}$) and the residual error for the adaptive procedure ($\eta$). The mass of the order parameter is computed using Eq.~(\ref{mass_def}).
The number of degrees of freedom at a time $t^\mathrm{n}$ is the total number of element nodes of the mesh in the last nonlinear iteration of the time step. The cumulative elapsed time is the cumulative time taken by the solver to complete a time step on $1$ CPU. The simulation is run for $5$ cycles of the oscillation of the interface inside the tank with a time period of $T_{oscill} = 3.6$s. All the tests are simulated using 4 core Intel Xeon $3.50$GHz $\times$ 1 CPU with $16$GB memory. Furthermore, no bounds on the maximum number of elements are considered for the sloshing tank problem, i.e., lines 21-25 of the Algorithm \ref{algorithm_1} are neglected. The refinement and coarsening parameters are selected as $\theta=0.5$ and $\theta_c=0.05$, respectively for all the cases.

\subsubsection{Mesh convergence}
\begin{figure}
		\centering
		\includegraphics[trim={0cm 0.3cm 0cm 0cm},clip,width=15cm]{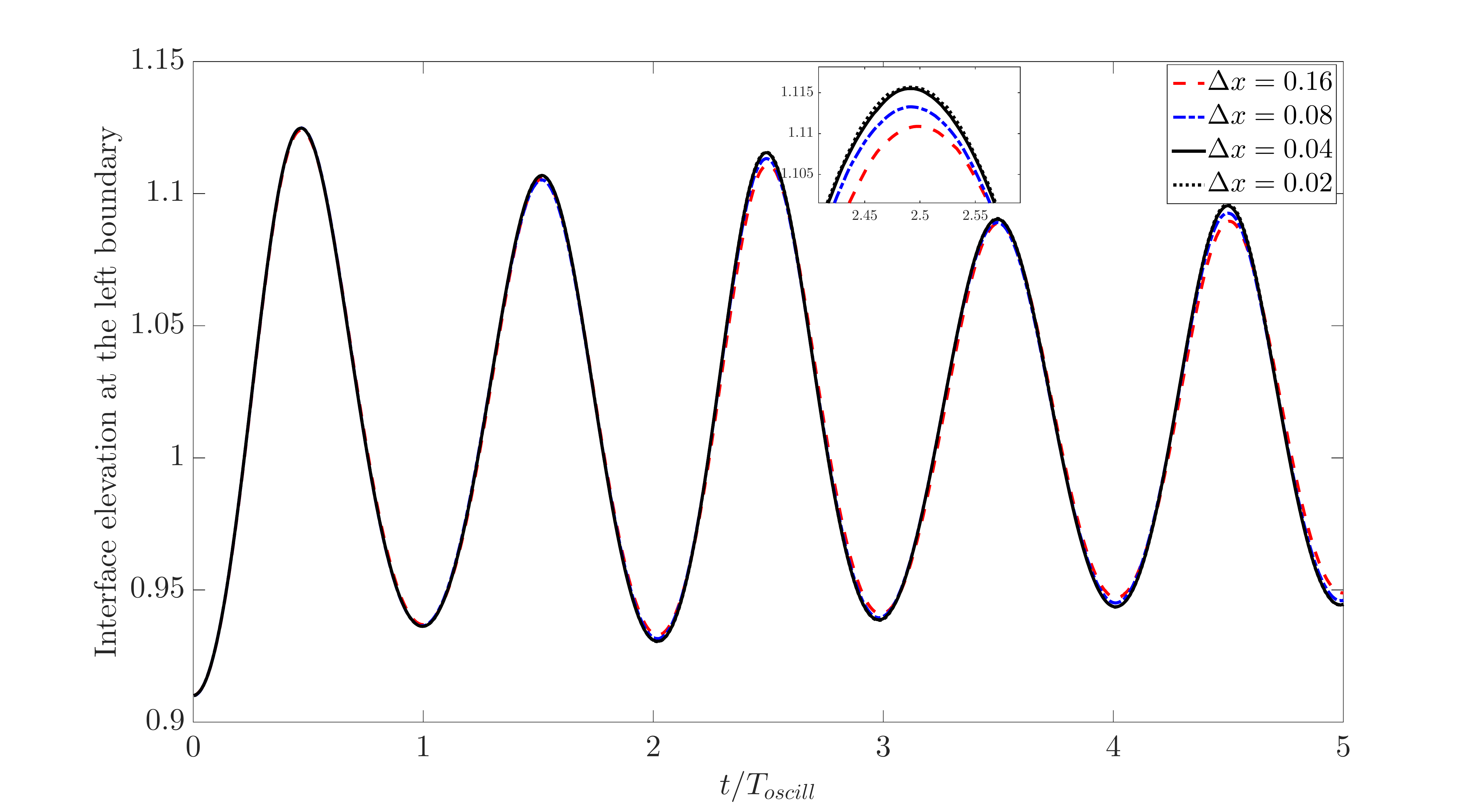}
\caption{Background mesh convergence for the sloshing tank problem: evolution of the interface at the left boundary with $\varepsilon = 0.01$ and different background mesh sizes.} 
\label{ST_2}
\end{figure}

\begin{figure}
\centering
	\hspace{-0.7cm}
	\begin{subfigure}[b]{0.3\textwidth}
		\includegraphics[trim={0.5cm 0.2cm 9cm 0.2cm},clip,width=4cm]{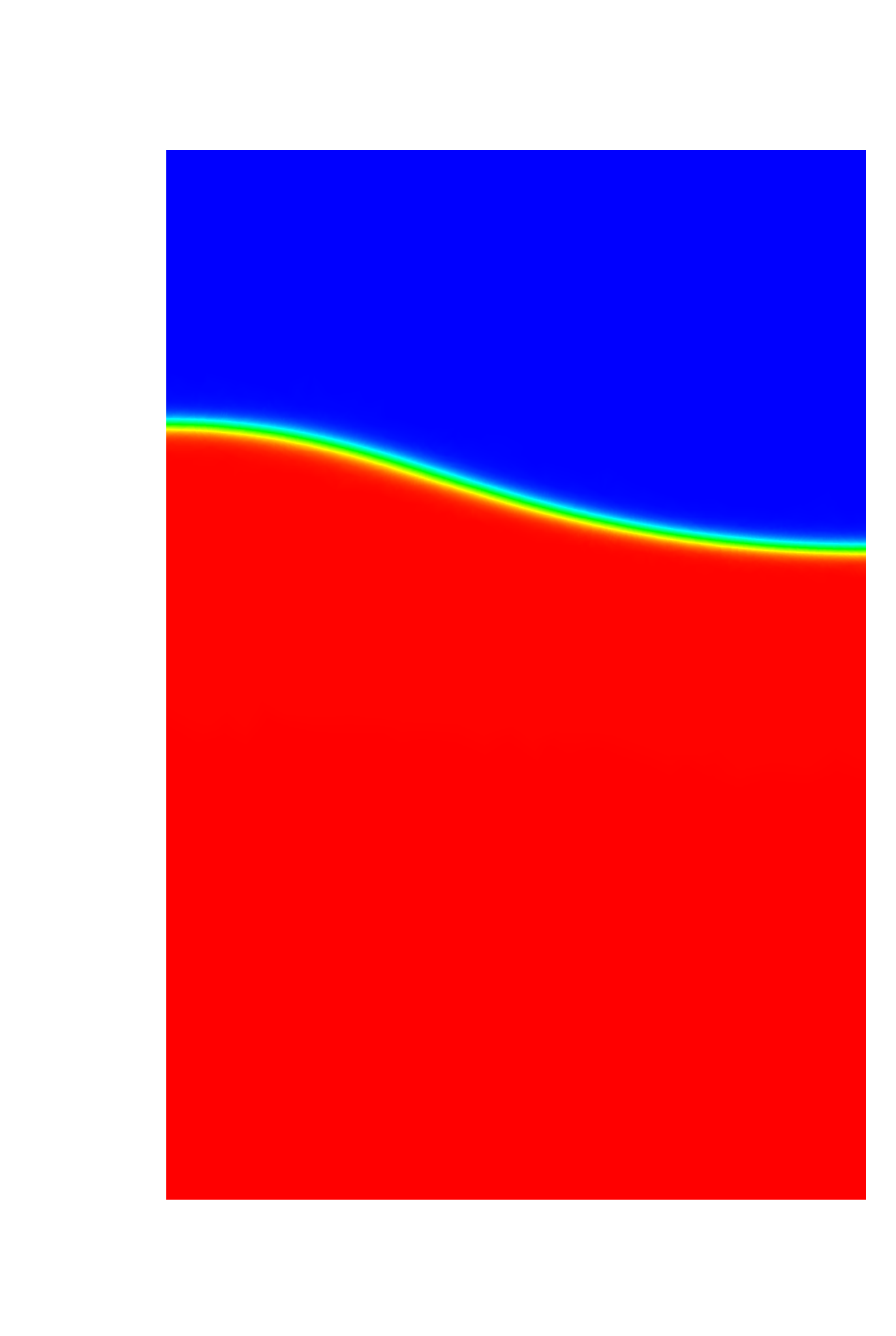}
	\end{subfigure}
	\hspace{-1cm}
	\begin{subfigure}[b]{0.3\textwidth}
		\includegraphics[trim={0.5cm 0.2cm 9cm 0.2cm},clip,width=4cm]{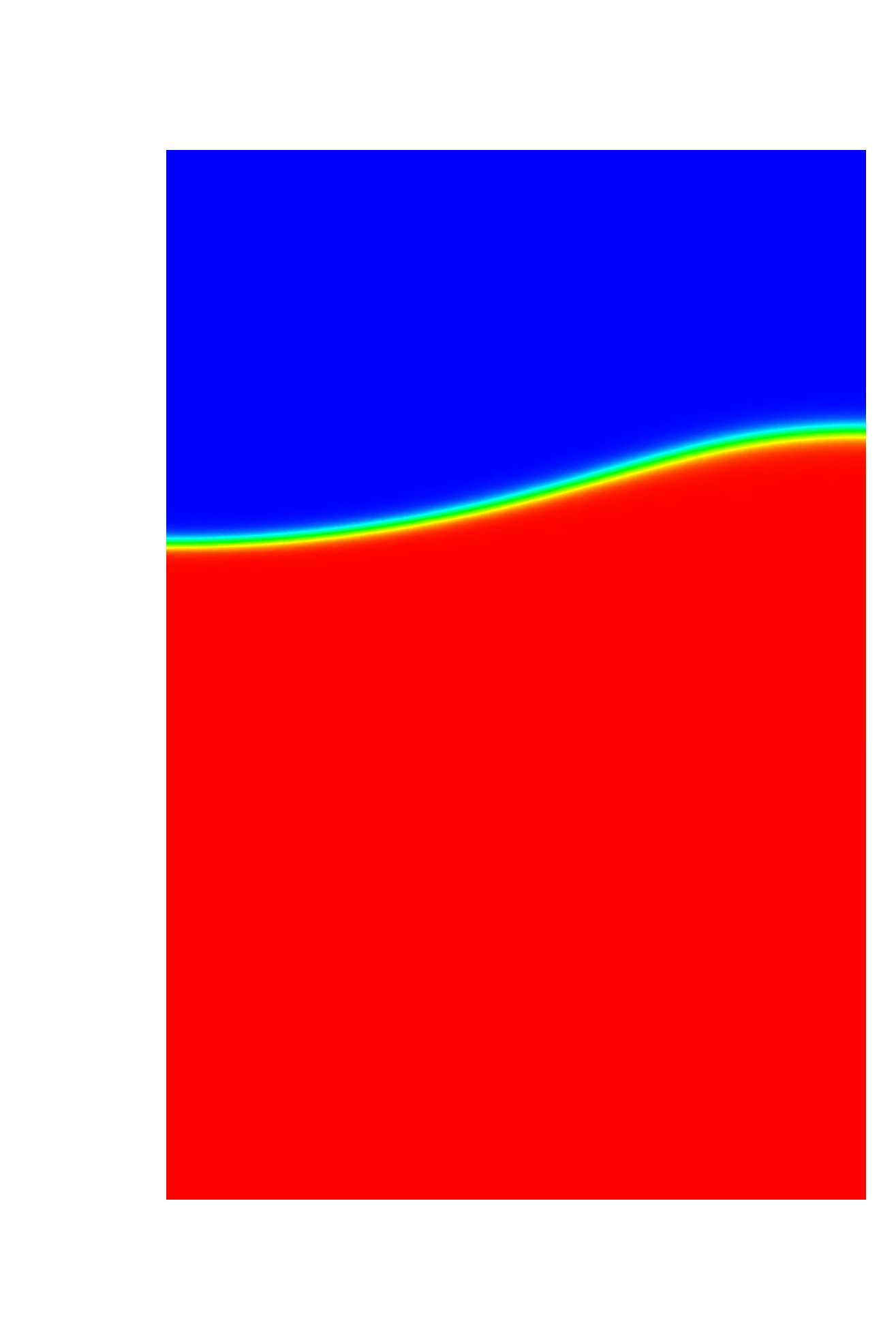}
	\end{subfigure}
	\hspace{-1cm}
	\begin{subfigure}[b]{0.3\textwidth}
		\includegraphics[trim={0.5cm 0.2cm 9cm 0.2cm},clip,width=4cm]{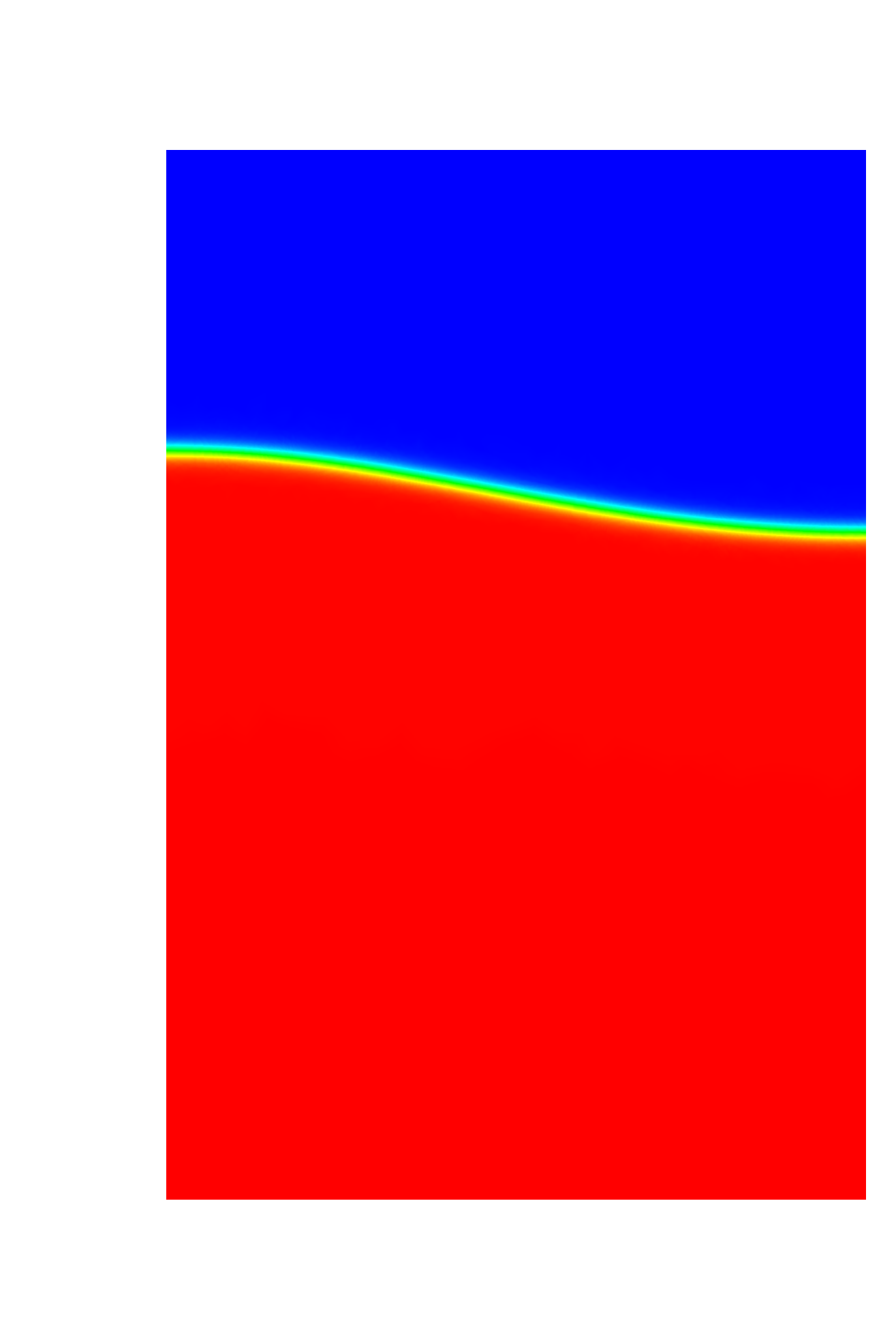}
	\end{subfigure}
	\hspace{-1cm}
	\begin{subfigure}[b]{0.3\textwidth}
		\includegraphics[trim={0.5cm 0.2cm 9cm 0.2cm},clip,width=4cm]{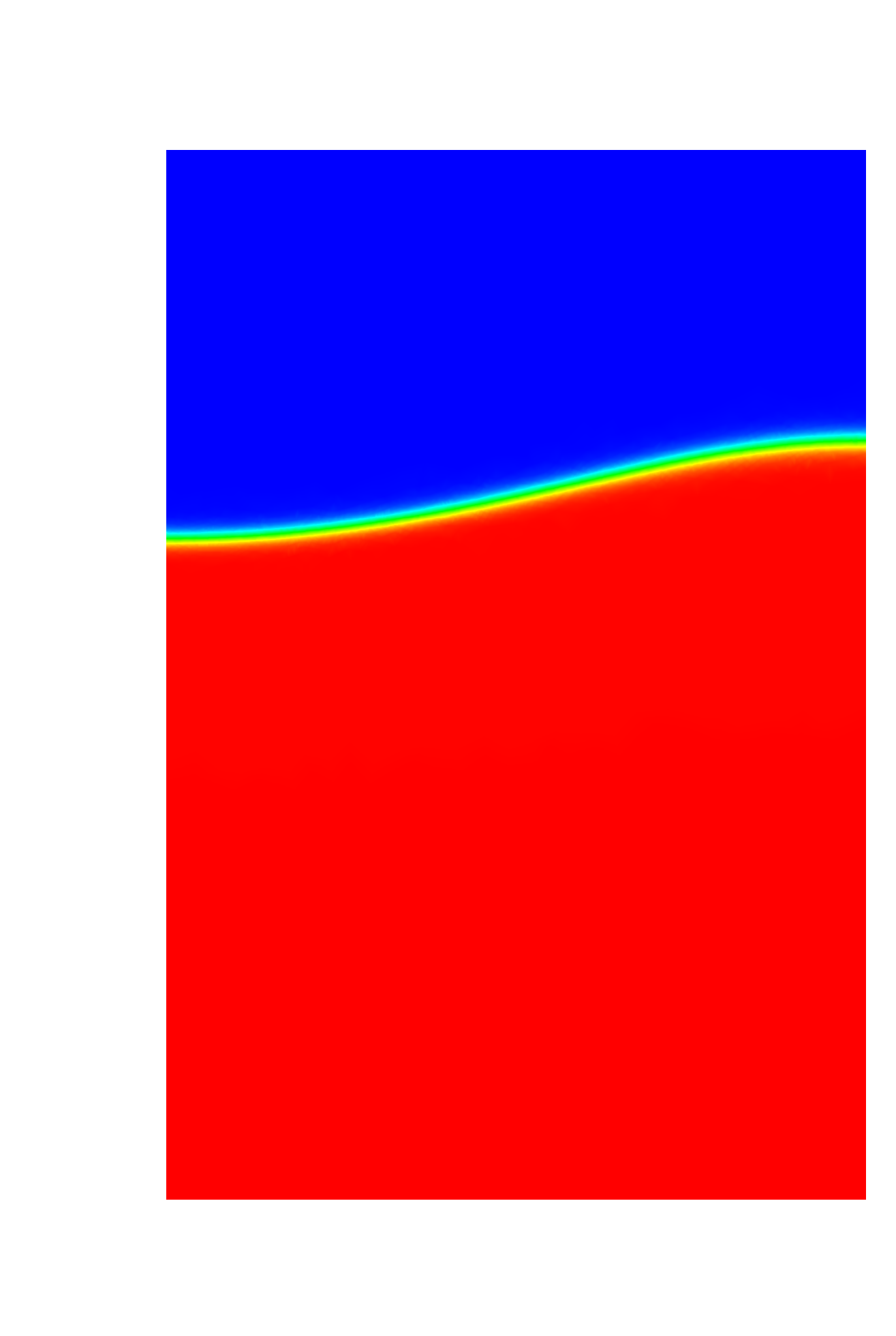}
	\end{subfigure}
	\hspace{-1cm}
	\vspace{-1cm}
	
	\hspace{-0.7cm}
	\begin{subfigure}[b]{0.3\textwidth}
		\includegraphics[trim={0.5cm 2cm 9cm 0.2cm},clip,width=4cm]{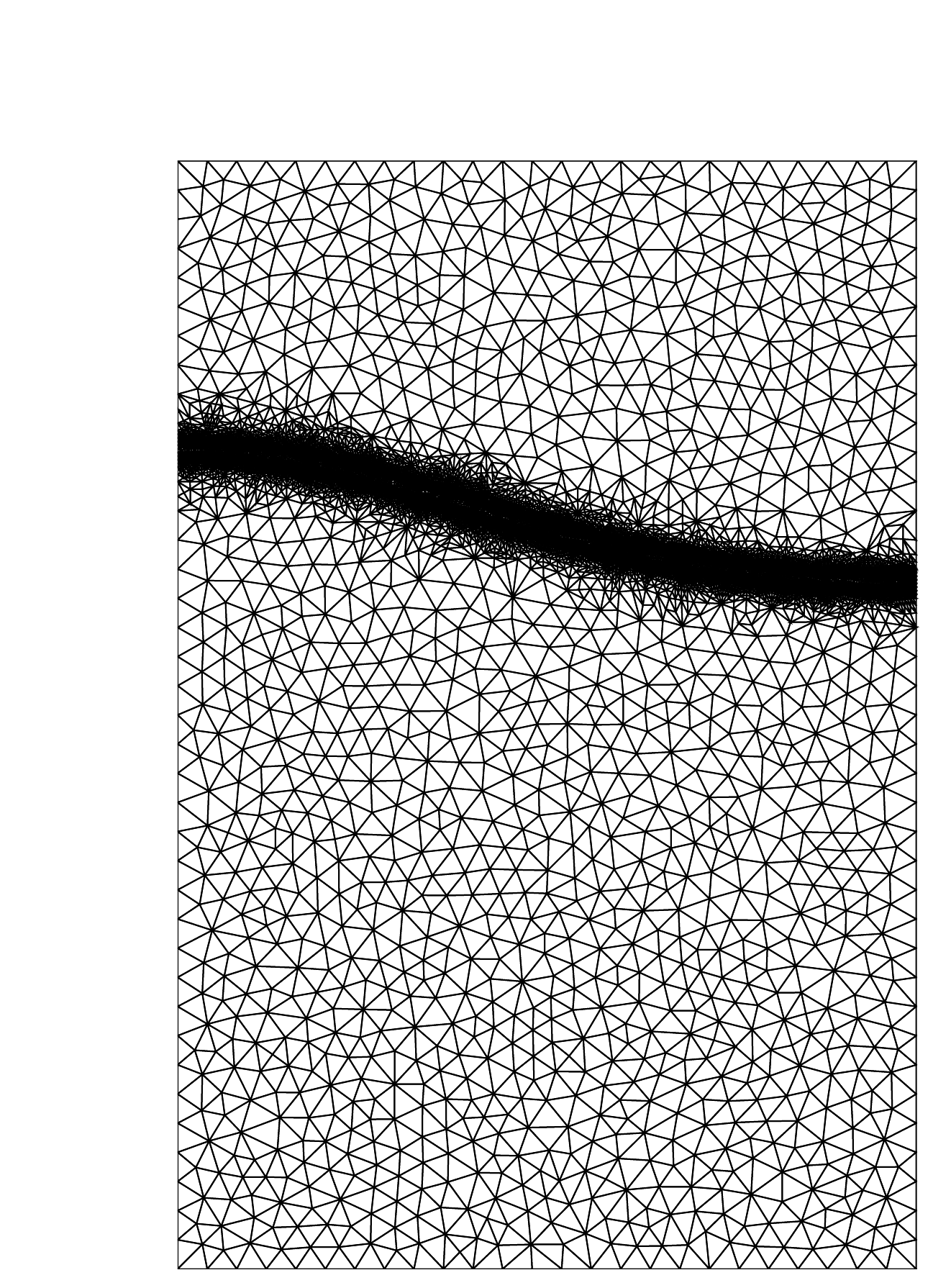}
	\caption{}
	\end{subfigure}
	\hspace{-1cm}
	\begin{subfigure}[b]{0.3\textwidth}
		\includegraphics[trim={0.5cm 2cm 9cm 0.2cm},clip,width=4cm]{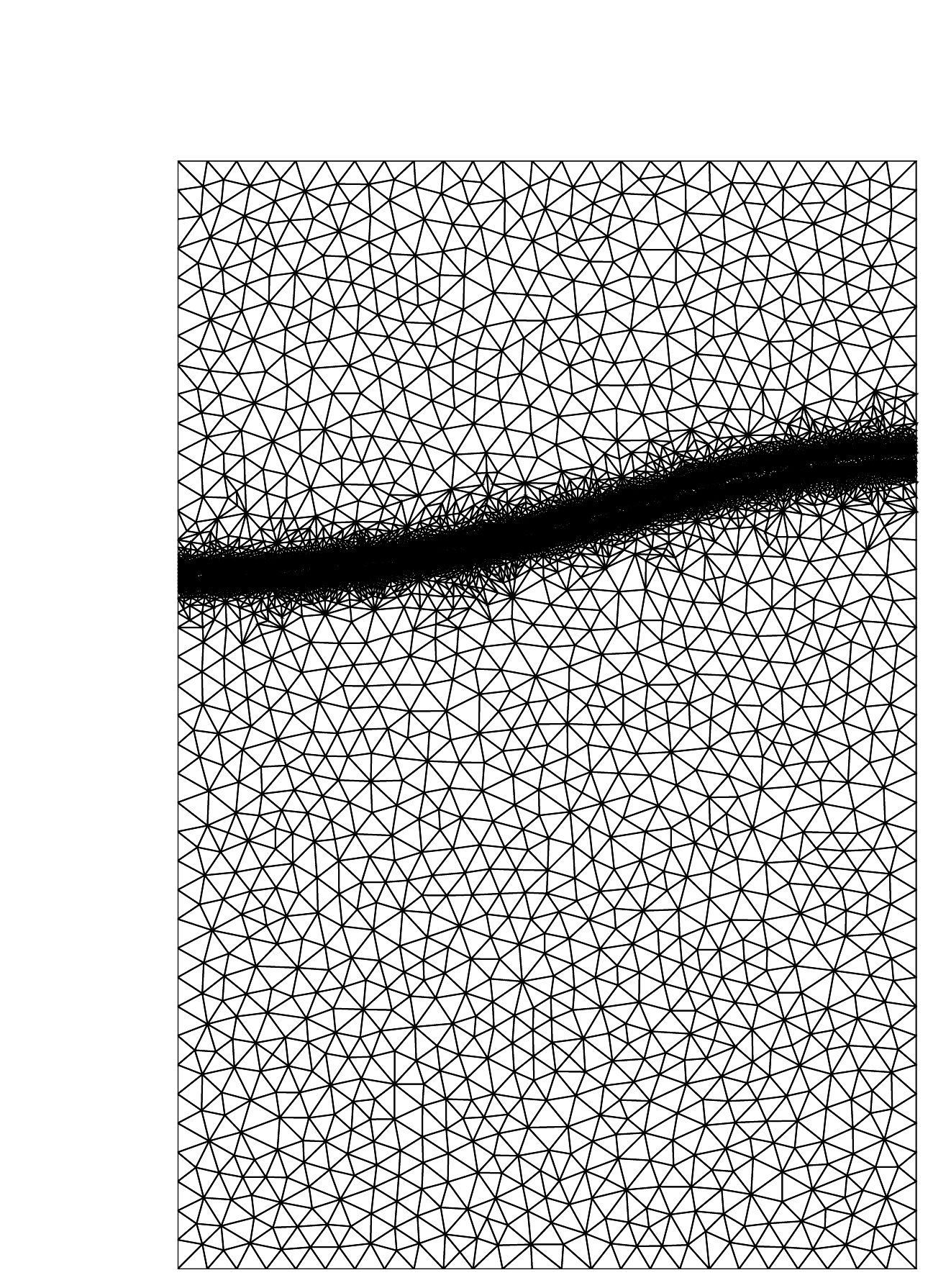}
	\caption{}
	\end{subfigure}
	\hspace{-1cm}
	\begin{subfigure}[b]{0.3\textwidth}
		\includegraphics[trim={0.5cm 2cm 9cm 0.2cm},clip,width=4cm]{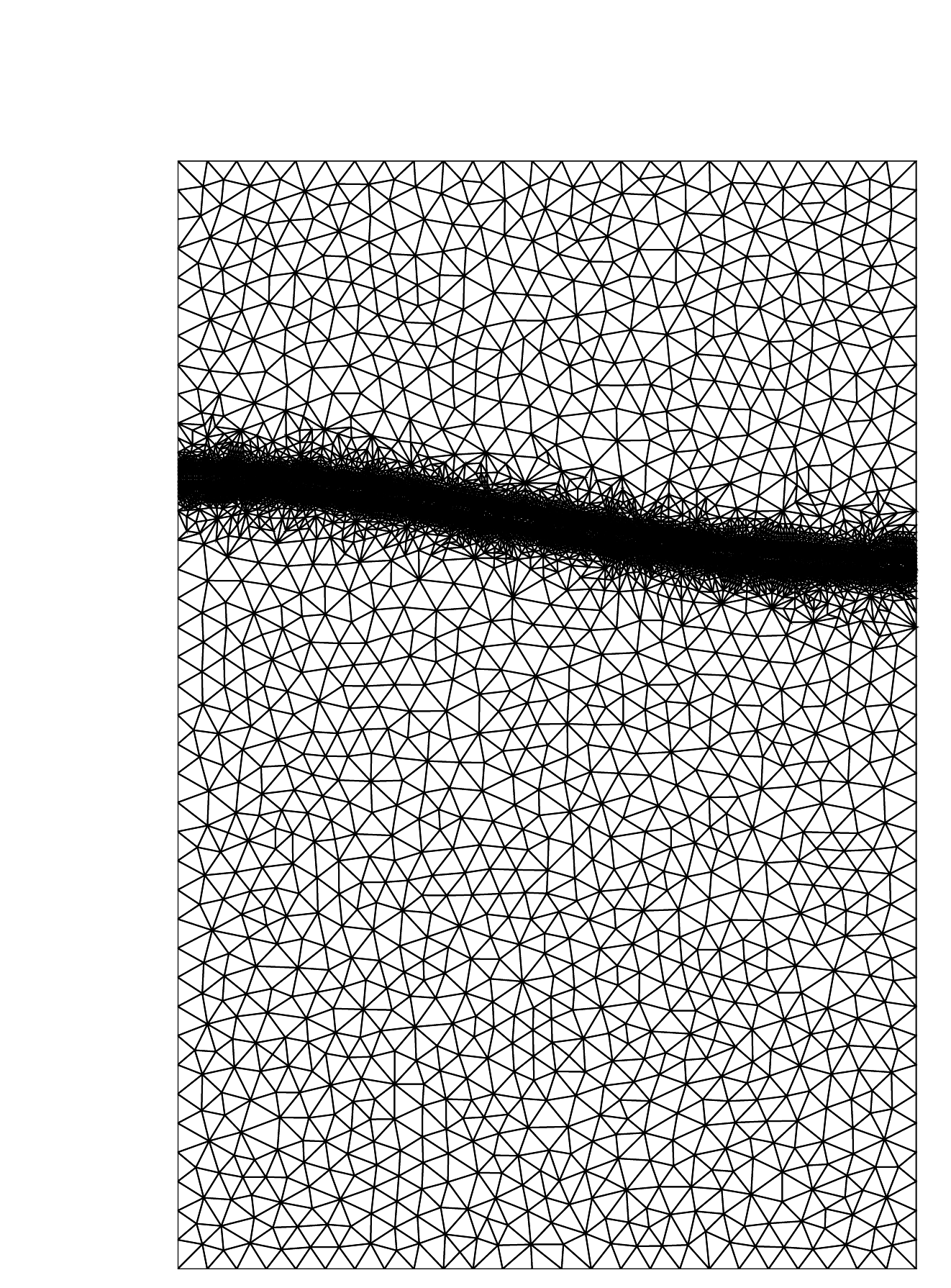}
	\caption{}
	\end{subfigure}
	\hspace{-1cm}
	\begin{subfigure}[b]{0.3\textwidth}
		\includegraphics[trim={0.5cm 2cm 9cm 0.2cm},clip,width=4cm]{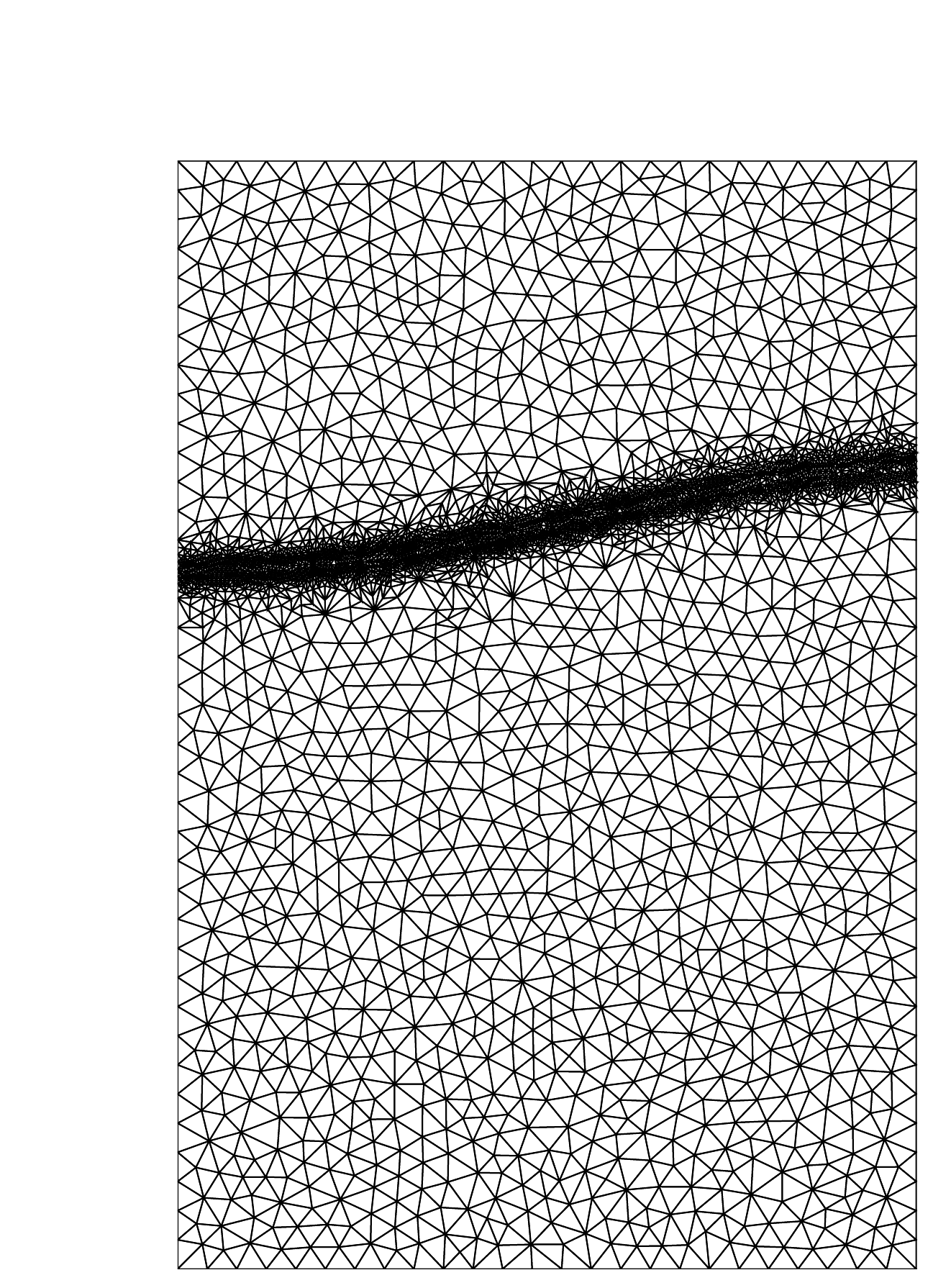}
	\caption{}
	\end{subfigure}
	\hspace{-1cm}
\caption{Contours of the order parameter $\phi$ (top) and the adaptive mesh (bottom) for the  sloshing tank problem with background mesh $\Delta x=0.04$ and $\varepsilon=0.01$ at $t/T_{oscill}$: (a) $0.55$, (b) $1.94$, (c) $3.6$ and (d) $5$.} 
\label{ST_3}
\end{figure}

Since the error estimates for the adaptive procedure are evaluated only for the Allen-Cahn equation, we need to have a sufficiently refined background mesh to capture the flow physics of the Navier-Stokes equations such as the high gradients in the velocity contours near the interface region. Therefore, a systematic mesh convergence study for the background mesh is performed for $\varepsilon=0.01$ by considering the initial background mesh of different sizes $\Delta x \in [0.02,0.16]$. The interface elevation at the left boundary throughout the simulation is recorded in Fig. \ref{ST_2}. The error is quantified as:
\begin{align}
	e_{bg} = \frac{||\Phi - \Phi_\mathrm{ref}||_2}{||\Phi_\mathrm{ref}||_2}
\end{align} 
where $\Phi$ is the temporal solution of the interface elevation at the left boundary for different background meshes and $\Phi_\mathrm{ref}$ is the finest mesh solution of the interface elevation ($\Delta x=0.02$). It is observed that the percentage error in $e_{bg}$ for $\Delta x = 0.16, 0.08$ and $0.04$ is $0.32\%$, $0.12\%$ and $0.023\%$ respectively. A similar error variation is obtained for $\varepsilon=0.005$ by considering $\Delta x \in [0.01,0.08]$ with error percentage as $0.14\%$, $0.17\%$ and $0.007\%$ for $\Delta x=0.08$, $0.04$ and $0.02$ respectively with $\Delta x=0.01$ as the reference mesh. We choose $\Delta x=0.04$ as the background mesh for further studies. The evolution of the adaptive grid with the contours of the order parameter $\phi$ at four time units corresponding to the extreme amplitudes of the sloshing is shown in Fig. \ref{ST_3}.

\subsubsection{Temporal convergence}
\begin{figure}
		\centering
		\includegraphics[trim={0cm 0.3cm 0cm 0cm},clip,width=15cm]{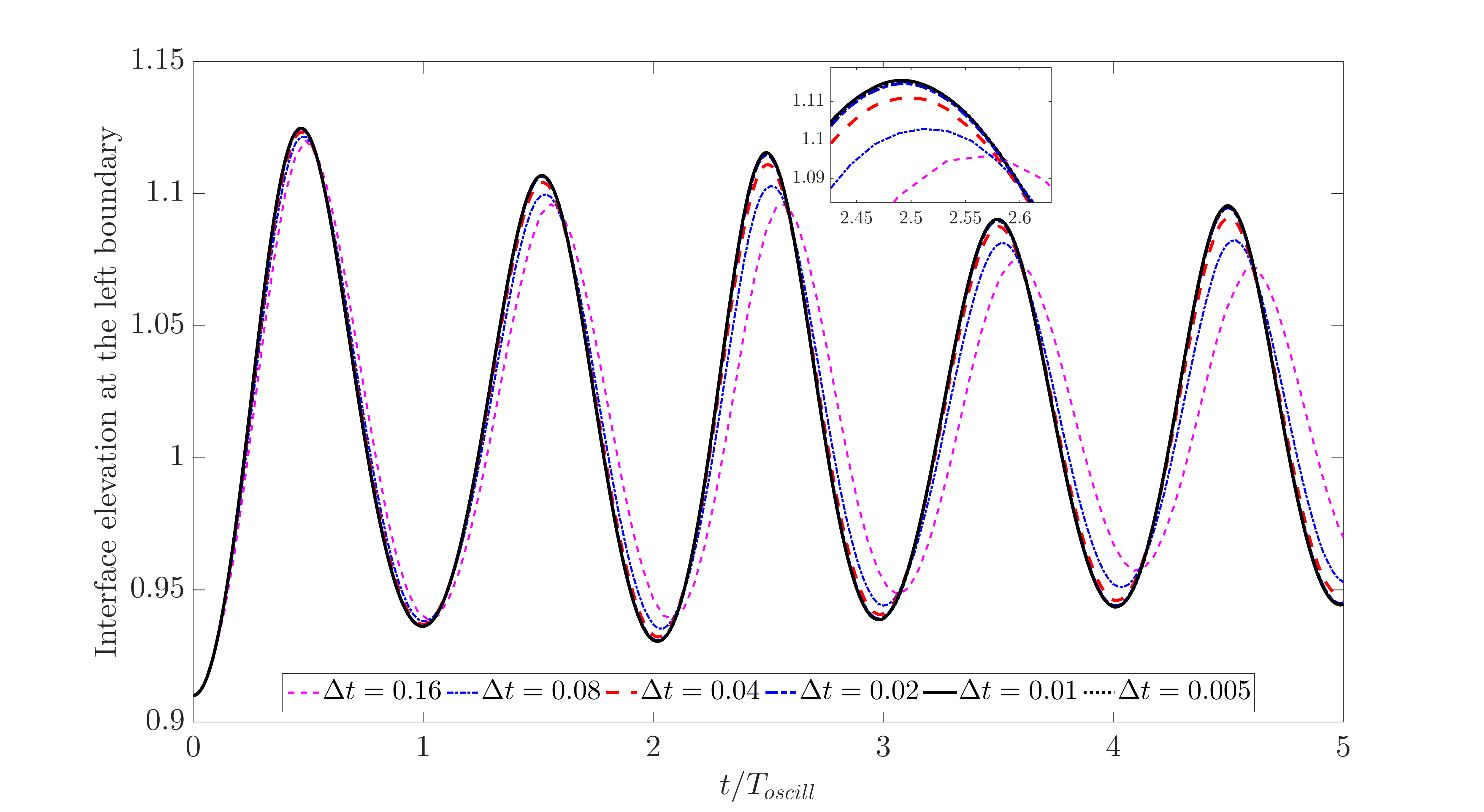}
\caption{Temporal convergence for the sloshing tank problem: evolution of the interface at the left boundary with $\varepsilon = 0.01$ and different time step sizes.} 
\label{timeConv_1}
\end{figure}

\begin{figure}
		\centering
		\includegraphics[trim={0cm 0.3cm 0cm 0cm},clip,width=15cm]{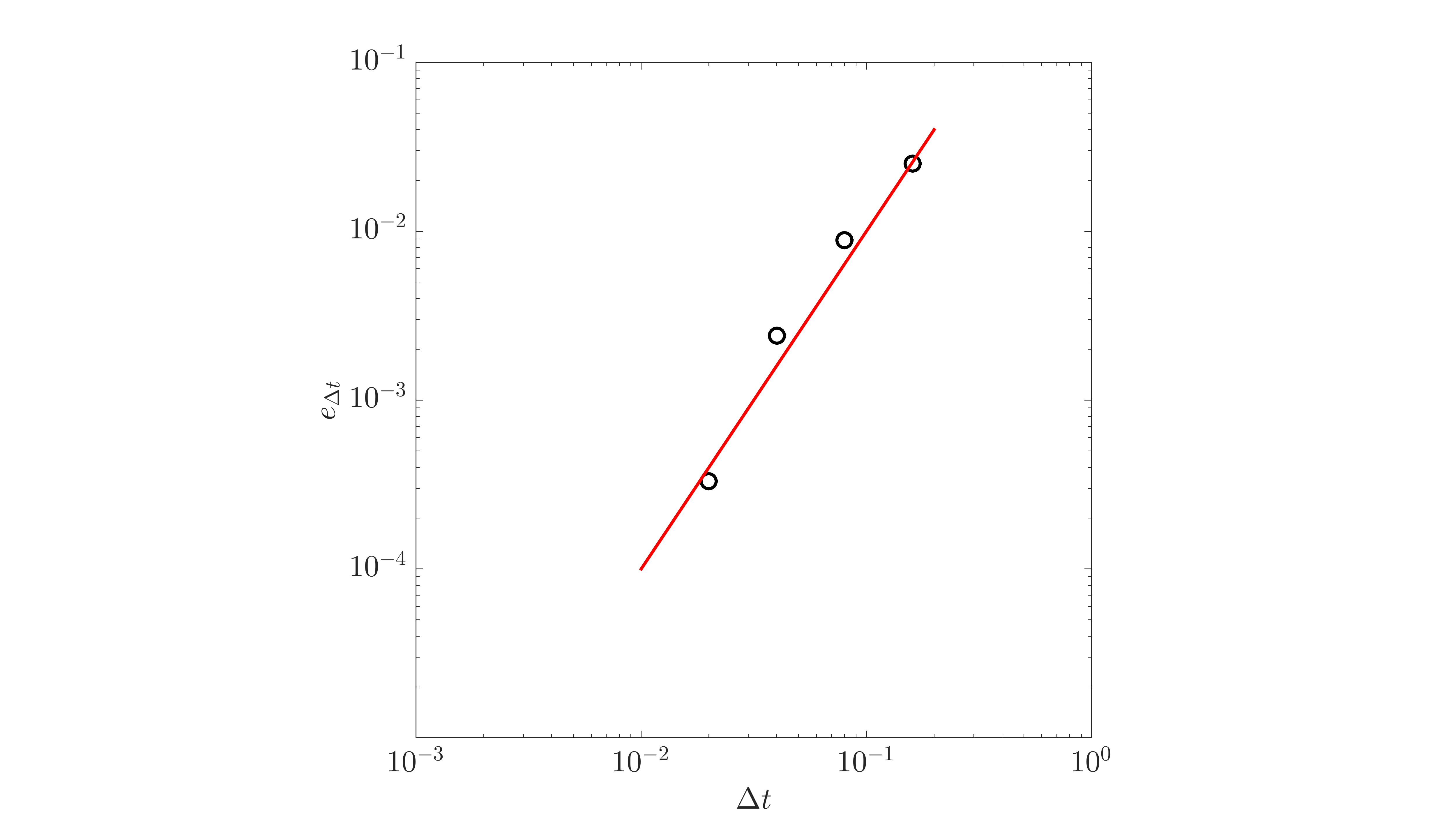}
\caption{Temporal convergence for the sloshing tank problem: plot for the error $e_\mathrm{\Delta t}$ with the time step size. The line represents the second-order convergence.} 
\label{timeConv_2}
\end{figure}

We evaluate the convergence properties of the adaptive procedure with respective to the time step $\Delta t$. The time step size is decreased by a factor of $2$ from $\Delta t=0.16$ to $\Delta t=0.005$. The evolution of the interface at the left boundary is shown in Fig. \ref{timeConv_1}. There seems to be no difference in the solution for $\Delta t \leq 0.02$. Moreover, the error in the temporal convergence is evaluated as
\begin{align}
	e_{\Delta t} = \frac{||{\Phi} - {\Phi}_\mathrm{ref}||_2}{||{\Phi}_\mathrm{ref}||_2},
\end{align}
where $\Phi$ and $\Phi_\mathrm{ref}$ are the temporal evolution of the interface at the left boundary for corresponding time step and the reference solution (the finest time step) respectively. Figure \ref{timeConv_2} shows the temporal convergence for the different time steps employed in the present study for the adaptive procedure. 
From the convergence plot, it is evident that the line plotted has a slope of two which confirms the second-order temporal accuracy of the NAVP scheme. 

\subsubsection{Mass conservation and divergence-free condition}
\label{effect_adaptive}
\begin{figure}
\centering
	\begin{subfigure}[b]{0.5\textwidth}
		\includegraphics[trim={10cm 0.2cm 11cm 0.2cm},clip,width=7cm]{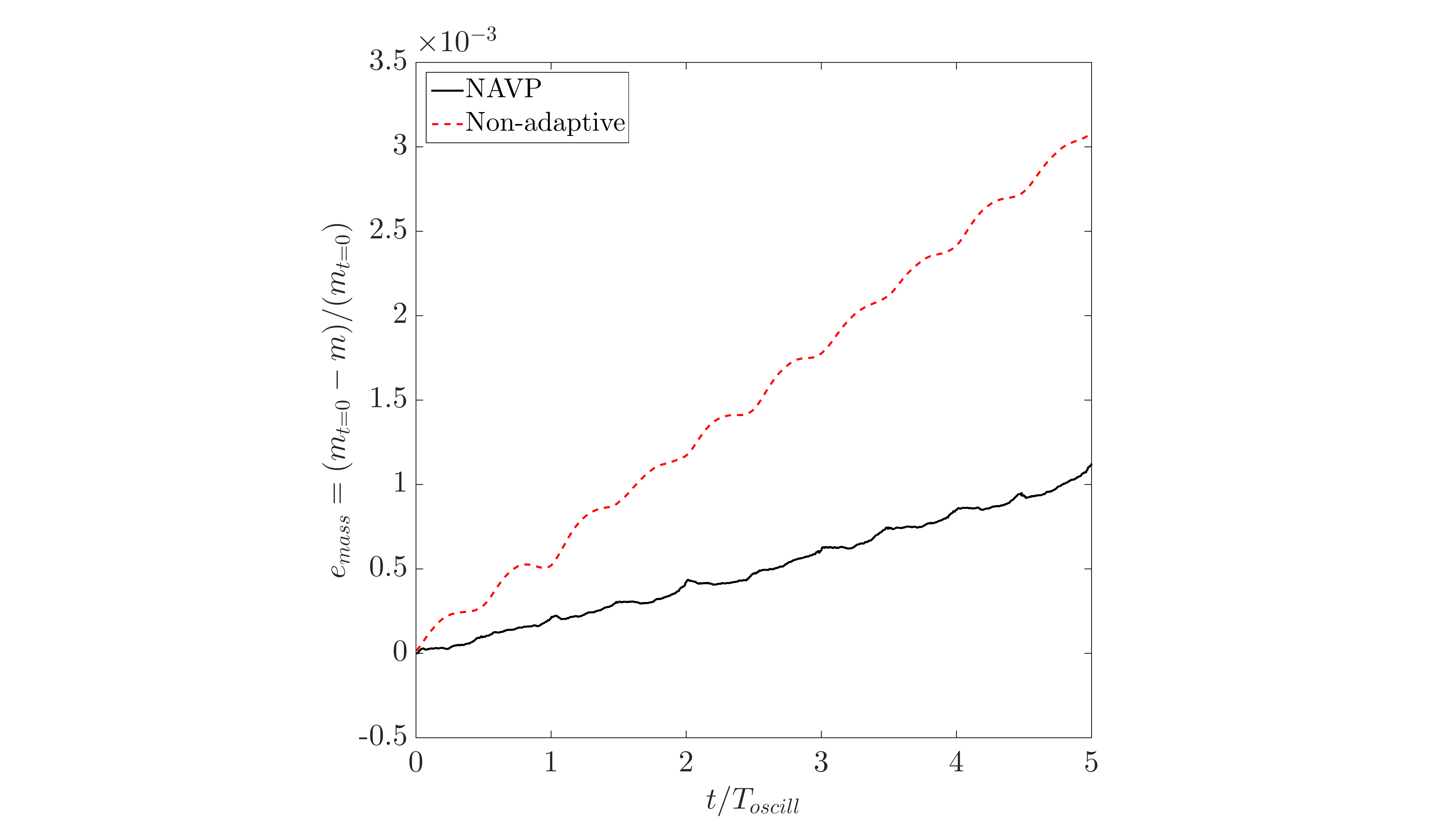}
	\caption{}
	\end{subfigure}%
	\begin{subfigure}[b]{0.5\textwidth}
		\includegraphics[trim={10cm 0.2cm 11cm 0.2cm},clip,width=7cm]{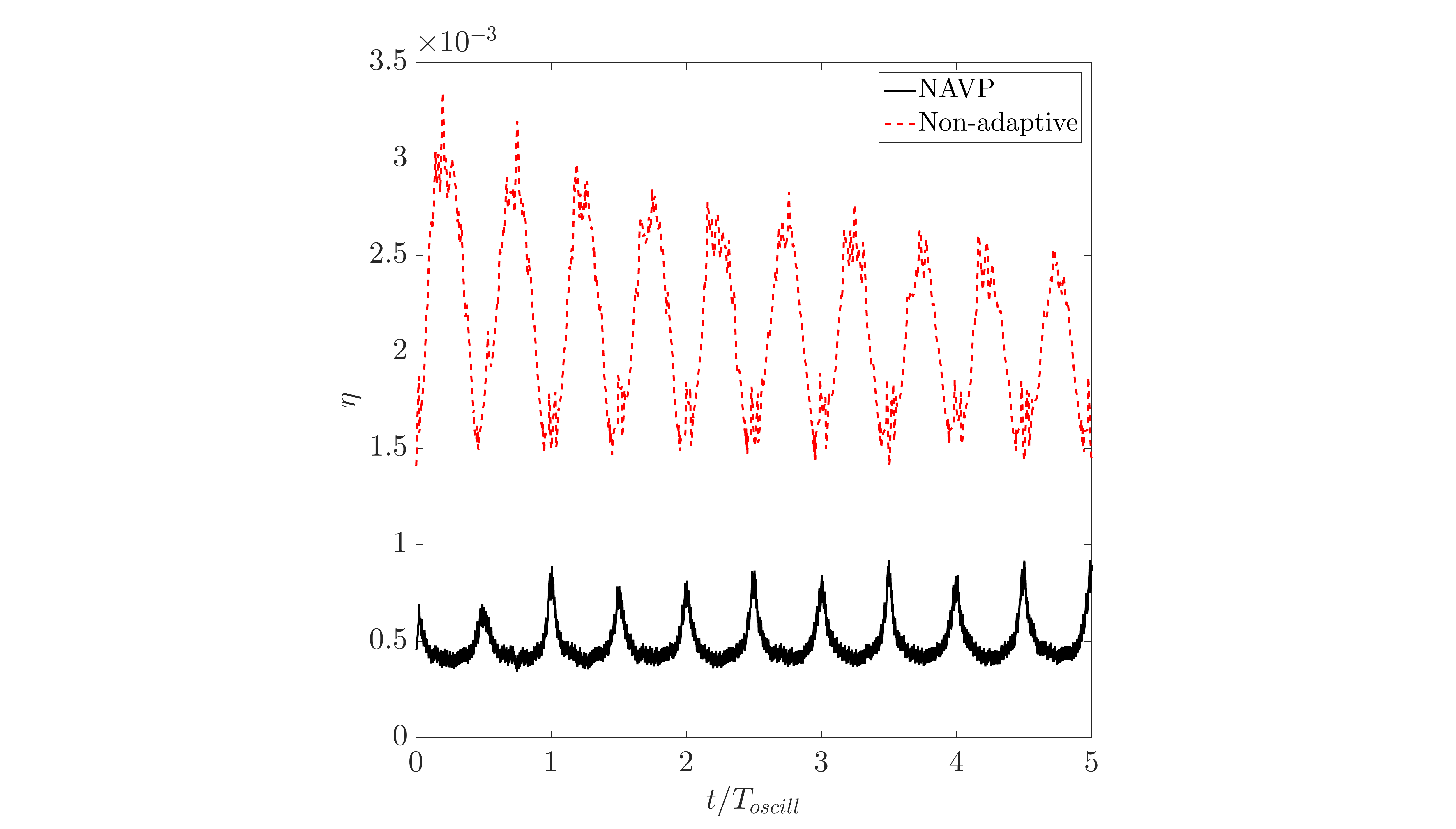}
	\caption{}
	\end{subfigure}
	
	\begin{subfigure}[b]{0.5\textwidth}
		\includegraphics[trim={10cm 0.2cm 11cm 0.2cm},clip,width=7cm]{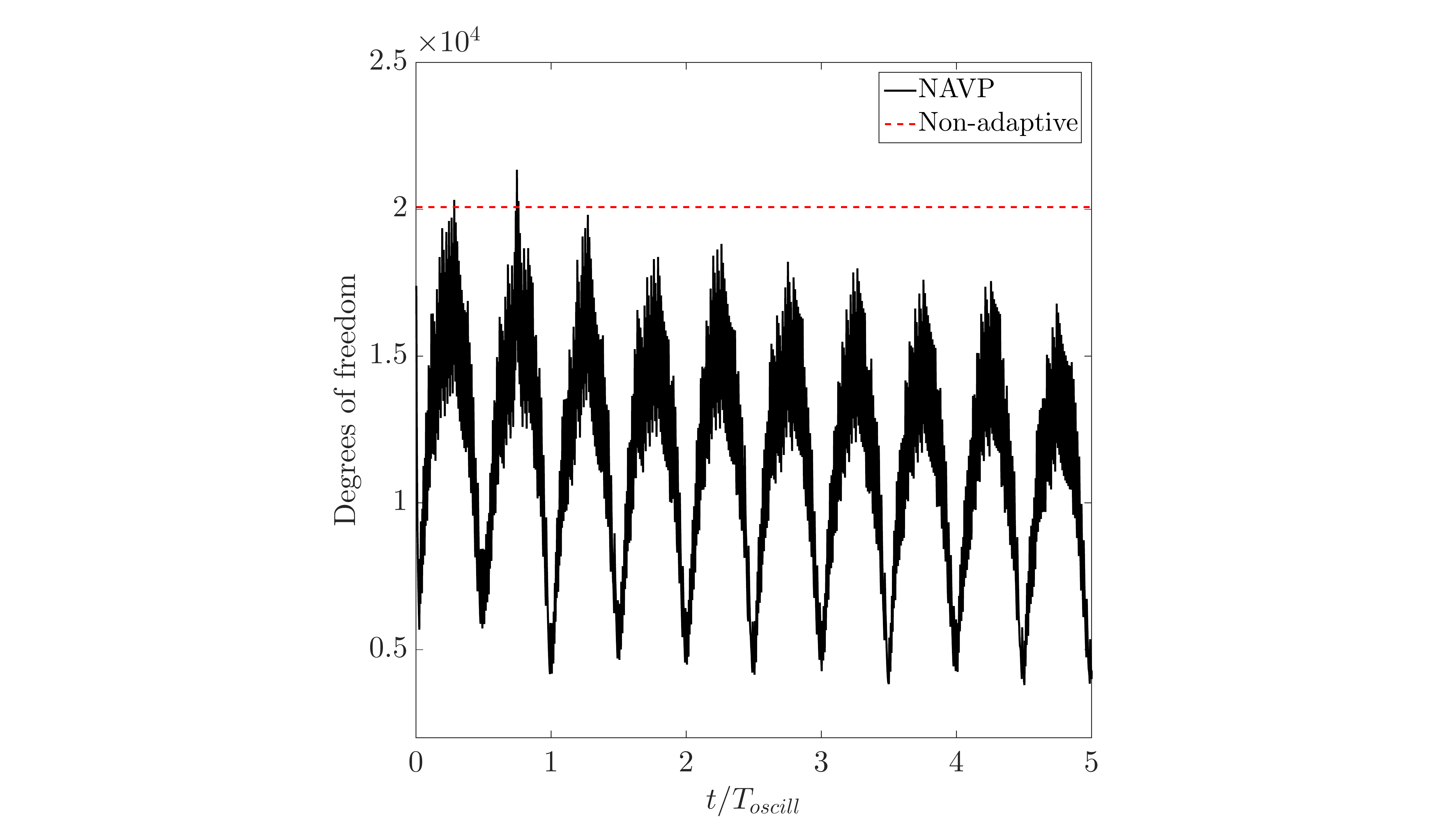}
	\caption{}
	\end{subfigure}%
	\begin{subfigure}[b]{0.5\textwidth}
		\includegraphics[trim={10cm 0.2cm 11cm 0.2cm},clip,width=7cm]{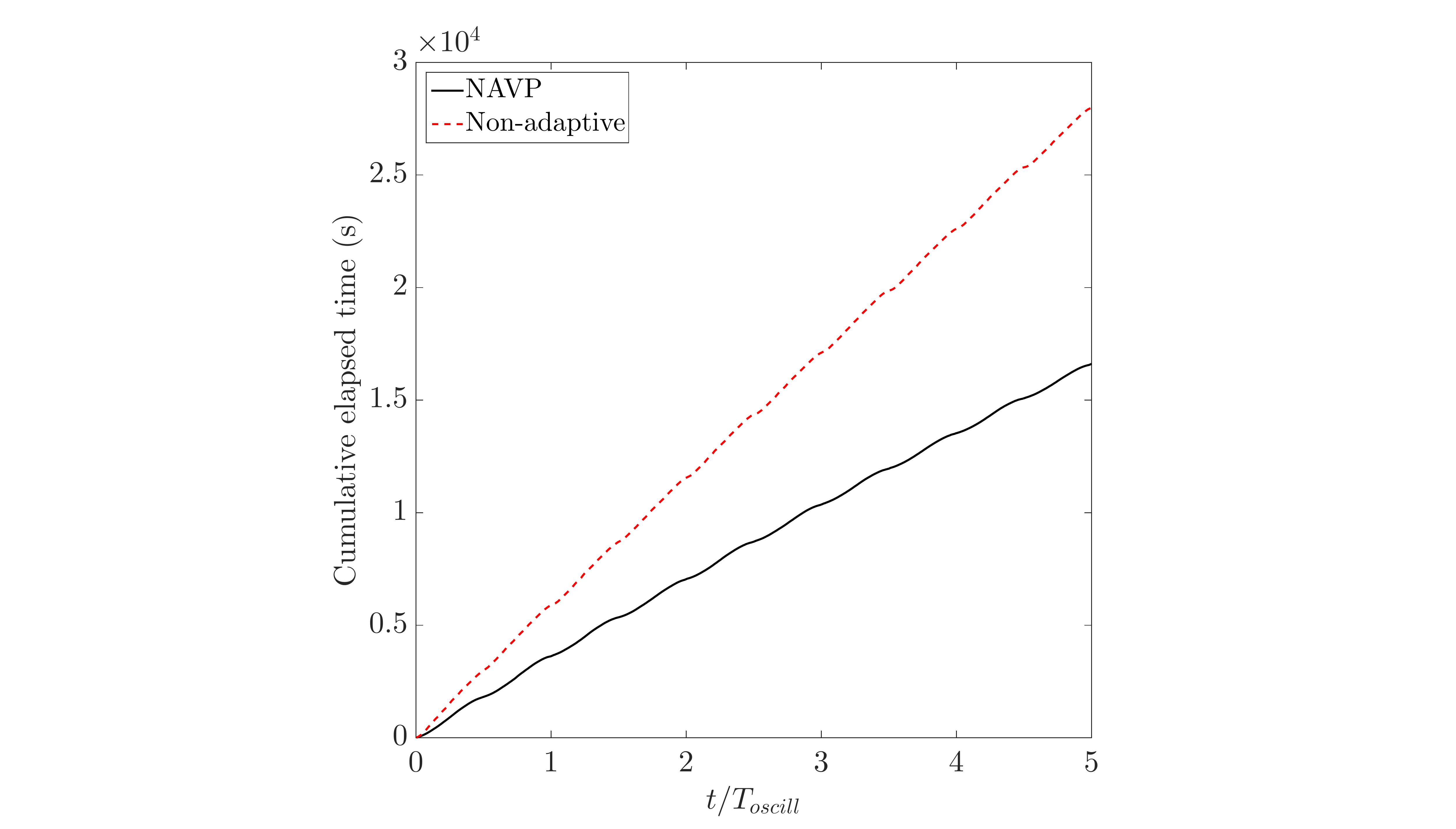}
	\caption{}
	\end{subfigure}
\caption{Variation of the measured quantities with time in the sloshing tank problem: (a) the error in mass conservation evaluated using Eq.~(\ref{mass_error}), (b) the residual error $\eta$, (c) the number of degrees of freedom or unknowns, and (d) the cumulative elapsed time.} 
\label{ST_4}
\end{figure}

In this subsection, we examine the effectiveness of the NAVP on the total mass conservation of $\phi$, the variation of residual error $\eta$ and the computational cost (i.e., degrees of freedom and elapsed time). Fig. \ref{ST_4} shows the variation of the measured quantities as a function of time on an adaptive grid employing the NAVP procedure ($\Delta x=0.04$) and a non-adaptive grid with $\Delta x=0.01$. Based on the previous analysis, the interfacial thickness parameter $\varepsilon$ is selected as $0.01$. The tolerances are set to $tol_{NS}=tol_{AC}=5\times 10^{-4}$ and $tol_R = 10^{-3}$. As shown in Fig. \ref{ST_4}(a), the error in the mass conservation is reduced approximately $3$ times for the adaptive grid in contrast to the non-adaptive Eulerian fixed grid. The error in the total mass conservation is evaluated as
\begin{align} \label{mass_error}
	e_{mass} = \frac{m_{t=0}-m}{m_{t=0}},
\end{align}
where $m$ is defined by Eq.~(\ref{mass_def}) and $m_{t=0}$ is the mass at $t=0$.
 This is the consequence of increased density of elements near the interfacial region which helps to capture the interface more accurately. The increase in the density of elements near the interfacial region leads to reduced residual error $\eta$ compared to the non-adaptive method (Fig. \ref{ST_4}(b)). The computational cost has been quantified as the variation of the number of degrees of freedom with non-dimensional time and the cumulative elapsed time of the simulation in Fig. \ref{ST_4}(c-d). The adaptive procedure reduces the number of degrees of freedom, which leads to a lesser elapsed time for the simulation.

\begin{figure}
\centering
	\begin{subfigure}[b]{0.5\textwidth}
		\includegraphics[trim={10cm 0 11.8cm 0cm},clip,width=7cm]{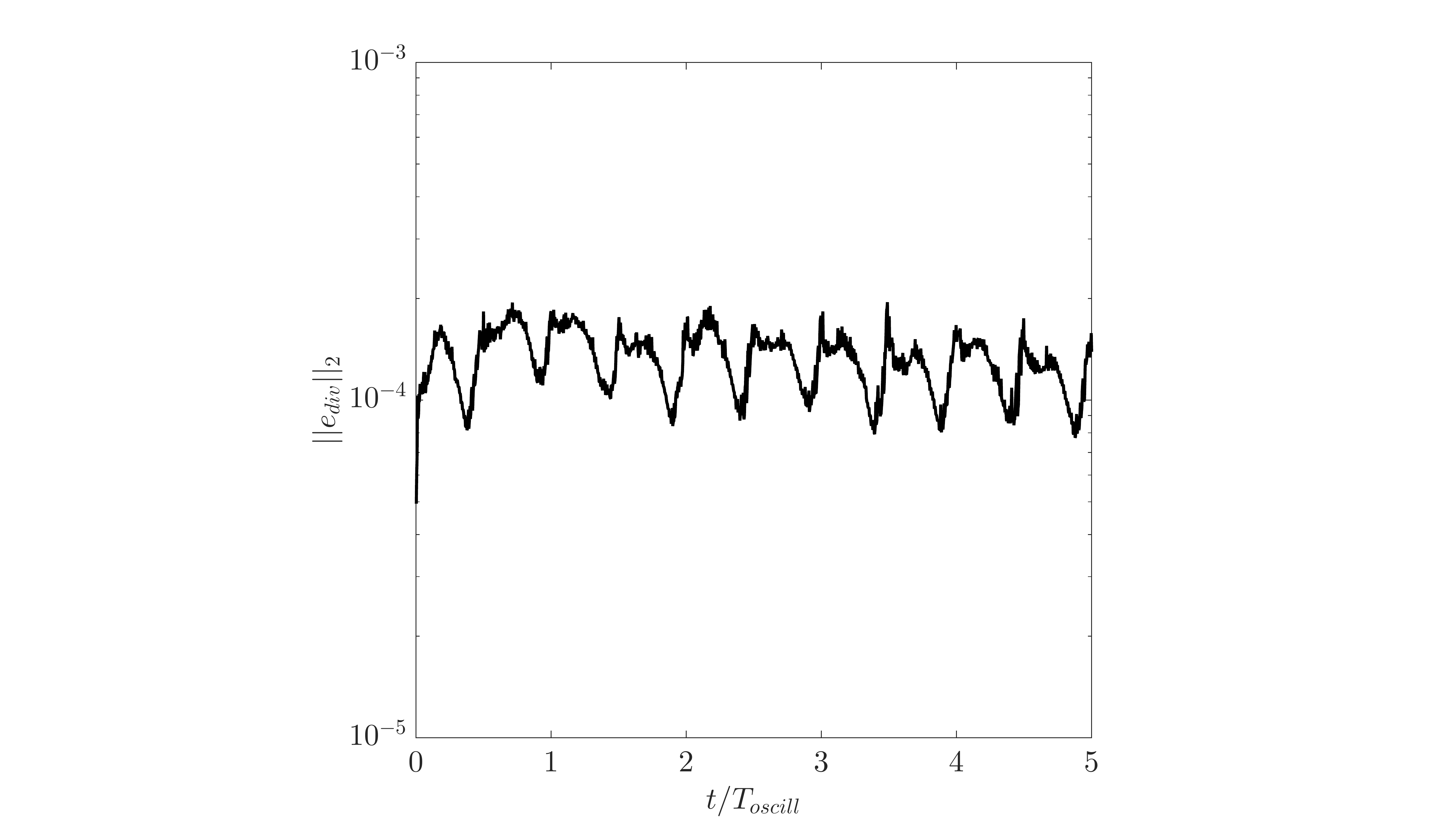}	
	\caption{}
	\end{subfigure}%
	\begin{subfigure}[b]{0.5\textwidth}
		\includegraphics[trim={0.5cm 0.5cm 0.5cm 0.5cm},clip,width=8.5cm]{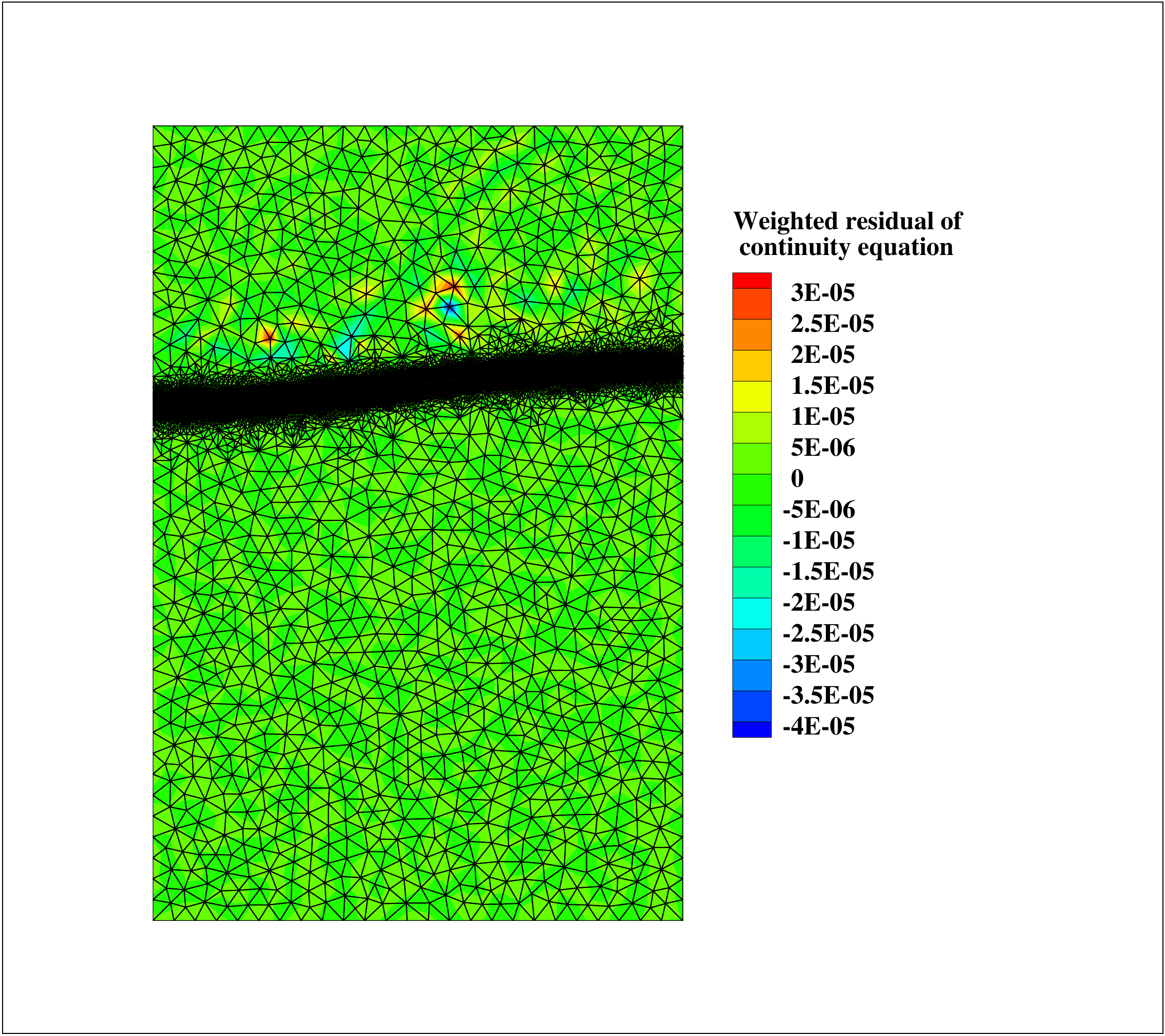}	
	\caption{}
	\end{subfigure}%
\caption{Quantification of the satisfaction of the incompressibility constraint: (a) the variation of the $L^2$ norm of the weighted residual of the continuity equation with time, and (b) the contour of the weighted residual of the continuity equation at $t/T_{oscill}=4.17$.} 
\label{div_plot}
\end{figure}

It is known that the divergence-free condition is not exactly satisfied pointwise in the stabilized finite element methods based on the continuous Petrov-Galerkin approximation \cite{HUGHES20051141}.
In the present fully-discrete variational formulation, the incompressibility constraint (the continuity equation) is approximately satisfied within the convergence tolerance. 
We quantify the error in the weighted residual of the continuity equation as
\begin{align}
	e_{div} =  \int_\Omega q_\mathrm{h} (\nabla\cdot\boldsymbol{u}_\mathrm{h}^\mathrm{n+\alpha}) \mathrm{d}\Omega,
\end{align}
where the integral provides the weighted residual at each nodal point. The contour of this quantity has been shown in Fig. \ref{div_plot}(b) where we observe that the extreme values of the weighted residual is within the convergence tolerance limit of $5\times 10^{-4}$. We also evaluate the $L^2$ norm of the vector containing the weighted residuals of the nodal points  using equal-order linear elements. Fig. \ref{div_plot}(a) shows the error in the divergence of the velocity field as a function of time for several sloshing oscillations. We observe that the residual of the continuity equation is within the nonlinear tolerance of $tol_{NS}=tol_{AC}=5\times 10^{-4}$ and $tol_R = 10^{-3}$. Using higher order elements or finer grid, the error in the divergence of the velocity field can be further improved \cite{doi:10.1137/100818583}.

\subsubsection{Advantage of the PPV technique}
To observe the effect of the addition of the positivity preserving nonlinear terms in Eq.~(\ref{AC_variational}), we compare the extreme values of the order parameter $\phi$ at $t/T_{oscill}=5$ for the adaptive method with and without the PPV terms. The setup is the same as that of Section \ref{effect_adaptive}.
\renewcommand{\arraystretch}{0.5}
\begin{table}
\caption{The role of positivity condition on the bound of the evolving interface solution over the whole domain}
\centering
\begin{tabular}{  M{3cm}  M{1.5cm}  M{1.5cm}  N }
	\hline
\centering
	\textbf{Method} & $\mathrm{min(\phi)}$ & $\mathrm{max(\phi)}$ &\\[10pt]
	\hline
\centering
	Adaptive (no PPV)  & -1.00044 & 1.00027  &\\[10pt]
	\hline
\centering
	NAVP  & -1.0 & 1.0  &\\[10pt]
	\hline
\end{tabular}
\label{table_ppv_comp}
\end{table}
The bounds of the solution of $\phi$ are summarized in Table \ref{table_ppv_comp}. It can be observed that the solution remains bounded when the nonlinear positivity preserving terms are incorporated in the formulation. However, the difference is very minor in the sloshing tank problem, it may be higher for different problems due to the restriction criterion for the number of elements in the algorithm. These nonlinear PPV terms provide a necessary stabilization in highly convection and reaction dominated flows.

\subsubsection{Relationship between $tol_R$ and $\varepsilon$}
\begin{figure}
\centering
	\hspace{-1cm}
	\begin{subfigure}[b]{0.33\textwidth}
		\includegraphics[trim={10cm 0.2cm 10cm 0.2cm},clip,width=6cm]{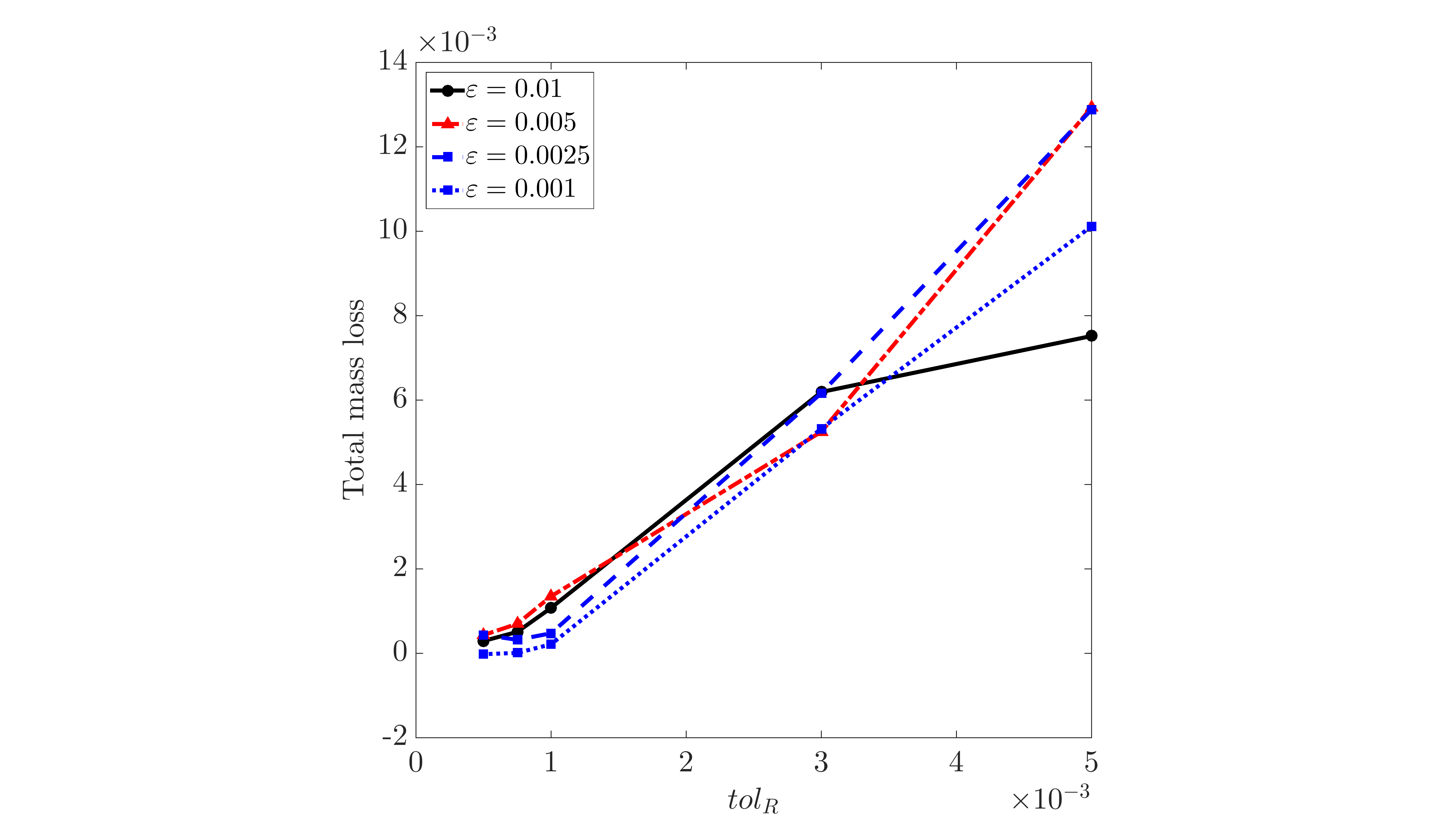}
	\caption{}
	\end{subfigure}%
		\hspace{0.25cm}
	\begin{subfigure}[b]{0.33\textwidth}
		\includegraphics[trim={10cm 0.2cm 10cm 0.2cm},clip,width=6cm]{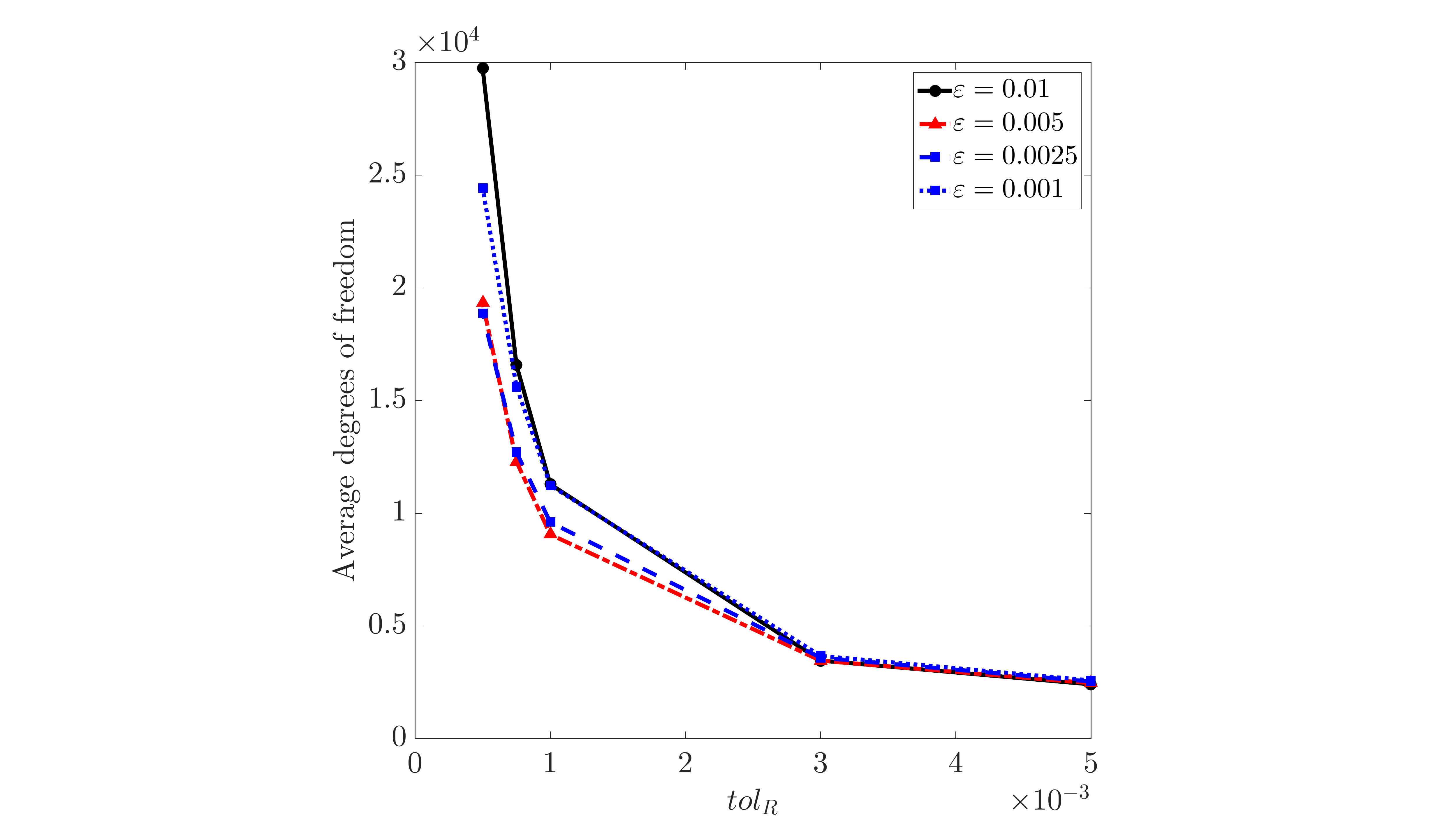}
	\caption{}
	\end{subfigure}
		\hspace{0.25cm}
	\begin{subfigure}[b]{0.34\textwidth}
		\includegraphics[trim={10cm 0.2cm 10cm 0.2cm},clip,width=6cm]{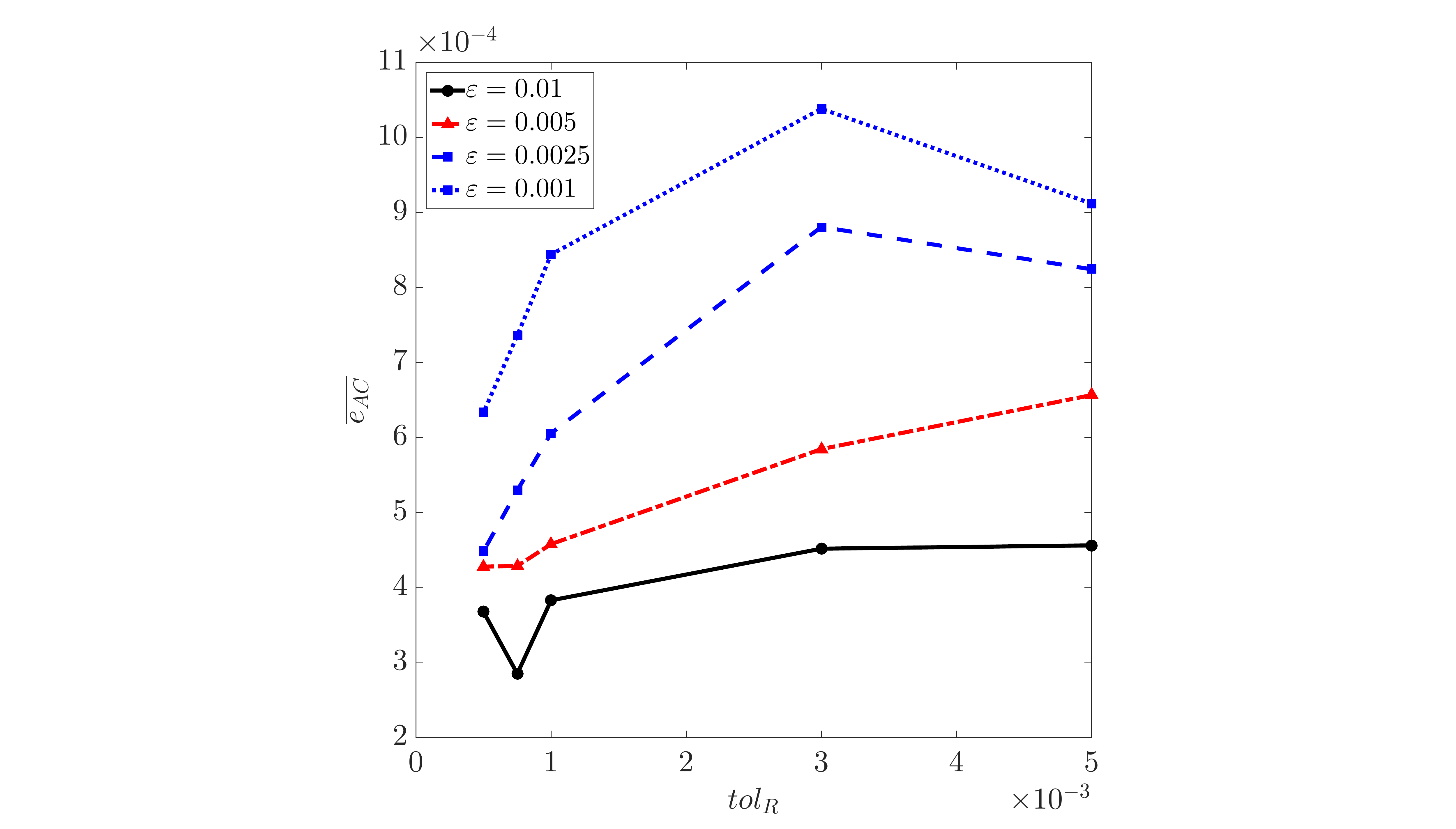}
	\caption{}
	\end{subfigure}%
		\hspace{-0.5cm}
\caption{Variation of the measured quantities with $tol_R$ in the sloshing tank problem: (a) the total mass loss, (b) the average number of degrees of freedom and (c) the average error in solving the Allen-Cahn equation ($\overline{e}_{AC}$).} 
\label{ST_7}
\end{figure}

\begin{figure}
		\centering
		\hspace{-1cm}
	\begin{subfigure}[b]{0.8\textwidth}
		\includegraphics[trim={0cm 0.3cm 0cm 0cm},clip,width=12cm]{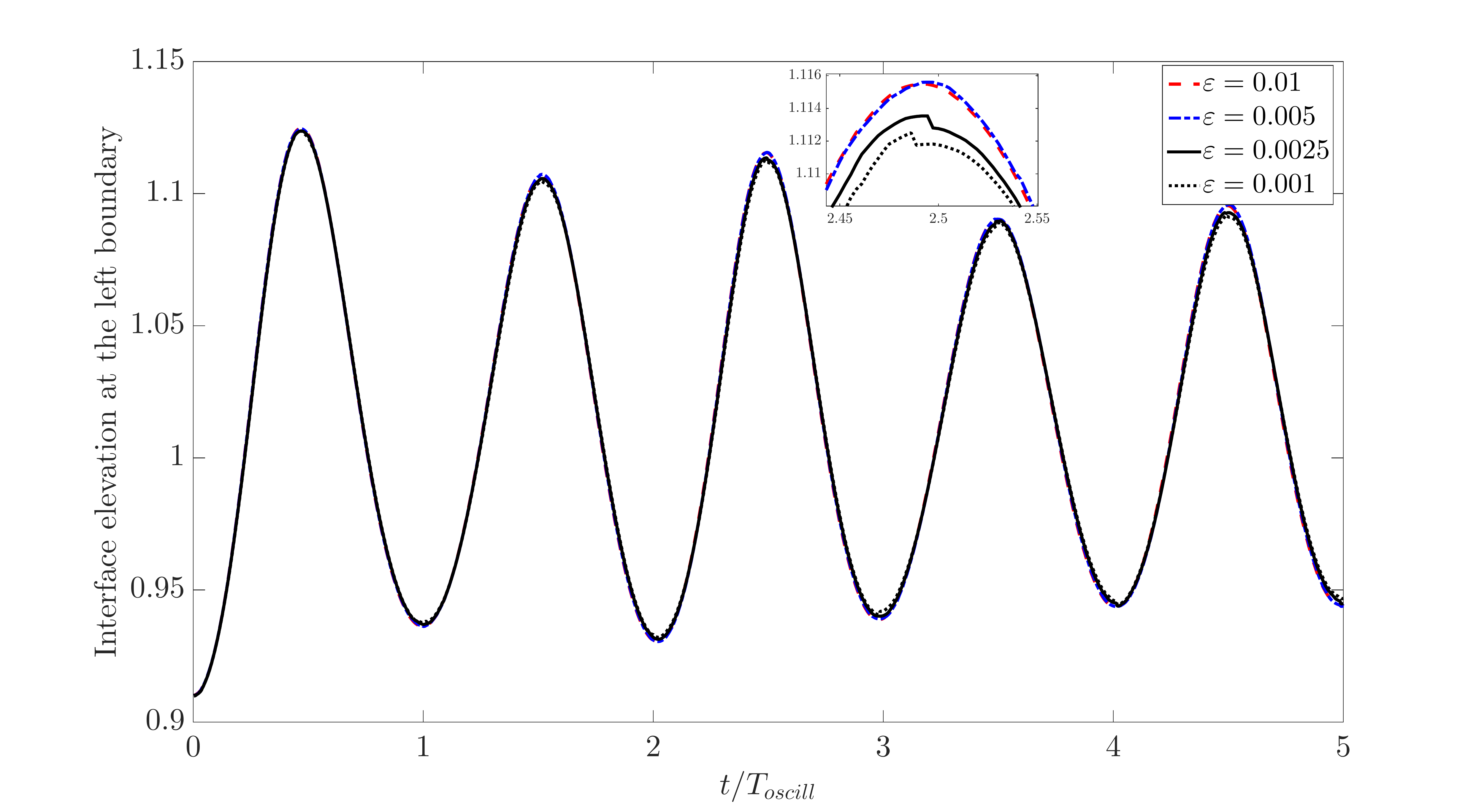}
	\caption{}
	\end{subfigure}%
	\hspace{-1cm}
	\begin{subfigure}[b]{0.28\textwidth}
	\begin{subfigure}[b]{\textwidth}
		\includegraphics[trim={15cm 14cm 16cm 0.2cm},clip,width=5cm]{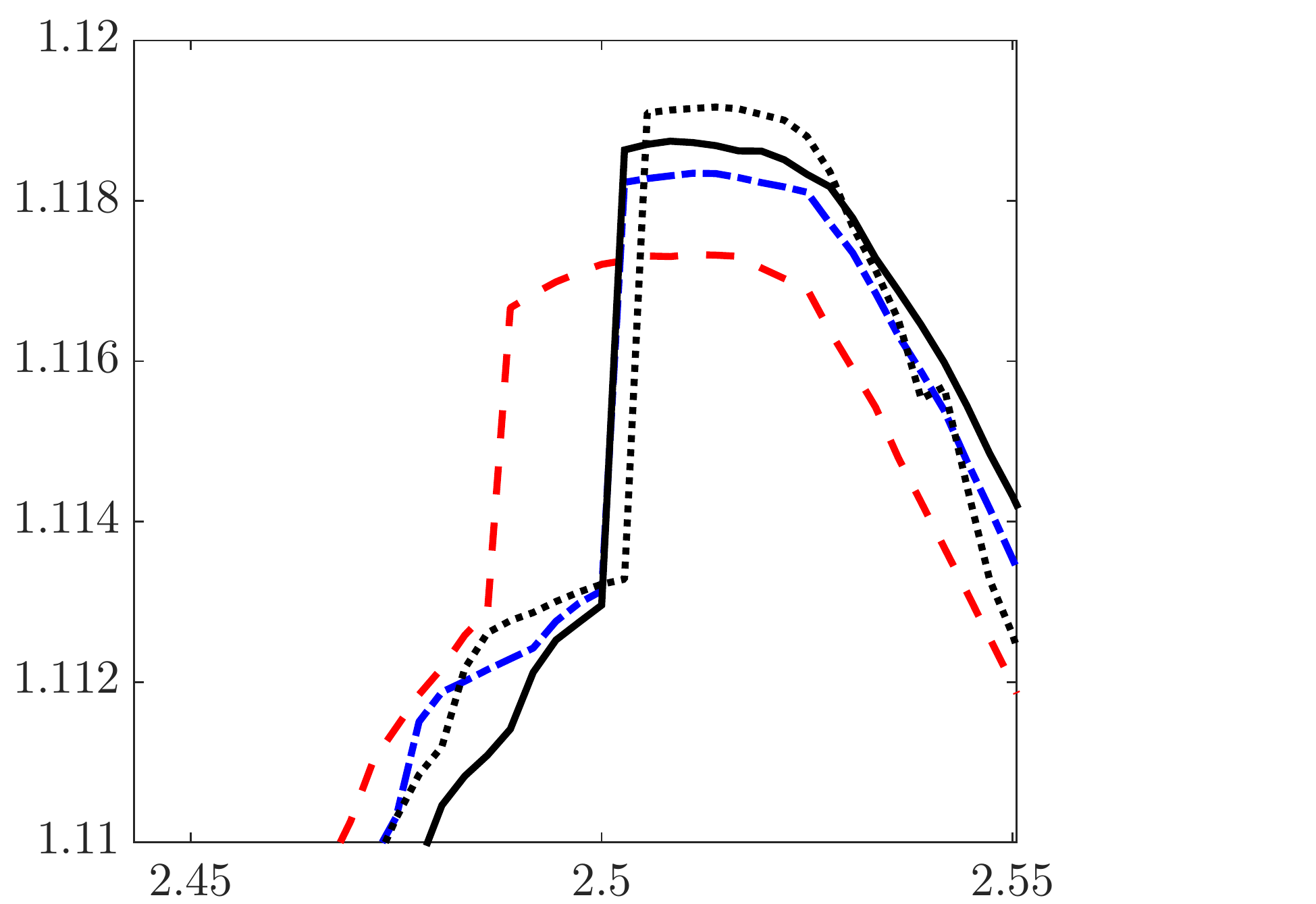}
	\caption{}
	\end{subfigure}\\
	\begin{subfigure}[b]{\textwidth}
		\includegraphics[trim={15cm 14cm 16cm 0.2cm},clip,width=5cm]{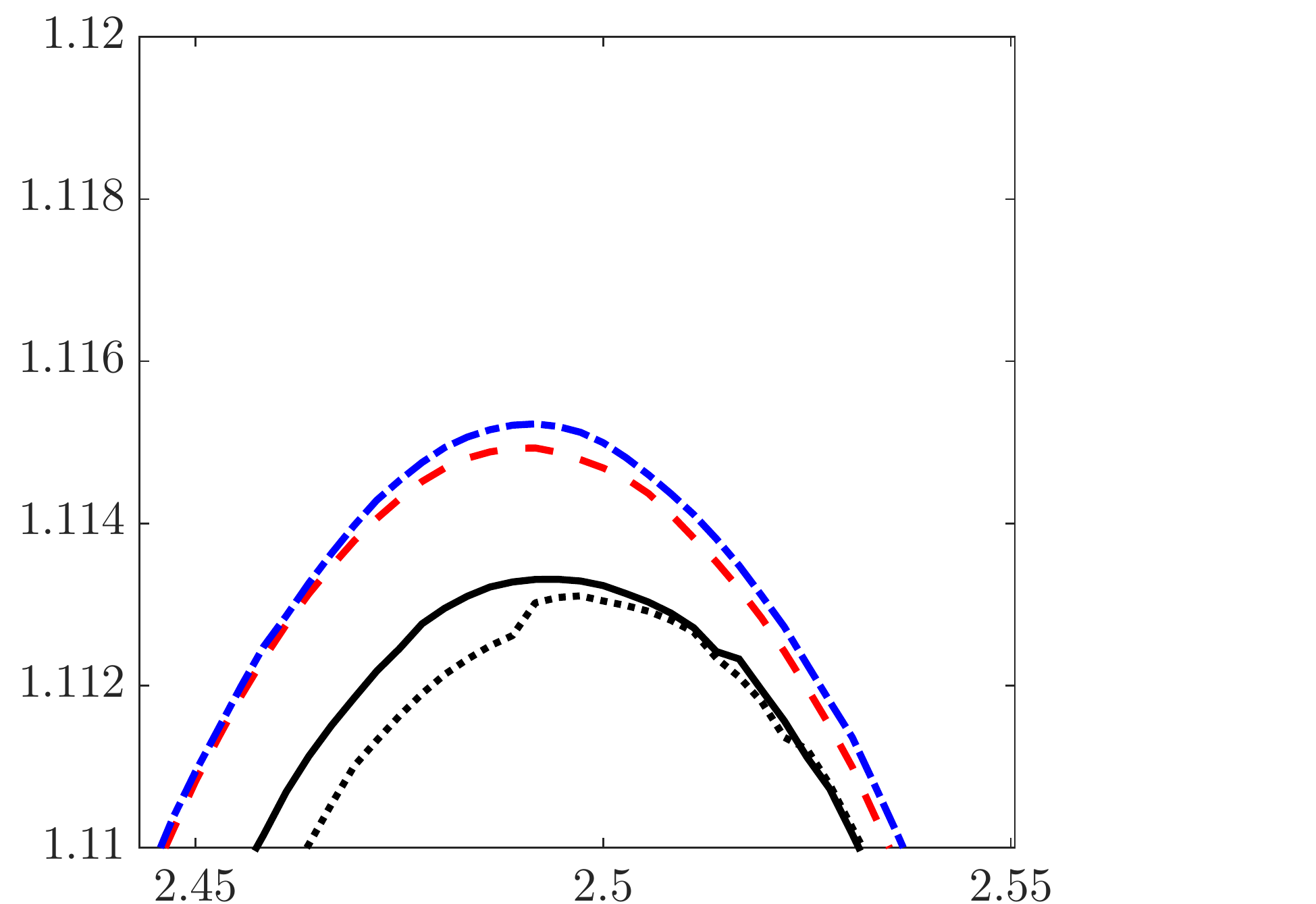}
	\caption{}
	\end{subfigure}%
	\end{subfigure}
\caption{Variation of the interface at the left boundary in the sloshing tank problem: (a) $tol_R=1\times 10^{-3}$, (b) $tol_R=5\times 10^{-3}$ and (c) $tol_R=5\times 10^{-4}$.} 
\label{ST_8}
\end{figure}

In this subsection, we briefly discuss the relationship between the tolerance set for the residual error $\eta$ ($tol_R$) and the interfacial thickness parameter $\varepsilon$. We measure the total mass loss, the average number of degrees of freedom and the average error in solving the Allen-Cahn equation ($\overline{e}_{AC}$). The total mass loss is evaluated using Eq.~(\ref{mass_error}) with $m$ taking the mass at final time $t/T_{oscill}=5$. The variation of these measured quantities with $tol_R$ is plotted in Fig. \ref{ST_7}. Some of the observations are as follows. First, the total mass loss is higher for low $\varepsilon$ when large $tol_R$ is chosen as the convergence criteria for $\eta$, while it reduces as the tolerance is tightened. Second, the average number of unknowns to be solved decreases with $\varepsilon$ as $tol_R$ becomes smaller. Third, the average error in solving the Allen-Cahn equation is much higher for lower $\varepsilon$ at high $tol_R$. These observations seem to suggest that selecting a smaller $\varepsilon$ value for large $tol_R$ does not improve the solution. This is evident from the time history of the surface elevation at the left boundary for varying $\varepsilon$ and $tol_R$ in Fig. \ref{ST_8}. Figure \ref{ST_8}(a) shows the variation of the free-surface elevation for $tol_R=1\times 10^{-3}$, while Figs. \ref{ST_8}(b) and (c) depict the close-up view of the same plot for $tol_R=5\times 10^{-3}$ and $tol_R=5\times 10^{-4}$ respectively. It can be seen that for higher $tol_R$, all values of $\varepsilon$ have a kink in the plot. This kink is due to excessive coarsening near the interface area because of high $tol_R$. Consequently, the error $\overline{e}_{AC}$ is higher because the given $tol_R$ is not enough to capture the crucial interfacial physics. As the tolerance $tol_R$ is tightened, lower $\varepsilon$ values tend to approach the converged solution. Therefore, the selection of $\varepsilon$ is directly dependent on the convergence criteria for $tol_R$. In order to exploit the benefits of lower computational cost, less error and good mass conservation properties, the tolerance $tol_R$ has to be reduced for small $\varepsilon$ values. In the next subsection, we present the results corresponding to $\varepsilon=0.005$ and $0.0025$ with $tol_R = 5\times 10^{-4}$.

\subsubsection{Role of the interfacial thickness parameter $\varepsilon$}
\begin{figure}
\centering
	\hspace{-1cm}
	\begin{subfigure}[b]{0.33\textwidth}
		\includegraphics[trim={10cm 0.2cm 11cm 0.2cm},clip,width=5.5cm]{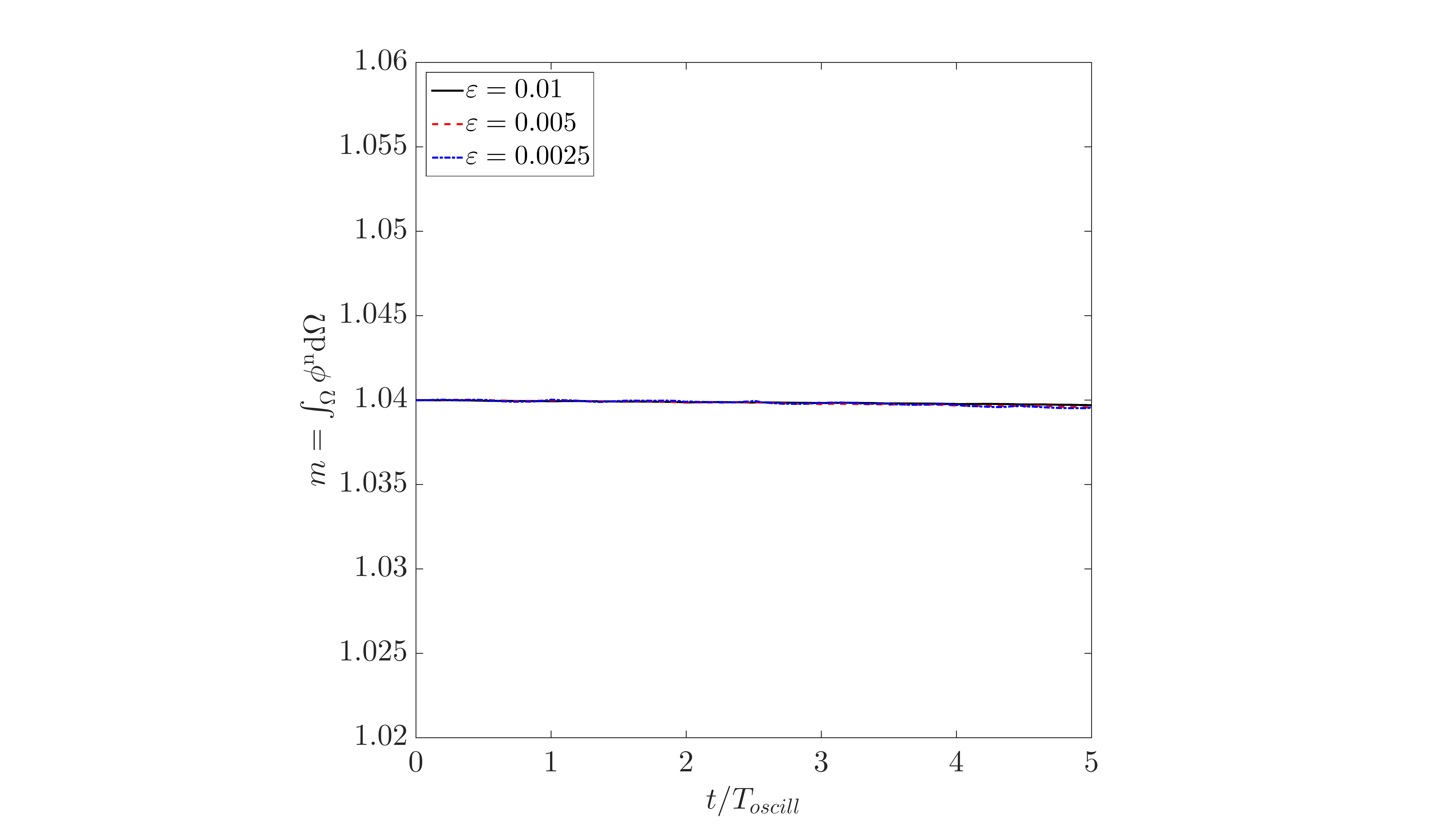}
	\caption{}
	\end{subfigure}%
	\begin{subfigure}[b]{0.33\textwidth}
		\includegraphics[trim={10cm 0.2cm 11cm 0.2cm},clip,width=5.5cm]{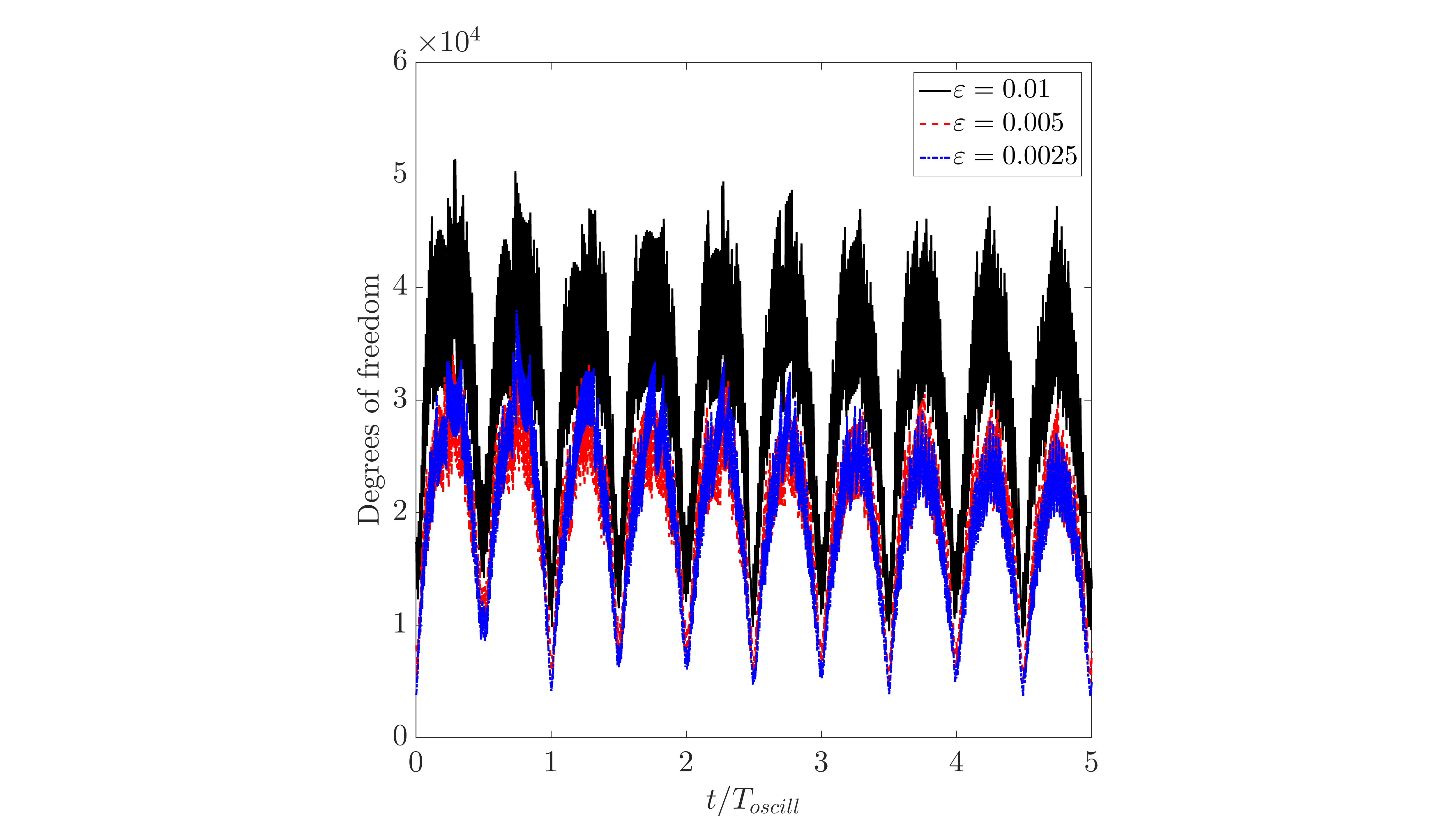}
	\caption{}
	\end{subfigure}
	\begin{subfigure}[b]{0.34\textwidth}
		\includegraphics[trim={10cm 0.2cm 11cm 0.2cm},clip,width=5.5cm]{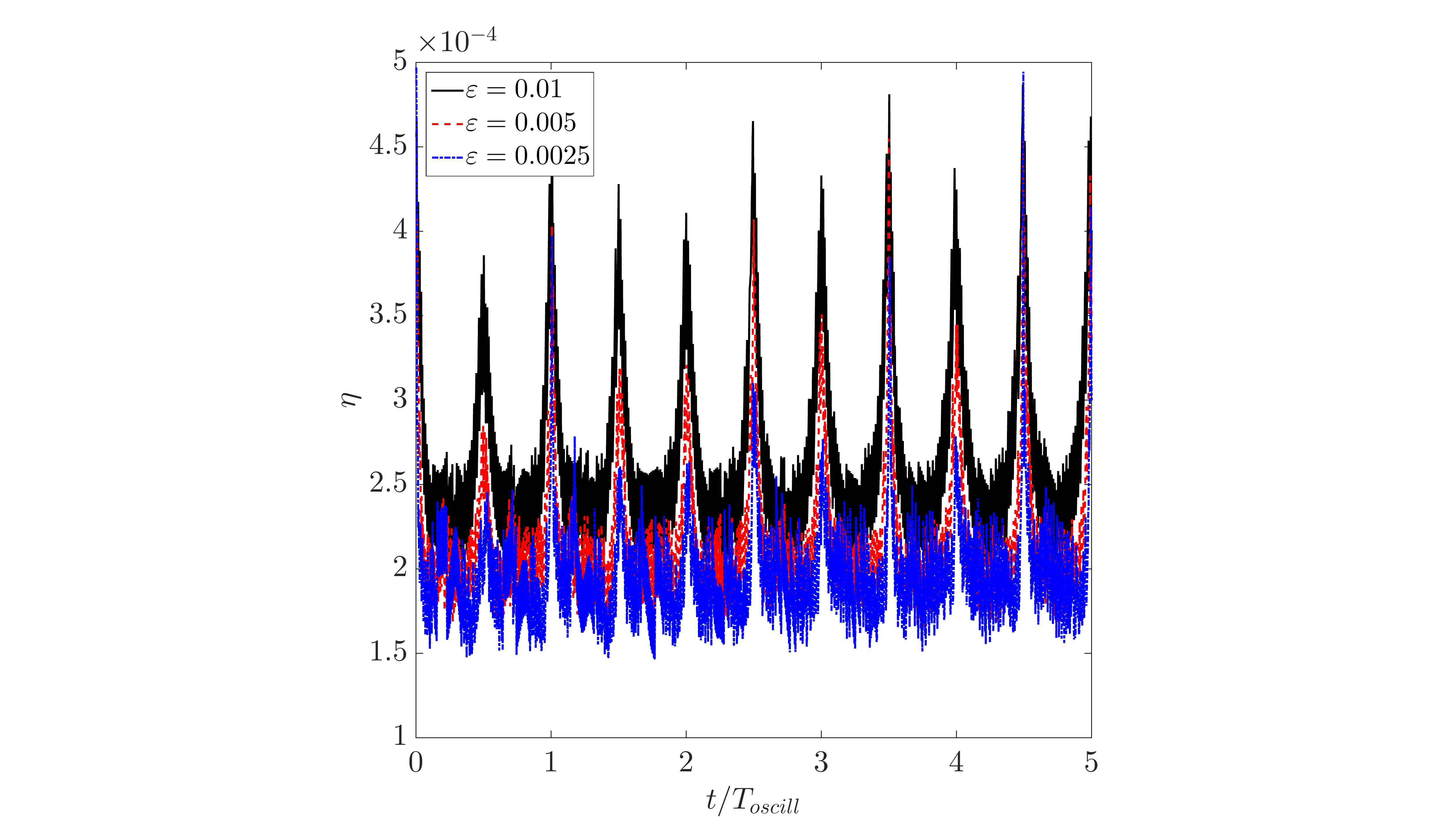}
	\caption{}
	\end{subfigure}%
\caption{Variation of the measured quantities with time in the sloshing tank problem for varying interface thickness parameter $\varepsilon$: (a) mass evaluated using Eq.~(\ref{mass_def}), (b) degrees of freedom or unknowns, and (c) residual error indicator $\eta$.} 
\label{ST_5}
\end{figure}

\begin{figure}
\centering
	\hspace{-0.7cm}
	\begin{subfigure}[b]{0.3\textwidth}
		\includegraphics[trim={0.5cm 2cm 9cm 0.2cm},clip,width=4cm]{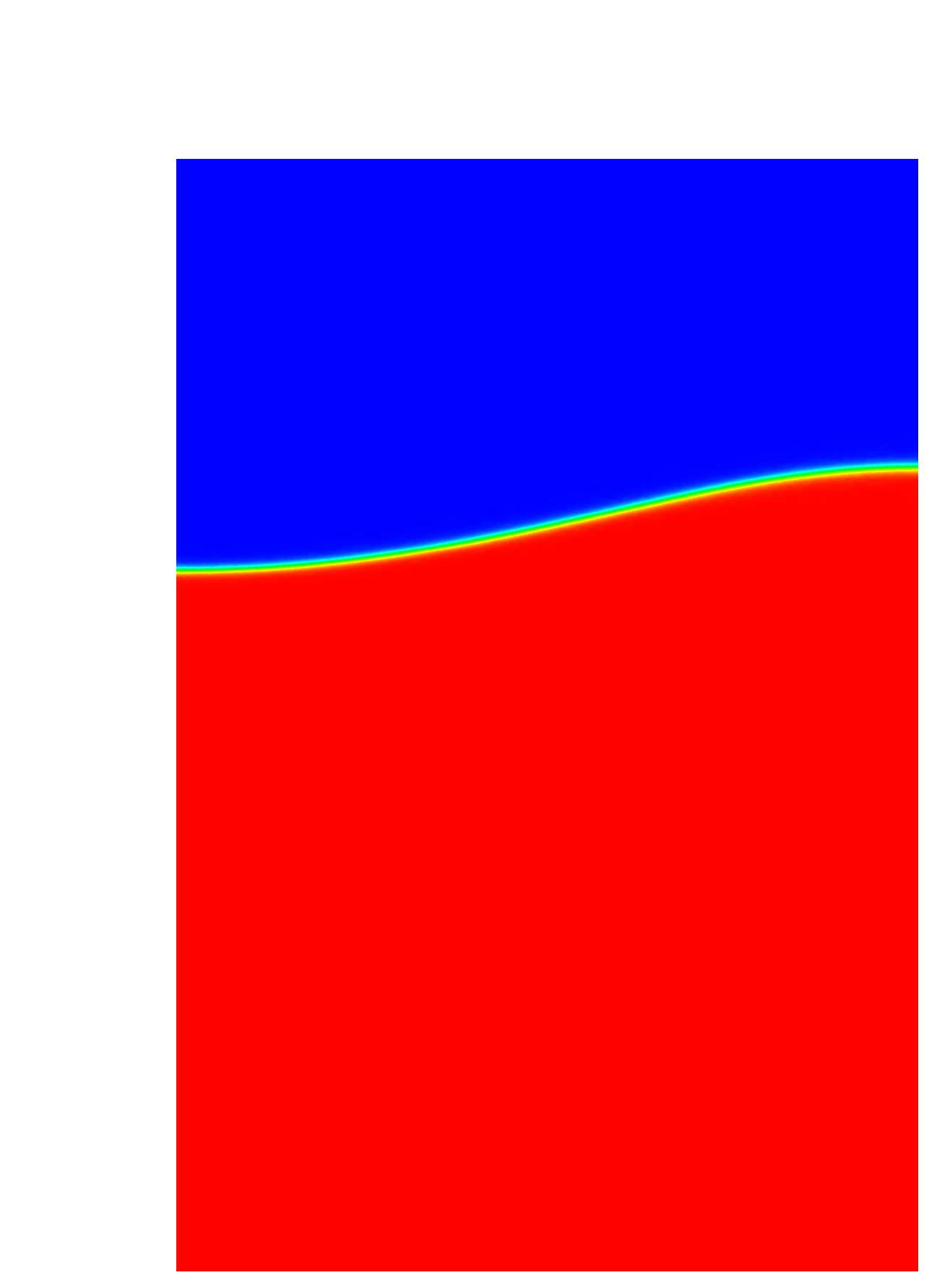}
	\caption{}
	\end{subfigure}
	\hspace{-1cm}
	\begin{subfigure}[b]{0.3\textwidth}
		\includegraphics[trim={0.5cm 2cm 9cm 0.2cm},clip,width=4cm]{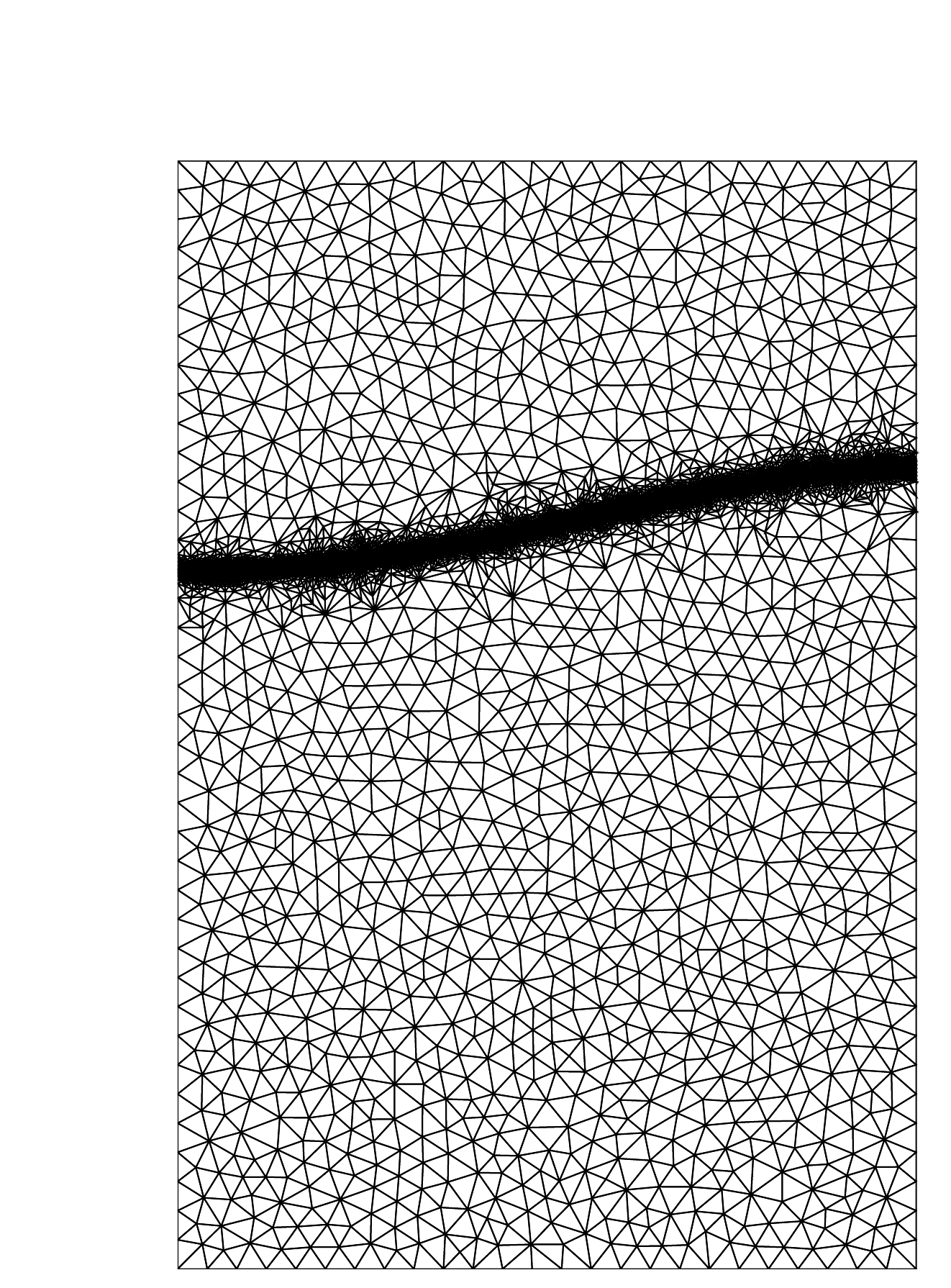}
	\caption{}
	\end{subfigure}
	\hspace{-1cm}
	\begin{subfigure}[b]{0.3\textwidth}
		\includegraphics[trim={0.5cm 2cm 9cm 0.2cm},clip,width=4cm]{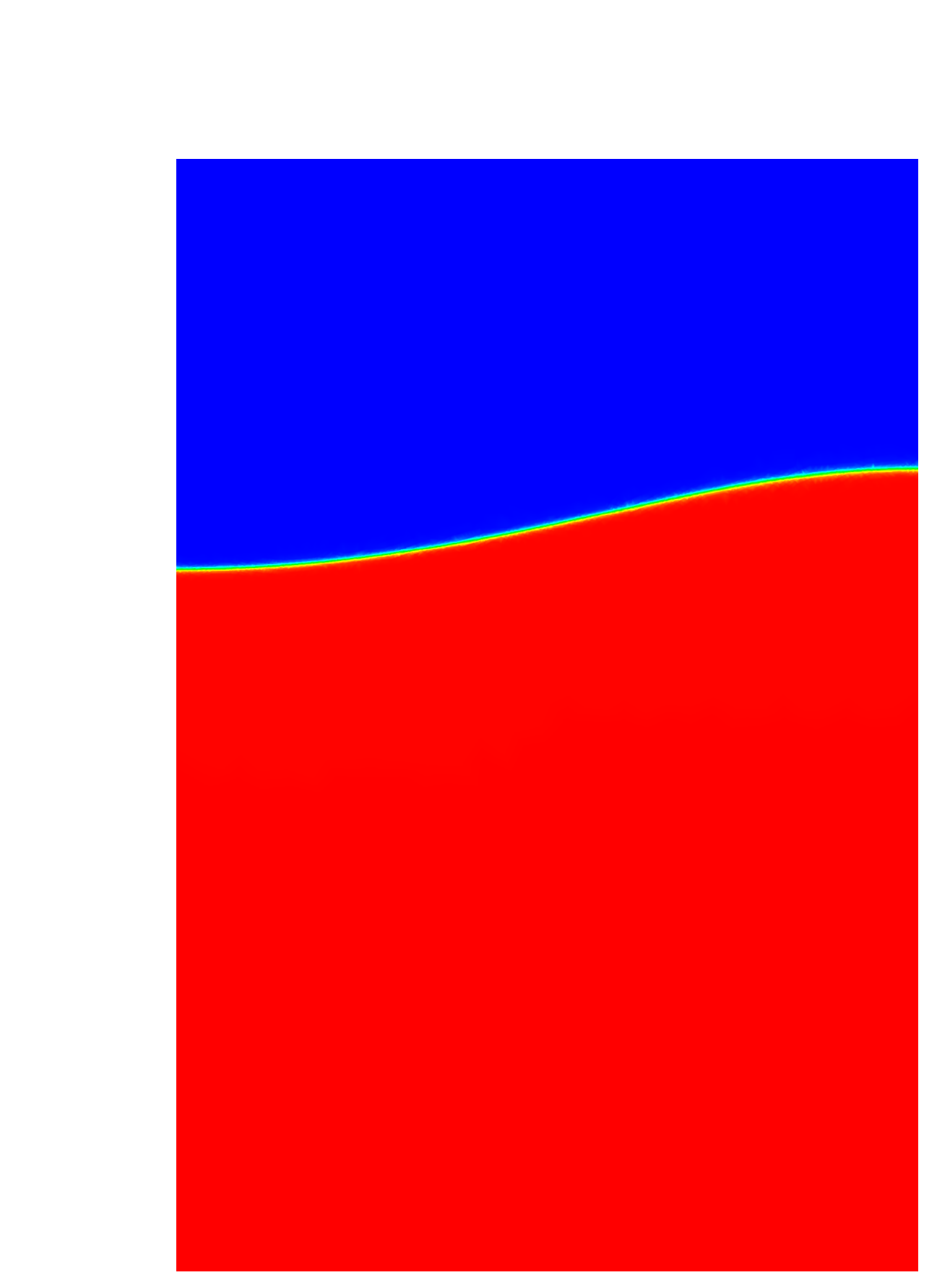}
	\caption{}
	\end{subfigure}
	\hspace{-1cm}
	\begin{subfigure}[b]{0.3\textwidth}
		\includegraphics[trim={0.5cm 2cm 9cm 0.2cm},clip,width=4cm]{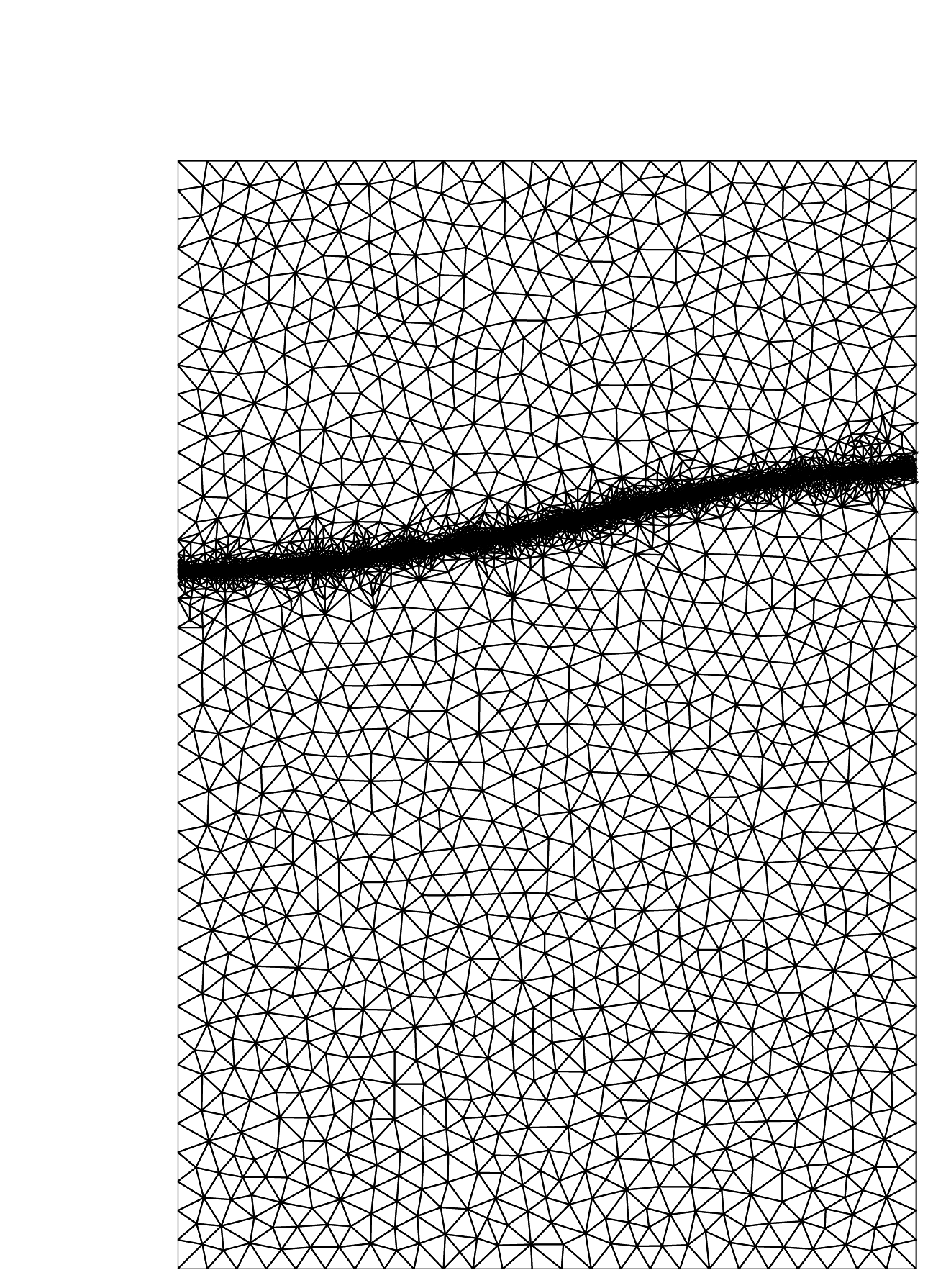}
	\caption{}
	\end{subfigure}
	\hspace{-1cm}
\caption{Contours of the order parameter $\phi$ (a and c) and the adaptive mesh (b and d) for the  sloshing tank problem at $t/T_{oscill} = 5$ for: $\varepsilon=0.005$ (a and b), and $\varepsilon=0.0025$ (c and d).} 
\label{ST_6}
\end{figure}

The effect of reducing the interfacial thickness parameter $\varepsilon$ is studied by performing some numerical experiments with $tol_{NS}=tol_{AC}=tol_R=5\times 10^{-4}$. The variation of mass, degrees of freedom and the residual error with $t/T_{oscill}$ is plotted in Fig. \ref{ST_5}. It is observed that reduction in the $\varepsilon$ leads to lesser degrees of freedom for the same $tol_{R}$. Therefore, the benefit of the adaptivity is noticeable while using the lower interface thickness parameter since the degrees of freedom would increase with lesser $\varepsilon$ for a fixed Eulerian grid. Moreover, the residual error $\eta$ also gets reduced with a decrease in $\varepsilon$. The contours of the order parameter and the respective mesh for $\varepsilon = 0.005$ and $0.0025$ at $t/T_{oscill}=5$ are shown in Fig. \ref{ST_6}. This completes the assessment of the proposed NAVP procedure. In the next section, we demonstrate the NAVP procedure for a classical dam break problem.

\section{Application to dam break problem}
\label{damBreak}
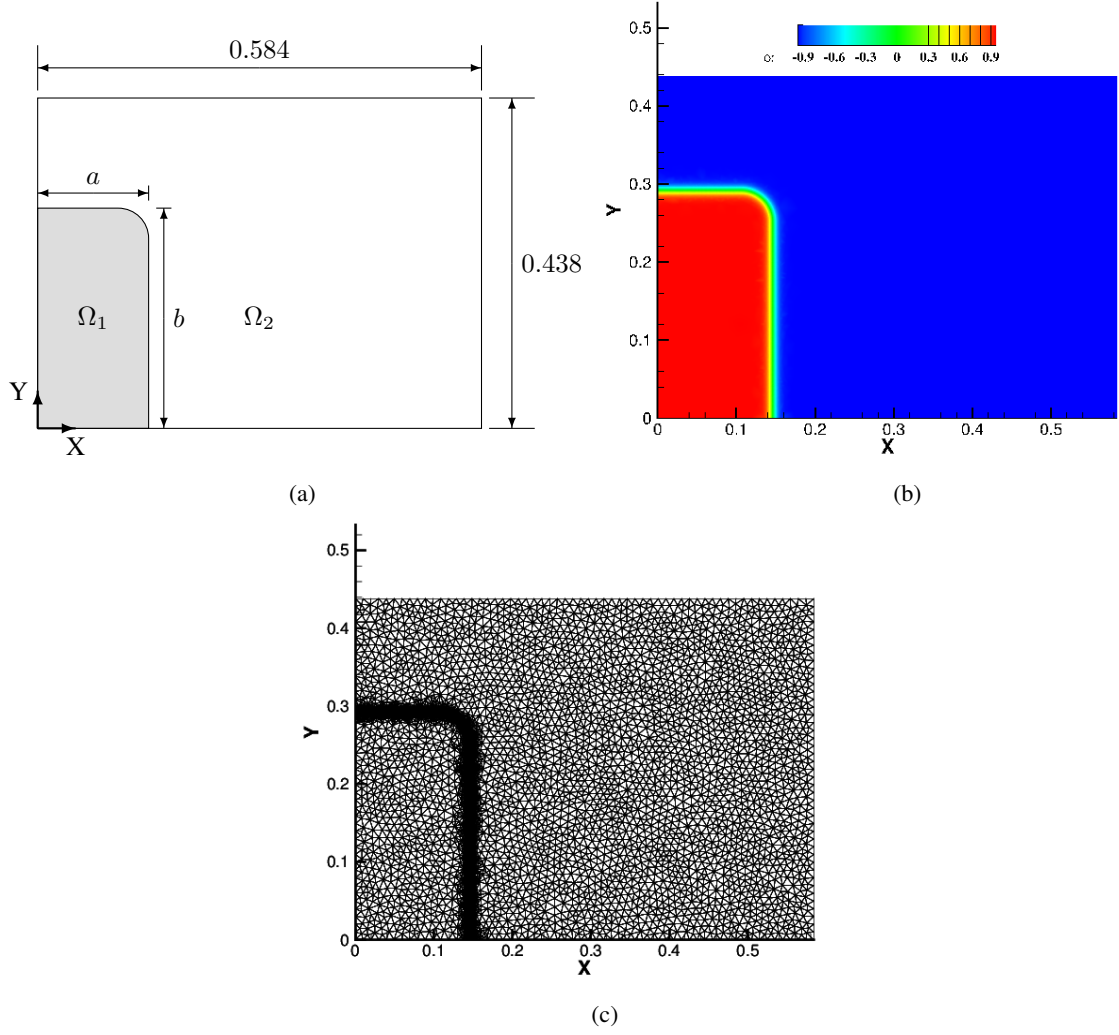
\begin{figure}
\centering
	\begin{subfigure}[b]{0.5\textwidth}
\begin{tikzpicture}[decoration={markings,mark=at position 1.0 with {\arrow{>}}},scale=10]
	\draw (0,0) -- (0.584,0)-- (0.584,0.438) -- (0,0.438) -- cycle;
	\draw[fill={rgb:black,1;white,2}, fill opacity=0.4] (0,0)--(0.146,0)--(0.146,0.252) arc (0:90:0.04)--(0,0.292)--(0,0);
	\draw (0.073,0.146) node(A){$\Omega_1$};
	\draw (0.292,0.146) node{$\Omega_2$};
	\draw[thick,postaction={decorate}] (0,0) to (0.05,0);
	\draw[thick,postaction={decorate}] (0,0) to (0,0.05);
	\draw (0.05,0) node[anchor=north]{X};
	\draw (0,0.05) node[anchor=east]{Y};
	\draw[postaction={decorate}] (0.624,0.219) to (0.624,0.438);
	\draw[postaction={decorate}] (0.624,0.219) to (0.624,0);
	\draw (0.594,0) -- (0.654,0);
	\draw (0.594,0.438) -- (0.654,0.438);
	\draw (0.624,0.219) node[anchor=west]{$0.438$};
	\draw[postaction={decorate}] (0.292,0.478) to (0.584,0.478);
	\draw[postaction={decorate}] (0.292,0.478) to (0,0.478);
	\draw (0,0.448) -- (0,0.508);
	\draw (0.584,0.448) -- (0.584,0.508);
	\draw (0.292,0.478) node[anchor=south]{$0.584$};
	\draw[postaction={decorate}] (0.166,0.146) to (0.166,0);
	\draw[postaction={decorate}] (0.166,0.146) to (0.166,0.292);
	\draw (0.156,0.292) -- (0.176,0.292);
	\draw (0.166,0.146) node[anchor=west]{$b$};
	\draw[postaction={decorate}] (0.073,0.312) to (0.146,0.312);
	\draw[postaction={decorate}] (0.073,0.312) to (0,0.312);
	\draw (0.146,0.302) -- (0.146,0.322);
	\draw (0.073,0.312) node[anchor=south]{$a$};
\end{tikzpicture}
	\caption{}
	\end{subfigure}%
	\begin{subfigure}[b]{0.5\textwidth}
	\includegraphics[trim={1cm 0.5cm 2cm 2cm},clip,width=7cm]{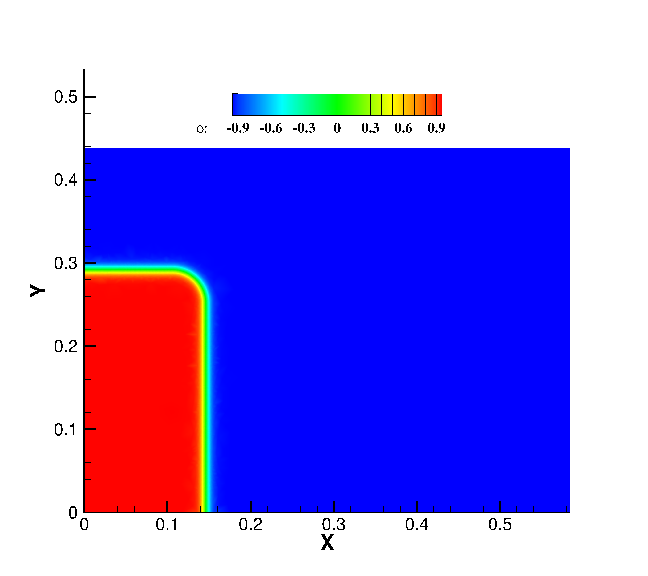}		
	\caption{}
	\end{subfigure}
	\begin{subfigure}[b]{0.5\textwidth}
	\includegraphics[trim={1cm 0.5cm 2cm 2cm},clip,width=7cm]{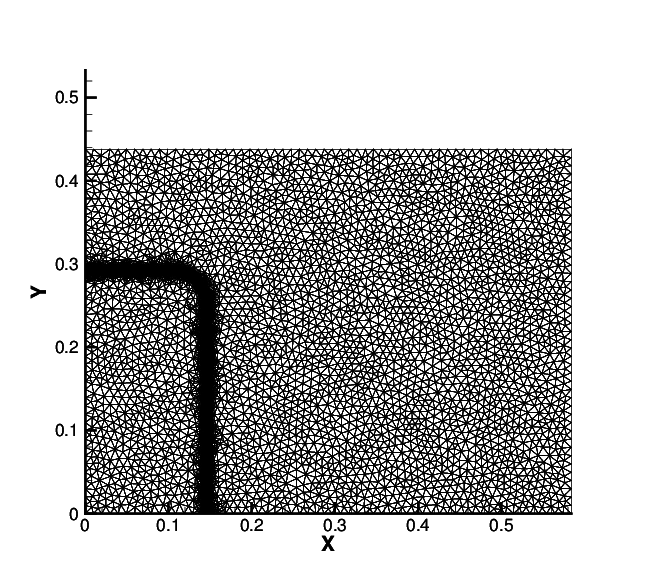}		
	\caption{}
	\end{subfigure}
\caption{Two-dimensional dam break problem: (a) schematic diagram showing the computational domain, (b) the contour plot of the order parameter $\phi$ at $t=0$, and (c) the initial refined mesh at $t=0$. In (a), $\Omega_1$ and $\Omega_2$ are the two phases with $\rho_1=1000$, $\rho_2=1$, $\mu_1=10^{-3}$ and $\mu_2=10^{-5}$, acceleration due to gravity is taken as $\boldsymbol{g}=(0,-9.81,0)$ and slip boundary condition is imposed on all the boundaries.} 
\label{db_2D}
\end{figure}

\begin{figure}
\centering
	\begin{subfigure}[b]{0.5\textwidth}
		\includegraphics[trim={10cm 0 11.8cm 0cm},clip,width=7cm]{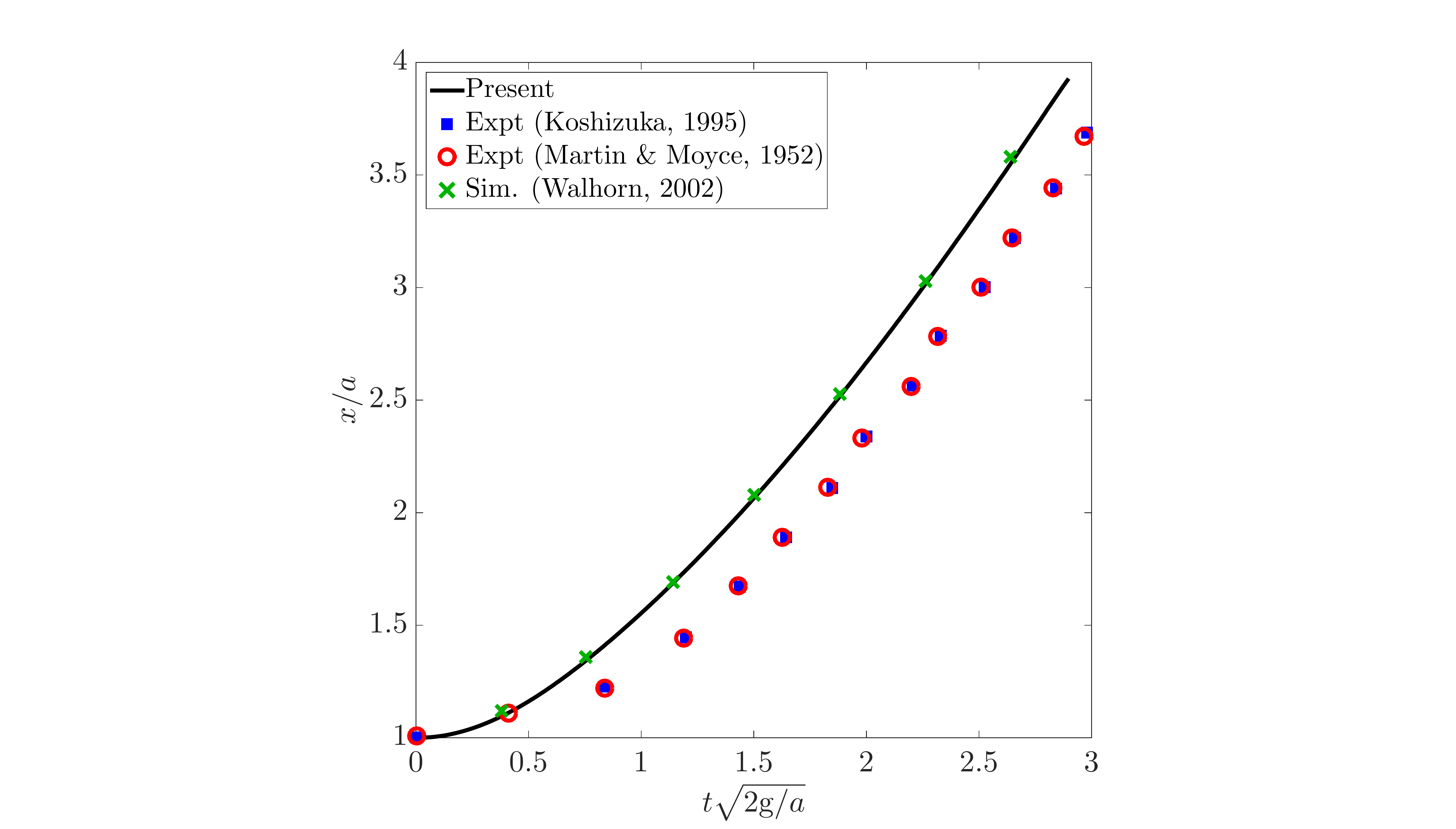}
	\caption{}
	\end{subfigure}%
	\begin{subfigure}[b]{0.5\textwidth}
		\includegraphics[trim={10cm 0 11.8cm 0cm},clip,width=7cm]{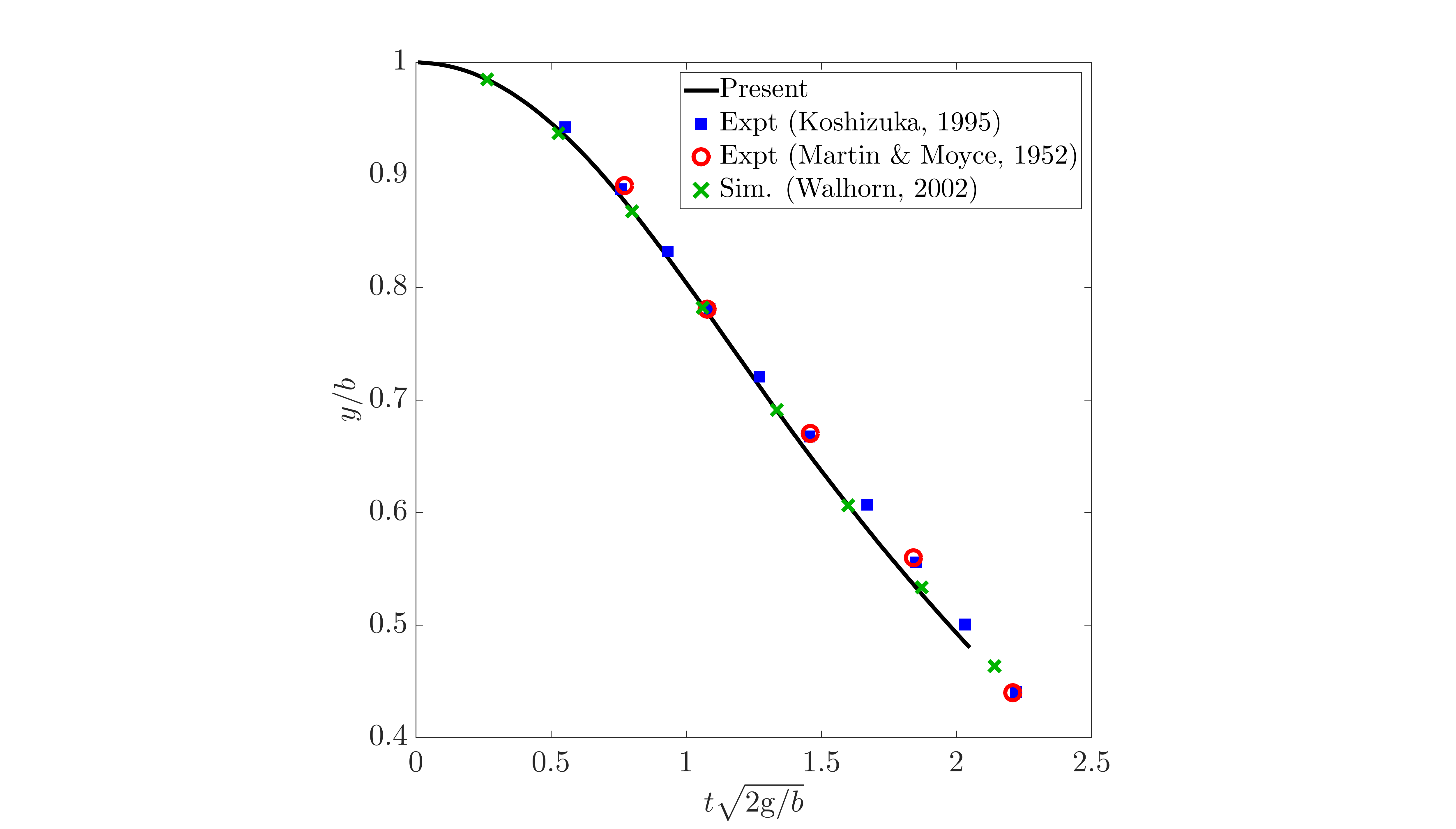}
	\caption{}
	\end{subfigure}		
\caption{Two-dimensional dam break problem: temporal evolution of non-dimensional water column (a) width and (b) height. The results from the present method are in good agreement with the literature.} 
\label{db_2D_val}
\end{figure}

\begin{figure}
\centering
\begin{subfigure}[b]{\textwidth}
	\begin{subfigure}[b]{0.5\textwidth}
		\includegraphics[trim={0.5cm 2cm 0.5cm 4cm},clip,width=8cm]{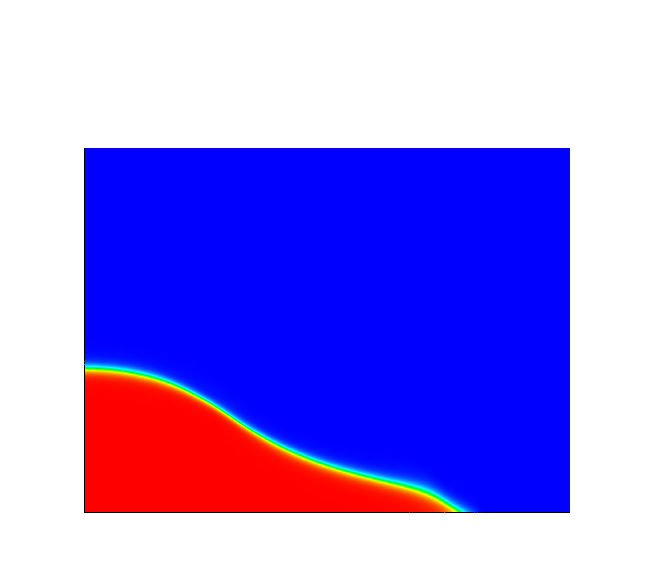}
	\end{subfigure}%
	\begin{subfigure}[b]{0.5\textwidth}
		\includegraphics[trim={0.5cm 2cm 0.5cm 4cm},clip,width=8cm]{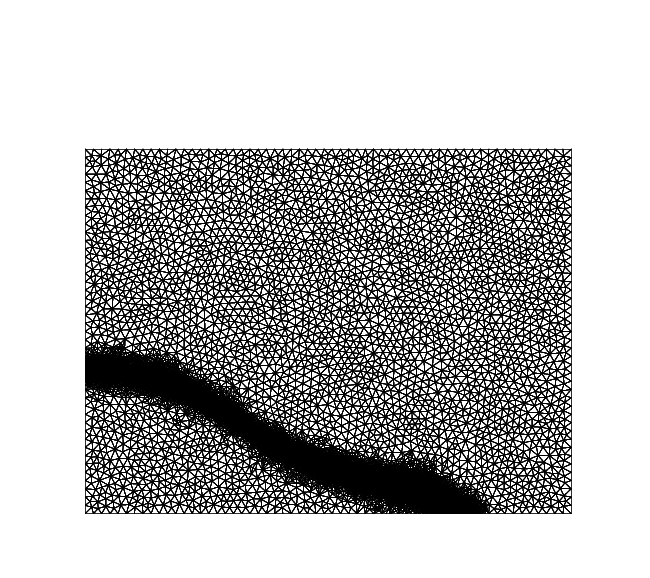}
	\end{subfigure}
	\caption{}
\end{subfigure}\\	
\begin{subfigure}[b]{\textwidth}
	\begin{subfigure}[b]{0.5\textwidth}
		\includegraphics[trim={0.5cm 2cm 0.5cm 4cm},clip,width=8cm]{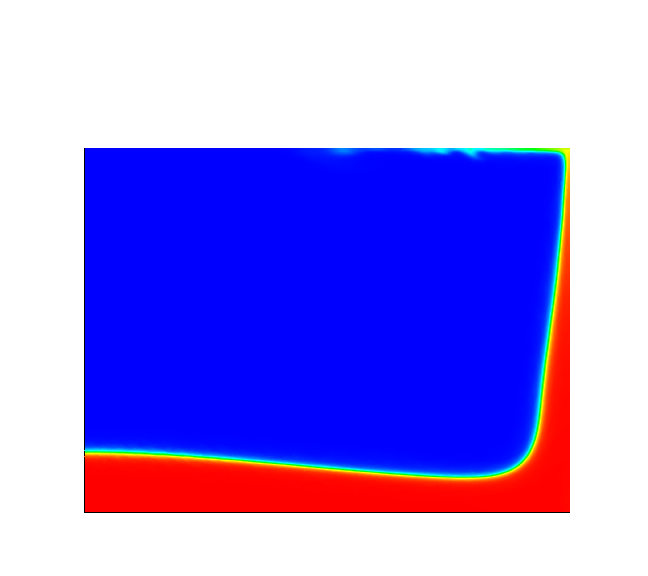}
	\end{subfigure}
	\begin{subfigure}[b]{0.5\textwidth}
		\includegraphics[trim={0.5cm 2cm 0.5cm 4cm},clip,width=8cm]{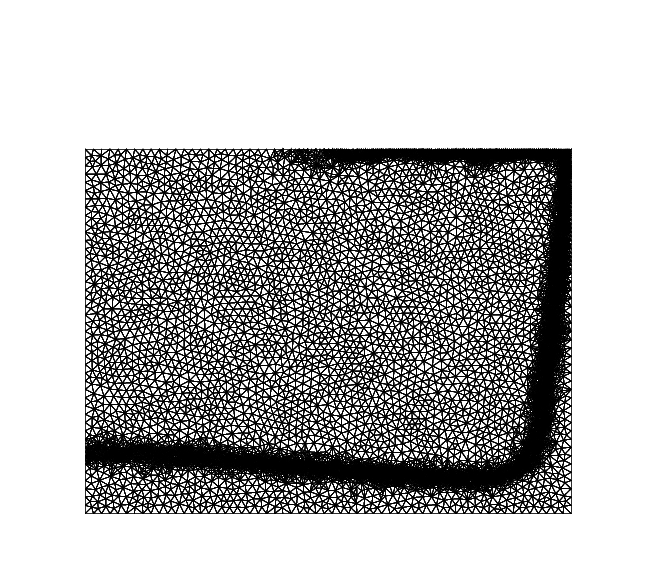}
	\end{subfigure}
	\caption{}
	\end{subfigure}\\
	\begin{subfigure}[b]{\textwidth}
	\begin{subfigure}[b]{0.5\textwidth}
		\includegraphics[trim={0.5cm 2cm 0.5cm 4cm},clip,width=8cm]{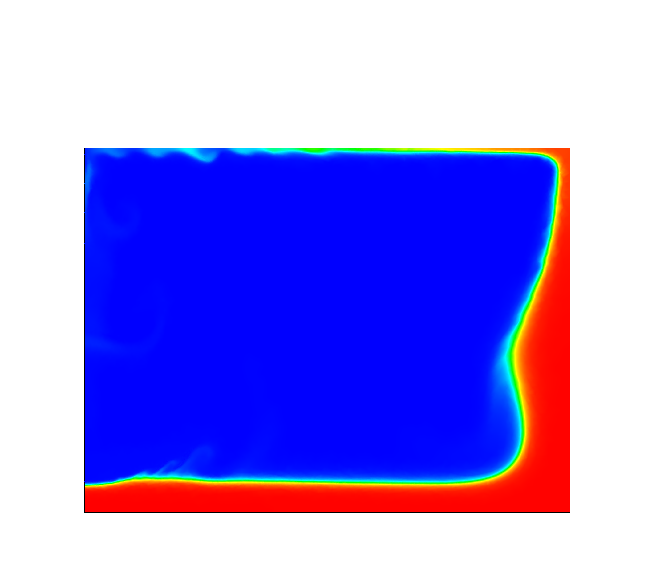}
	\end{subfigure}%
	\begin{subfigure}[b]{0.5\textwidth}
		\includegraphics[trim={0.5cm 2cm 0.5cm 4cm},clip,width=8cm]{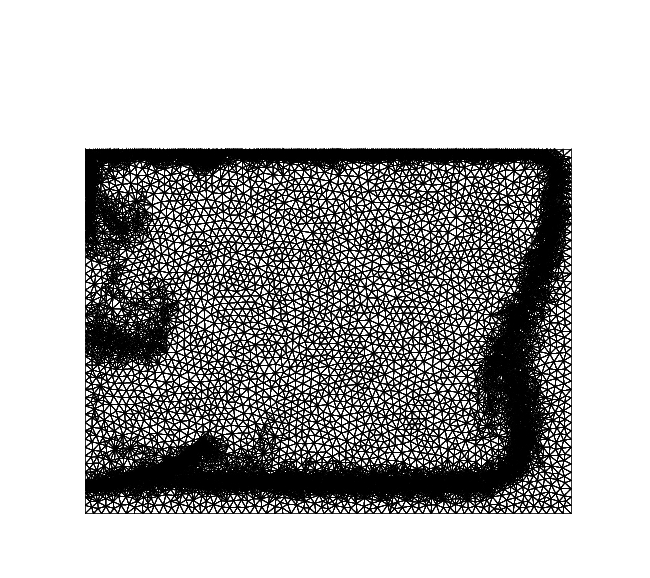}
	\end{subfigure}
	\caption{}
	\end{subfigure}\\
	\begin{subfigure}[b]{\textwidth}
	\begin{subfigure}[b]{0.5\textwidth}
		\includegraphics[trim={0.5cm 2cm 0.5cm 4cm},clip,width=8cm]{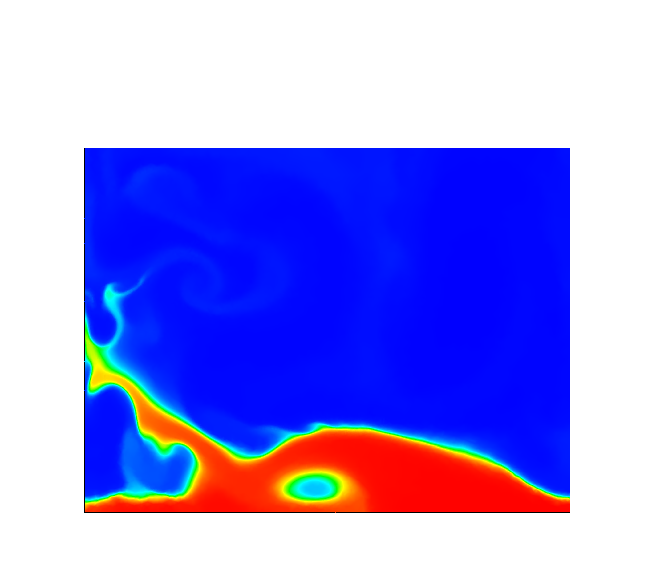}
	\end{subfigure}
	\begin{subfigure}[b]{0.5\textwidth}
		\includegraphics[trim={0.5cm 2cm 0.5cm 4cm},clip,width=8cm]{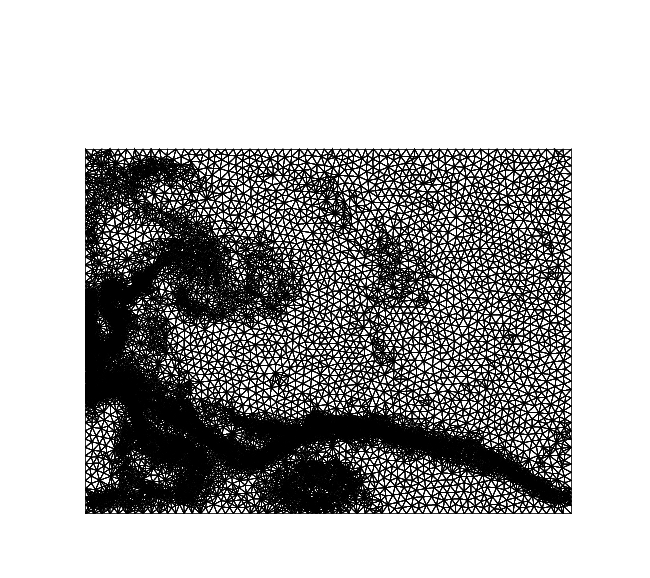}
	\end{subfigure}
	\caption{}
	\end{subfigure}
\caption{Contours of the order parameter $\phi$ (left) and the adaptive mesh (right) for the  dam break problem at $t\sqrt{(2g/a)}$: (a) $2.32$, (b) $4.64$, (c) $9.96$ and (d) $11.60$.} 
\label{db2D_2}
\end{figure}

We consider a two-dimensional dam break problem to demonstrate the  applicability and robustness of the proposed NAVP procedure. 
In this widely studied problem of dam break flow, there is a sudden collapse of a rectangular column of fluid onto a horizontal surface. 
This problem involves a large unsteady deformation of air-water interface and characterizes the dynamics associated with gravity and viscosity effects.
A rectangular computational domain $[0,0.584]\times [0,0.438]$ shown in Fig. \ref{db_2D}(a) is considered for the present study. A water column of size $0.146 \times 0.292$ units is placed at the left boundary of the domain at time $t=0$. 
The initial condition is given by:
\begin{align}
	\phi(x,y,0) = \begin{cases}
		-\mathrm{tanh}\big( \frac{y-b}{\sqrt{2}\varepsilon} \big),\ \mathrm{for}\ x \leq (a - r), y \geq (b-r) \\
		 \\
		-\mathrm{tanh}\big( \frac{x-a}{\sqrt{2}\varepsilon} \big),\ \mathrm{for}\ x > (a-r), y < (b-r) \\
		\\
		\mathrm{tanh}\big( \frac{r - \sqrt{(x-(a-r))^2 + (y-(b-r))^2} }{\sqrt{2}\varepsilon} \big),\ \mathrm{for}\ x \geq (a-r), y \geq (b-r) \\
		 \\
		 1,\ \mathrm{elsewhere}
		\end{cases}
\end{align}
where $a=0.146$, $b=0.292$ and $r=0.04$ are the width, height and the radius of the curve of the water column respectively, $\phi=1$ and $\phi=-1$ correspond to the order parameter on $\Omega_1$ and $\Omega_2$ respectively. A non-uniform triangular unstructured mesh  with background mesh size $\Delta x=0.01$ is employed for the demonstration purpose. The interfacial thickness parameter $\varepsilon$ is selected as $0.005$. The tolerances are set as $tol_{NS}=tol_{AC}=1\times 10^{-4}$ and $tol_R=5\times 10^{-4}$. A restriction of the maximum number of elements is kept in the convergence criteria as $25000$.
The interface location at the left and bottom boundaries are tracked with time for the validation with experiments \cite{Martin_Moyce, Ubbink_thesis} and the interface-tracking simulation of \cite{Walhorn_thesis}.  In Fig. \ref{db_2D_val}, the temporal variation of the interface location based on the non-dimensional water column width/height is compared 
with the results from the literature. Good agreement is found for our NAVP procedure based on the Navier-Stokes and the Allen-Cahn equations. The evolution of the height of the water column is captured quite well. However, the expansion of the water column (in width) in the experiment is slower than what is predicted from the simulation. 
In \cite{Fries_1}, such delay has been attributed to the time required to remove the partition which holds the water column to its initial profile in the experiment.
The contours of the order parameter $\phi$ and the evolving mesh are depicted in Fig. \ref{db2D_2} at $t\sqrt{(2g/a)} \in [2.32,11.60]$. In terms of the computational cost, the average number of degrees of freedom for the adaptive simulation was $15000$ which yielded an average $\eta$ of $3.36\times 10^{-3}$ compared to the non-adaptive grid with degrees of freedom $21355$ with an average $\eta$ of $7.57\times 10^{-3}$. This shows the decrease in the residual error indicators in the adaptive grid with lesser degrees of freedom, thus reducing the computational cost.

The unique feature of the presented algorithm is its generality in the sense that it can be extended to three dimensions, whereby the variational formulation poses  no restriction with respect to the extension. This was demonstrated by three dimensional simulations in \cite{AC_JCP}. However, some effort has to be put in to extend the adaptive algorithm of refining through bisection and coarsening to three-dimensions. The advantage of the adaptive algorithm is its unique feature of avoiding complicated data structures. The refinement algorithm is extended to three dimensions and some suggestions regarding the extension of the coarsening algorithm to three dimensions are provided in \cite{iFEM}. 

\section{Conclusions}
\label{conclusion}
In the present work, a stable, robust and general adaptive variational partitioned procedure has been proposed to solve the coupled Navier-Stokes 
and Allen-Cahn equations for two-phase fluid flows.
The Allen-Cahn equation is solved by the positivity preserving variational technique to ensure boundedness in the nonlinear convective-diffusion-reaction system. 
The newest vertex bisection algorithm is employed as the adaptive algorithm based on the residual error estimates of the Allen-Cahn equation. 
The procedure aims at coarsening of grid at the last nonlinear iteration while maintaining the convergence properties of the nonlinear Navier-Stokes and Allen-Cahn equations.
While satisfying the positivity preservation and the energy stability, 
we have shown a remarkable reduction in the computational cost with respect its non-adaptive counterpart on general unstructured adaptive meshes. 
The simplicity of the refinement/coarsening avoids complicated tree-type data structures, thus providing the ease of implementation.
Several test problems of increasing complexity are considered to demonstrate the various aspects of the proposed adaptive variational scheme. 
%
We have first assessed the effectiveness of our adaptive algorithm for the spinodal decomposition in a complex geometry and the volume conservation property for the two circles in a square domain. The generality of the present scheme for curved domains with complex geometry was successfully demonstrated. The adaptive algorithm was found to be energy stable with a decreasing free energy functional via the spinodal decomposition while the algorithm was found to be globally mass-conserving on the evolution of the radii of two circles in a square domain.
To further characterize the performance and the efficiency of the adaptive scheme, free-surface sloshing of water in a tank was considered for the systematic numerical experiments. Some of the salient findings from the sloshing tank problem are: (i) the adaptive procedure improved the mass conservation by three times compared to the non-adaptive grid, and reduced the computational time by 40 percent, (ii) the boundedness property of the scheme was also noticed in the extremum values of the order parameter, (iii) the proposed scheme evolved the fluid-fluid interface with less residual error ($\eta$) compared to the uniform non-adaptive grid, and (iv) the study of the dependency of the interface thickness parameter $\varepsilon$ on the error indicator tolerance concluded that sharper interfaces tend to require more stringent tolerance criterion to yield accurate results. Finally, the adaptive procedure was applied to the dam break problem to demonstrate its capability to handle practical problems with topological changes. 

{\bf{Acknowledgements}}

The first author would like to thank for the financial support from  National Research Foundation through Keppel-NUS Corporate Laboratory. The conclusions put forward reflect the views of the authors alone, and not necessarily those of the institutions.

\appendix
\setcounter{equation}{0} 
\setcounter{figure}{0}
\section{Residual error estimates for the Allen-Cahn equation}
\label{error_estimates}
Let the time interval $[0,T]$ be subdivided into intervals of length $\Delta t= t^\mathrm{n+1} - t^\mathrm{n}, \mathrm{n}=0,1,2,...$. For each time $t^\mathrm{n}$, an affine equivalent, admissible and shape-regular partition $\mathcal{T}$ of the domain $\Omega$ which consists of elements $\Omega_\mathrm{e}$, is chosen such that $\Omega = \cup_\mathrm{e=1}^\mathrm{n_{el}} \Omega_\mathrm{e}$ and $\emptyset = \cap_\mathrm{e=1}^\mathrm{n_{el}} \Omega_\mathrm{e}$ where $\mathrm{n}_\mathrm{el}$ is the number of elements. Let the finite element space of trial solution in the partition be denoted by $\mathcal{S}^\mathrm{h} \in \mathbb{V}^\mathrm{h}$ and the space of test functions be represented by $\mathcal{V}^\mathrm{h} \in \mathbb{V}^\mathrm{h}$, where $\mathbb{V}^\mathrm{h} \in \mathbb{V}$ is the finite element space on the partition $\mathcal{T}$.

The variational form of the semi-discrete Allen-Cahn equation under the generalized-$\alpha$ framework is given as (Eq.~(\ref{AC_variational})): find $\phi_\mathrm{h} \in \mathcal{S}^\mathrm{h}$ such that for all $w_\mathrm{h}\in \mathcal{V}^\mathrm{h}$,
\begin{align} \label{Res1}
	&\int_\Omega \bigg( w_\mathrm{h}\partial_t{\phi}_\mathrm{h} + w_\mathrm{h}(\boldsymbol{u}\cdot\nabla\phi_\mathrm{h}) + \nabla w_\mathrm{h}\cdot(k\nabla\phi_\mathrm{h} ) + w_\mathrm{h}s\phi_\mathrm{h} - w_\mathrm{h}f \bigg) \mathrm{d}\Omega \nonumber \\
	&+ \displaystyle\sum_\mathrm{e=1}^\mathrm{n_{el}}\int_{\Omega_\mathrm{e}}\bigg( \big(\boldsymbol{u}\cdot\nabla w_\mathrm{h} \big)\tau \big( \partial_t{\phi}_\mathrm{h} + \boldsymbol{u}\cdot\nabla\phi_\mathrm{h} - \nabla\cdot(k\nabla\phi_\mathrm{h}) + s\phi_\mathrm{h} -f \big) \bigg) \mathrm{d}\Omega_\mathrm{e} = 0, 	
\end{align}
For simplicity, we only consider the Galerkin discretization of the Allen-Cahn equation. We have followed a similar procedure as carried out in \cite{Verfurth}.
Equation~(\ref{Res1}) can be written as:
\begin{align}
&\int_\Omega w_\mathrm{h}\partial_t\phi_\mathrm{h} \mathrm{d}\Omega + \int_\Omega w_\mathrm{h}(\boldsymbol{u}\cdot\nabla\phi_\mathrm{h}) \mathrm{d}\Omega +\int_{\Omega} \nabla w_\mathrm{h}\cdot(k \nabla\phi_\mathrm{h}) \mathrm{d}\Omega + \int_\Omega w_\mathrm{h}s\phi_\mathrm{h} \mathrm{d}\Omega - \int_\Omega w_\mathrm{h}f\mathrm{d}\Omega, \\
	 =&\int_\Omega w_\mathrm{h}\partial_t\phi_\mathrm{h} \mathrm{d}\Omega + \int_\Omega w_\mathrm{h}(\boldsymbol{u}\cdot\nabla\phi_\mathrm{h}) \mathrm{d}\Omega
	 - \int_{\Omega} w_\mathrm{h}(k \nabla^2\phi_\mathrm{h}) \mathrm{d}\Omega + \displaystyle\sum_\mathrm{e=1}^\mathrm{n_{el}} \int_{\partial\Omega_\mathrm{e}} w_\mathrm{h}k (\mathbf{n}_{\partial\Omega_\mathrm{e}}\cdot \nabla\phi_\mathrm{h}) \mathrm{d}\Omega_\mathrm{e}\nonumber \\
	 +&\int_\Omega w_\mathrm{h}s\phi_\mathrm{h} \mathrm{d}\Omega - \int_\Omega w_\mathrm{h}f\mathrm{d}\Omega, \\
	 =&\int_\Omega w_\mathrm{h}( \partial_t\phi_\mathrm{h} + \boldsymbol{u}\cdot\nabla\phi_\mathrm{h} - k\nabla^2\phi_\mathrm{h} + s\phi_\mathrm{h} - f)\mathrm{d}\Omega + \displaystyle\sum_\mathrm{e=1}^\mathrm{n_{el}} \int_{\partial\Omega_\mathrm{e}} w_\mathrm{h}k (\mathbf{n}_{\partial\Omega_\mathrm{e}}\cdot \nabla\phi_\mathrm{h}) \mathrm{d}\Omega_\mathrm{e} \\
	 =& \displaystyle\sum_\mathrm{e=1}^\mathrm{n_{el}} \int_{\Omega_\mathrm{e}} w_\mathrm{h} \mathcal{R}_{\Omega_\mathrm{e}}(\phi_\mathrm{h}) \mathrm{d}\Omega_\mathrm{e} + \displaystyle\sum_{E\in \mathcal{E}} \int_{E} w_\mathrm{h} \mathcal{R}_{E}(\phi_\mathrm{h}) \mathrm{d}E  \label{Res2}
\end{align}
where $\mathcal{R}_{\Omega_\mathrm{e}}$ and $\mathcal{R}_{E}$ are the element and edge based residuals and $\mathbf{n}_{\partial\Omega_\mathrm{e}}$ is the unit normal to the element boundary. Using the property of Galerkin orthogonality in Eq.~(\ref{Res2}),
\begin{align}
	&\int_\Omega w_\mathrm{h}\partial_t\phi_\mathrm{h} \mathrm{d}\Omega + \int_\Omega w_\mathrm{h}(\boldsymbol{u}\cdot\nabla\phi_\mathrm{h}) \mathrm{d}\Omega +\int_{\Omega} \nabla w_\mathrm{h}\cdot(k \nabla\phi_\mathrm{h}) \mathrm{d}\Omega + \int_\Omega w_\mathrm{h}s\phi_\mathrm{h} \mathrm{d}\Omega - \int_\Omega w_\mathrm{h}f\mathrm{d}\Omega \nonumber \\
	=& \displaystyle\sum_\mathrm{e=1}^\mathrm{n_{el}} \int_{\Omega_\mathrm{e}} (w_\mathrm{h} - w_\mathcal{T}) \mathcal{R}_{\Omega_\mathrm{e}}(\phi_\mathrm{h}) \mathrm{d}\Omega_\mathrm{e} + \displaystyle\sum_{E\in \mathcal{E}} \int_{E} (w_\mathrm{h} - w_\mathcal{T})\mathcal{R}_{E}(\phi_\mathrm{h}) \mathrm{d}E.
\end{align}
Let the first line of the above equation be denoted as $\mathrm{LHS}$. Choosing $w_\mathcal{T} = \mathcal{I}_\mathcal{T}w_\mathrm{h}$ where $\mathcal{I}_\mathcal{T}$ is the quasi-interpolation operator and using the Cauchy-Schwarz inequality for integrals,
\begin{align}
	\mathrm{LHS} \leq &\displaystyle\sum_\mathrm{e=1}^\mathrm{n_{el}} ||\mathcal{R}_{\Omega_\mathrm{e}}||_{\Omega_\mathrm{e}}||w_\mathrm{h} - \mathcal{I}_\mathcal{T}w_\mathrm{h}||_{\Omega_\mathrm{e}} + \displaystyle\sum_{E\in \mathcal{E}} ||\mathcal{R}_{E}||_{E}||w_\mathrm{h} - \mathcal{I}_\mathcal{T}w_\mathrm{h}||_{E}
\end{align}
The properties of the interpolation operator $\mathcal{I}_\mathcal{T}$ further reduces the inequality to
 \begin{align}
	\mathrm{LHS}\leq \displaystyle\sum_\mathrm{e=1}^\mathrm{n_{el}} ||\mathcal{R}_{\Omega_\mathrm{e}}||_{\Omega_\mathrm{e}}c_1 h_{\Omega_\mathrm{e}} ||w_\mathrm{h}||_{H^1(\tilde{\omega}_{\Omega_\mathrm{e}})} + \displaystyle\sum_{E\in \mathcal{E}} ||\mathcal{R}_{E}||_Ec_2 h_{E}^{1/2} ||w_\mathrm{h}||_{H^1(\tilde{\omega}_{E})}
\end{align}
where $h_{\Omega_\mathrm{e}}$ is the diameter of the element $\Omega_\mathrm{e}$, $h_{E}$ is the length of the edge, $c_1$ and $c_2$ are constants which depend upon the shape of the partition $\mathcal{T}$, and $\tilde{\omega}_{\Omega_\mathrm{e}}$ and $\tilde{\omega}_{E}$ are defined as:
\begin{align}
	\tilde{\omega}_{\Omega_\mathrm{e}} = \bigcup_{N_{\Omega_\mathrm{e}}\bigcap N_{\Omega^{'}_\mathrm{e}} \neq \emptyset} \Omega^{'}_\mathrm{e},\qquad \tilde{\omega}_{E} = \bigcup_{N_{E}\bigcap N_{\Omega^{'}_\mathrm{e}} \neq \emptyset} \Omega^{'}_\mathrm{e}
\end{align}
where $N_{\Omega_\mathrm{e}}$ are the vertices of $\Omega_\mathrm{e}$ and $N_{E}$ are the vertices of an edge $E$. Again with the help of the Cauchy-Schwarz inequality for the sums,
\begin{align}
	\mathrm{LHS} \leq \mathrm{max}\{c_1,c_2\} \bigg\{ \displaystyle\sum_\mathrm{e=1}^\mathrm{n_{el}} h^2_{\Omega_\mathrm{e}}||\mathcal{R}_{\Omega_\mathrm{e}}||^2_{\Omega_\mathrm{e}} + \displaystyle\sum_{E\in \mathcal{E}} h_{E} ||\mathcal{R}_{E}||^2_{E}  \bigg\}^{1/2}\bigg\{ \displaystyle\sum_\mathrm{e=1}^\mathrm{n_{el}} ||w_\mathrm{h}||^2_{H^1(\tilde{\omega}_{\Omega_\mathrm{e}})} + \displaystyle\sum_{E\in \mathcal{E}} ||w_\mathrm{h}||^2_{H^1(\tilde{\omega}_{E})} \bigg\}^{1/2}
\end{align}
Moreover, using the shape regularity assumption for the partition $\mathcal{T}$,
\begin{align}
	\bigg\{ \displaystyle\sum_\mathrm{e=1}^\mathrm{n_{el}} ||w_\mathrm{h}||^2_{H^1(\tilde{\omega}_{\Omega_\mathrm{e}})} + \displaystyle\sum_{E\in \mathcal{E}} ||w_\mathrm{h}||^2_{H^1(\tilde{\omega}_{E})} \bigg\}^{1/2} \leq c||w_\mathrm{h}||_{H^1(\Omega)}
\end{align}
Therefore for some constant $c^*$,
\begin{align}
	\mathrm{LHS} \leq c^* \bigg\{ \displaystyle\sum_\mathrm{e=1}^\mathrm{n_{el}} h^2_{\Omega_\mathrm{e}}||\mathcal{R}_{\Omega_\mathrm{e}}||^2_{\Omega_\mathrm{e}} + \displaystyle\sum_{E\in \mathcal{E}} h_{E} ||\mathcal{R}_{E}||^2_{E}  \bigg\}^{1/2}||w||_{H^1(\Omega)}.
\end{align} 
Finally,  we use the residual-error equivalence that the norm of the error in $\mathbb{V}$ is bounded from above and below by the norm of the residual in the dual space $\mathbb{V}'$ which is established using the Friedrichs and Cauchy-Schwarz inequalities, which gives
\begin{align}
	||\phi - \phi_\mathrm{h}|| \leq c^* \bigg\{ \displaystyle\sum_\mathrm{e=1}^\mathrm{n_{el}} h^2_{\Omega_\mathrm{e}}||\mathcal{R}_{\Omega_\mathrm{e}}||^2_{\Omega_\mathrm{e}} + \displaystyle\sum_{E\in \mathcal{E}} h_{E} ||\mathcal{R}_{E}||^2_{E}  \bigg\}^{1/2}||w||_{H^1(\Omega)}.
\end{align} 

\section{The positivity condition for convective transport equation}
\label{positivity_derivation}
We present a brief description of the positivity condition for the stabilized finite element method.
Consider a simplified form of Eq.~(\ref{CAC}) with only convection effects:
 \begin{align}
	\partial_t\phi + \boldsymbol{u}\cdot\nabla\phi = 0.
\end{align} 
The positivity preserving variational formulation of the above problem will lead to the following finite element based stencil form:
\begin{align}
	\phi^\mathrm{n+1}_{i} = \phi^\mathrm{n}_{i} + C^+ (\phi^\mathrm{n}_{i+1} - \phi^\mathrm{n}_{i}) - C^- (\phi^\mathrm{n}_{i}-\phi^\mathrm{n}_{i-1}),
\end{align}
where 
\begin{align}
	C^+ &= -\frac{u\Delta t}{2h} + \frac{u^2\tau\Delta t}{h^2} + \frac{\chi\Delta t}{h^2}\frac{|\mathcal{R}(\phi)|}{|\nabla\phi|}\mathrm{max}\bigg\{ \frac{uh}{2} - \tau u^2,0 \bigg\}, \\
	C^- &= \frac{u\Delta t}{2h} + \frac{u^2\tau\Delta t}{h^2} + \frac{\chi\Delta t}{h^2}\frac{|\mathcal{R}(\phi)|}{|\nabla\phi|}\mathrm{max}\bigg\{ \frac{uh}{2} - \tau u^2,0 \bigg\},
\end{align}
where $h$ is the element length in one-dimension, $\Delta t$ is the time step, $\tau$ is the linear stabilization parameter given by Eq.~(\ref{tau_eqn}) and $\chi = 1/u$. All the parameters are defined by taking the diffusion and reaction coefficients to be null in the transient convection-diffusion-reaction equation. For the scheme to be positivity preserving, the coefficients $C^+$ and $C^-$ need to satisfy
\begin{align}
	C^+\geq 0,\qquad C^-\geq 0,\qquad C^++C^- \leq 1.
\end{align}
Next, we consider two scenarios as follows:

\textbf{Case 1}: Element nodes far from the discontinuity or less convection-dominated regime, i.e., either $\mathcal{R}(\phi) \to 0$ or $uh/2 - \tau u^2 < 0$. For these conditions, the coefficients will be
\begin{align}
	C^+ &= -\frac{u\Delta t}{2h} + \frac{u^2\tau\Delta t}{h^2},\\
	C^- &= \frac{u\Delta t}{2h} + \frac{u^2\tau\Delta t}{h^2},\\
	C^+ + C^- &= \frac{2u^2\tau\Delta t}{h^2} = \bigg( \frac{u\Delta t}{h}\bigg)\bigg( \frac{2u\tau}{h}\bigg),
\end{align}
Now we satisfy the positivity condition for this case.
\begin{align}
	\frac{uh}{2} - \tau u^2 &< 0 \implies \frac{u\Delta t}{2h} - \frac{\tau u^2\Delta t}{h^2} < 0 \implies -\frac{u\Delta t}{2h} + \frac{\tau u^2\Delta t}{h^2} > 0 \implies C^+ > 0
\end{align}
Since $u$, $\Delta t$, $h$ and $\tau$ are all positive, $C^->0$. Furthermore, the stabilization parameter $\tau$ can be expressed for the one-dimensional problem as
\begin{align}
	\tau = \bigg[ \bigg( \frac{2}{\Delta t} \bigg)^2 + \bigg( \frac{2u}{h}\bigg)^2 \bigg]^{-1/2} \implies \frac{2u\tau}{h} = \bigg[ \bigg(\frac{h}{u\Delta t}\bigg)^2 + 1\bigg]^{-1/2}.
\end{align}
The explicit scheme employed here satisfies the Courant-Friedrichs-Lewy condition for stability, i.e., $Co = u\Delta t/h \leq 1$. Therefore,
\begin{align}
	\frac{2u\tau}{h} = \bigg[ \bigg(\frac{1}{Co}\bigg)^2 + 1\bigg]^{-1/2} < 1.
\end{align}
which implies that $C^++C^- \leq 1$ and the scheme satisfies the positivity condition. The definition of the stabilization parameter $\tau$ puts a bound to the term $2u\tau/h$ while satisfying the positivity condition.

\textbf{Case 2}: Element nodes near the discontinuity with optimal discrete upwind operation, i.e., $uh/2 - \tau u^2 > 0$ and $\chi|\mathcal{R}(\phi)|/|\nabla\phi|\approx 1$. The coefficients can be written as
\begin{align}
	C^+ &= -\frac{u\Delta t}{2h} + \frac{u^2\tau\Delta t}{h^2} +\frac{u\Delta t}{2h} - \frac{u^2\tau\Delta t}{h^2} = 0,\\
	C^- &= \frac{u\Delta t}{2h} + \frac{u^2\tau\Delta t}{h^2} +\frac{u\Delta t}{2h} - \frac{u^2\tau\Delta t}{h^2} = \frac{u\Delta t}{h} > 0,\\
	C^+ + C^- &= \frac{u\Delta t}{h} \leq 1.
\end{align}
We have employed the Courant-Friedrichs-Lewy condition and the fact that $Co >0$ for satisfying the inequalities above. The above two cases show the behaviour of the presented variational scheme to maintain the positivity property. The nonlinear residual term acts as a limiter function to regulate the stabilization terms.

\bibliographystyle{unsrt}
\bibliography{citations}

\begin{thebibliography}{10}

\bibitem{Khatavkar}
V.~V. Khatavkar, P.~D. Anderson, P.~C. Duineveld, and H.~H.~E. Meijer.
\newblock Diffuse interface modeling of droplet impact on a pre-patterned solid
  surface.
\newblock {\em Macromolecular Rapid Communications}, 26(4):298--303, 2005.

\bibitem{Fried}
E.~Fried and M.~E. Gurtin.
\newblock Dynamic solid-solid transitions with phase characterized by an order
  parameter.
\newblock {\em Physica D: Nonlinear Phenomena}, 72(4):287 -- 308, 1994.

\bibitem{jfm_phasefield_mit}
Xie F, X.~Zheng, M.~S. Triantafyllou, Y.~Constantinides, Y.~Zheng, and G.~E.
  Karniadakis.
\newblock Direct numerical simulations of two-phase flow in an inclined pipe.
\newblock {\em Journal of Computational Physics}, 825:189 -- 207, 2017.

\bibitem{jcp_breakingwave}
T.~Li, P.~Troch, and J.~De~Rouck.
\newblock Interactions of breaking waves with a current over cut cells.
\newblock {\em Journal of Computational Physics}, 223:865 -- 897, 2007.

\bibitem{Unverdi}
S.~O. Unverdi and G.~Tryggvason.
\newblock A front-tracking method for viscous, incompressible, multi-fluid
  flows.
\newblock {\em Journal of Computational Physics}, 100(1):25 -- 37, 1992.

\bibitem{donea}
J.~Donea.
\newblock Arbitrary {L}agrangian-{E}ulerian finite element methods.
\newblock 192:4195--4215, 1983.

\bibitem{Osher}
S.~Osher and J.~A. Sethian.
\newblock Fronts propagating with curvature-dependent speed: {A}lgorithms based
  on {H}amilton-{J}acobi formulations.
\newblock {\em Journal of Computational Physics}, 79(1):12 -- 49, 1988.

\bibitem{Hirt}
C.~W. Hirt and B.~D. Nichols.
\newblock Volume of fluid ({VOF}) method for the dynamics of free boundaries.
\newblock {\em Journal of Computational Physics}, 39(1):201 -- 225, 1981.

\bibitem{Kim}
J.~Kim.
\newblock Phase-field models for multi-component fluid flows.
\newblock {\em Communications in Computational Physics}, 12(3):613--661, 009
  2012.

\bibitem{Lohner_adaptivity}
R.~L\"{o}hner.
\newblock {\em Applied Computational Fluid Dynamics Techniques}, pages
  269--297.
\newblock John Wiley \& Sons, Ltd, 2008.

\bibitem{Bank}
R.~E. Bank and R.~K. Smith.
\newblock A posteriori error estimates based on hierarchical bases.
\newblock {\em SIAM Journal on Numerical Analysis}, 30(4):921--935, 1993.

\bibitem{ZZ}
O.~C. Zienkiewicz and J.~Z. Zhu.
\newblock A simple error estimator and adaptive procedure for practical
  engineerng analysis.
\newblock {\em International Journal for Numerical Methods in Engineering},
  24(2):337--357, 1987.

\bibitem{Verfurth_1}
R.~Verf\"{u}rth.
\newblock A posteriori error estimation and adaptive mesh-refinement
  techniques.
\newblock {\em J. Comput. Appl. Math.}, 50(1-3):67--83, May 1994.

\bibitem{Verfurth_2}
R.~Verf\"{u}rth.
\newblock {\em A review of a posteriori error estimation and adaptive
  mesh-refinement techniques}.
\newblock Wiley-Teubner, 1996.

\bibitem{Verfurth}
R.~Verf\"{u}rth.
\newblock Adaptive finite element methods.
\newblock Lecture notes, Fakult$\mathrm{\ddot{a}}$t f$\mathrm{\ddot{u}}$r
  Mathematik, Ruhr-Universit$\mathrm{\ddot{a}}$t, Bochum, Germany.

\bibitem{Schmidt}
A.~Schmidt and K.~G. Siebert.
\newblock {\em Design of adaptive finite element software}.
\newblock Springer, 2005.

\bibitem{Kessler}
D.~Kessler, R.~H. Nochetto, and A.~Schmidt.
\newblock A posteriori error control for the {A}llen--{C}ahn problem:
  {C}ircumventing {G}ronwall's inequality.
\newblock {\em ESAIM: Mathematical Modelling and Numerical Analysis},
  38(1):129–142, 2004.

\bibitem{Bartels_1}
S.~Bartels, R.~M\"{u}ller, and C.~Ortner.
\newblock Robust a priori and a posteriori error analysis for the approximation
  of allen–cahn and {G}inzburg-{L}andau equations past topological changes.
\newblock {\em SIAM Journal on Numerical Analysis}, 49(1):110--134, 2011.

\bibitem{Georgoulis}
E.~H. Georgoulis and C.~Makridakis.
\newblock On a posteriori error control for the {A}llen-{C}ahn problem.
\newblock {\em Mathematical Methods in the Applied Sciences}, 37(2):173--179,
  2014.

\bibitem{Bartels_2}
S.~Bartels.
\newblock A posteriori error analysis for time-dependent {G}inzburg-{L}andau
  type equations.
\newblock {\em Numerische Mathematik}, 99(4):557--583, Feb 2005.

\bibitem{Feng2005}
X.~Feng and H.~Wu.
\newblock A posteriori error estimates and an adaptive finite element method
  for the {A}llen-{C}ahn equation and the mean curvature flow.
\newblock {\em Journal of Scientific Computing}, 24(2):121--146, Aug 2005.

\bibitem{Zhang_2}
J.~Zhang and Q.~Du.
\newblock Numerical studies of discrete approximations to the {A}llen-{C}ahn
  equation in the sharp interface limit.
\newblock {\em SIAM Journal on Scientific Computing}, 31(4):3042--3063, 2009.

\bibitem{Vasconcelos}
D.~F.~M. Vasconcelos, A.~L. Rossa, and A.~L. G.~A. Coutinho.
\newblock A residual-based {A}llen-{C}ahn phase field model for the mixture of
  incompressible fluid flows.
\newblock {\em International Journal for Numerical Methods in Fluids},
  75(9):645--667, 2014.

\bibitem{Zhang}
Z.~Zhang and H.~Tang.
\newblock An adaptive phase field method for the mixture of two incompressible
  fluids.
\newblock {\em Computers \& Fluids}, 36(8):1307 -- 1318, 2007.

\bibitem{libMesh}
B.~S. Kirk, J.~W. Peterson, R.~H. Stogner, and G.~F. Carey.
\newblock libmesh: a {C}++ library for parallel adaptive mesh
  refinement/coarsening simulations.
\newblock {\em Engineering with Computers}, 22(3):237--254, Dec 2006.

\bibitem{Chen_3}
L.~Chen and C.~Zhang.
\newblock A coarsening algorithm on adaptive grids by newest vertex bisection
  and its applications.
\newblock {\em Journal of Computational Mathematics}, 28(6):767--789, 2010.

\bibitem{AC_JCP}
V.~Joshi and R.~K. Jaiman.
\newblock A positivity preserving and conservative variational scheme for
  phase-field modeling of two-phase flows.
\newblock {\em Journal of Computational Physics}, 360:137 -- 166, 2018.

\bibitem{PPV}
V.~Joshi and R.~K. Jaiman.
\newblock A positivity preserving variational method for multi-dimensional
  convection-diffusion-reaction equation.
\newblock {\em Journal of Computational Physics}, 339:247 -- 284, 2017.

\bibitem{Chen_2}
L.~Chen and C.~Zhang.
\newblock {AFEM}@{MATLAB}: {A} {MATLAB} package of adaptive finite element
  methods.
\newblock Technical report, University of Maryland.

\bibitem{Brackbill}
J.~U. Brackbill, D.~B. Kothe, and C.~Zemach.
\newblock A continuum method for modeling surface tension.
\newblock {\em Journal of Computational Physics}, 100(2):335 -- 354, 1992.

\bibitem{Gen_alpha}
J.~Chung and G.~M. Hulbert.
\newblock A time integration algorithm for structural dynamics with improved
  numerical dissipation: {T}he generalized-$\alpha$ method.
\newblock {\em Journal of Applied Mechanics}, 60(2):371--375, 1993.

\bibitem{Hughes_X}
F.~Shakib, T.~J.~R. Hughes, and Z.~Johan.
\newblock A new finite element formulation for computational fluid dynamics:
  {X}. {T}he compressible {E}uler and {N}avier-{S}tokes equations.
\newblock {\em Computer Methods in Applied Mechanics and Engineering},
  89:141--219, 1991.

\bibitem{France_II}
L.~Franca and S.~Frey.
\newblock Stabilized finite element methods: {II}. {T}he incompressible
  {N}avier-{S}tokes equations.
\newblock {\em Computer Methods in Applied Mechanics and Engineering},
  99:209--233, 1992.

\bibitem{Hughes_inv_est}
I.~Harari and T.~J.~R. Hughes.
\newblock What are {C} and h?: {I}nequalities for the analysis and design of
  finite element methods.
\newblock {\em Computer Methods in Applied Mechanics and Engineering},
  97(2):157--192, 1992.

\bibitem{Johnson}
C.~Johnson.
\newblock {\em Numerical solutions of partial differential equations by the
  finite element method}.
\newblock Cambridge {U}niversity {P}ress, 1987.

\bibitem{Hsu}
M.~Hsu, Y.~Bazilevs, Calo V., T.~Tezduyar, and T.~J.~R. Hughes.
\newblock Improving stability of multiscale formulations of fluid flow at small
  time steps.
\newblock {\em Computer Methods in Applied Mechanics and Engineering},
  199:828--840, 2010.

\bibitem{Tezduyar_1}
T.~E. Tezduyar, S.~Mittal, S.~Ray, and R.~Shih.
\newblock Incompressible flow computations with stabilized bilinear and linear
  equal-order interpolation velocity-pressure elements.
\newblock {\em Computer Methods in Applied Mechanics and Engineering},
  95:221--242, 1992.

\bibitem{Hughes_V}
T.~J.~R. Hughes, L.~P. Franca, and M.~A. Balestra.
\newblock A new finite element formulation for computational fluid dynamics:
  {V}. {C}ircumventing the {B}abuska-{B}rezzi condition: {A} stable
  {P}etrov-{G}alerkin formulation of the {S}tokes problem accommodating
  equal-order interpolations.
\newblock {\em Computer Methods in Applied Mechanics and Engineering},
  59:85--99, 1986.

\bibitem{Tezduyar_stab}
T.~E. Tezduyar.
\newblock Finite elements in fluids: {S}tabilized formulations and moving
  boundaries and interfaces.
\newblock {\em Computers and Fluids}, 36:191--206, 2007.

\bibitem{allen_cahn}
S.~M. Allen and J.~W. Cahn.
\newblock A microscopic theory for antiphase boundary motion and its
  application to antiphase domain coarsening.
\newblock {\em Acta Metallurgica}, 27(6):1085 -- 1095, 1979.

\bibitem{Du}
Q.~Du and R.~A. Nicolaides.
\newblock Numerical analysis of a continuum model of phase transition.
\newblock {\em SIAM Journal on Numerical Analysis}, 28(5):1310--1322, 1991.

\bibitem{Harten}
A.~Harten.
\newblock High resolution schemes for hyperbolic conservation laws.
\newblock {\em Journal of Computational Physics}, 49:357--393, 1983.

\bibitem{Kuzmin}
D.~Kuzmin.
\newblock {\em A {G}uide to {N}umerical {M}ethods for {T}ransport {E}quations}.
\newblock Friedrich-Alexander-Universit$\ddot{\mathrm{a}}$t,
  Erlangen-N$\ddot{\mathrm{u}}$rnberg, 2010.

\bibitem{FCT}
D.~Kuzmin, R.~L$\mathrm{\ddot{o}}$hner, and S.~Turek.
\newblock {\em Flux-corrected {T}ransport: {P}rinciples, {A}lgorithms and
  {A}pplications}.
\newblock Springer, 2005.

\bibitem{shin_kim2014}
J.~Shin, Park~S. K., and J.~Kim.
\newblock A hybrid {FEM} for solving the {A}llen-{C}ahn equation.
\newblock {\em Applied Mathematics and Computation}, 244:606 -- 612, 2014.

\bibitem{Dorfler}
W.~D\"{o}rfler.
\newblock A convergent adaptive algorithm for {P}oisson's equation.
\newblock {\em SIAM Journal on Numerical Analysis}, 33(3):1106--1124, 1996.

\bibitem{H_G_Lee}
Hyun~Geun Lee.
\newblock High-order and mass conservative methods for the conservative
  {A}llen-{C}ahn equation.
\newblock {\em Computers \& Mathematics with Applications}, 72(3):620 -- 631,
  2016.

\bibitem{HUGHES20051141}
T.~J.~R. Hughes and G.~N. Wells.
\newblock Conservation properties for the {G}alerkin and stabilised forms of
  the advection-diffusion and incompressible {N}avier-{S}tokes equations.
\newblock {\em Computer Methods in Applied Mechanics and Engineering},
  194(9):1141 -- 1159, 2005.

\bibitem{doi:10.1137/100818583}
R.~J. Labeur and G.~N. Wells.
\newblock Energy stable and momentum conserving hybrid finite element method
  for the incompressible {N}avier-{S}tokes equations.
\newblock {\em SIAM Journal on Scientific Computing}, 34(2):A889--A913, 2012.

\bibitem{Martin_Moyce}
J.~C. Martin and W.~J. Moyce.
\newblock Part {IV}. {A}n experimental study of the collapse of liquid columns
  on a rigid horizontal plane.
\newblock {\em Philosophical Transactions of the Royal Society of London.
  Series A, Mathematical and Physical Sciences}, 244(882):312--324, 1952.

\bibitem{Ubbink_thesis}
O.~Ubbink.
\newblock {\em Numerical prediction of two fluid systems with sharp
  interfaces}.
\newblock PhD thesis, Imperial College of Science, Technology \& Medicine,
  1997.

\bibitem{Walhorn_thesis}
E.~Walhorn.
\newblock {\em Ein simultanes Berechnungsverfahren f${\ddot{u}}$r
  Fluid-Struktur-Wechselwirkungen mit finiten Raum-Zeit-Elementen}.
\newblock PhD thesis, Technische Universit${\ddot{a}}$t Braunschweig, 2002.

\bibitem{Fries_1}
H.~Sauerland and T-P. Fries.
\newblock The extended finite element method for two-phase and free-surface
  flows: {A} systematic study.
\newblock {\em Journal of Computational Physics}, 230(9):3369 -- 3390, 2011.

\bibitem{iFEM}
L.~Chen.
\newblock i{FEM}: {A}n innovative finite element method package in {MATLAB}.
\newblock Technical report, University of California at Irvine.

\end{thebibliography}

\end{document}